\documentclass[twocolumn,showpacs,
 amsmath,amssymb,
 aps,
pra,
standalone]{revtex4-1}

\usepackage{graphicx}
\usepackage{dcolumn}
\usepackage{bm}
\usepackage{subeqnarray}
\usepackage{cases}
\usepackage{hyperref}

\begin{document}

\preprint{APS/123-QED}

\title{Hysteretic Excitation in Non-collinear Antiferromagnetic Spin-Torque Oscillators: A Terminal Velocity Motion Perspective}

\author{Hao-Hsuan Chen}
\email{d95222014@ntu.edu.tw}
\email{haohsuan02@gmail.com}
\affiliation{Graduate School of Materials Science, National Yunlin University of Science and Technology, Douliou, 64002, Taiwan}

\author{Ching-Ming Lee}
\email{cmlee@yuntech.edu.tw}
\affiliation{Graduate School of Materials Science, National Yunlin University of Science and Technology, Douliou, 64002, Taiwan}





\date{\today}

\begin{abstract}
  We present a novel theoretical framework for non-collinear anti-ferromagnetic spin torque oscillators (NC-AFM STO) by unifying spin dynamics under the Poisson Bracket formalism. By shifting from traditional torque-based descriptions to a continuous operational symmetry perspective, we develop two complementary viewpoints: a vector perspective that identifies infinite degenerate Rigid Body Precession (RBP) states-arising from the fact that exchange energy is solely a function of the Rigid Body Rotational Transformation (RBRT) generator (i.e., total magnetic momentum)-and a particle perspective that decomposes the dynamics into Center-of-Mass (CM) translation and Relative Motion (RM) oscillation. Utilizing time-dependent RBRT and Rigid Body Uniform Translational Transformation (RBUTT) techniques, we analytically resolve the rapid ($\sim 10$ ps) transient evolution into a stable RBP state driven by SOT and damping. We further demonstrate that the out-of-plane anisotropy (OPA) component of the uniaxial crystalline anisotropy lifts the exchange degeneracy, triggering a long-term ($\sim 1$ ns) oscillatory decay toward a finalized steady state \textbf{s}, characterized by uniform spin $z$-components and a $120^\circ$ inter-spin locking angle. The dynamics of this state are accurately governed by our proposed Terminal Velocity Motion (TVM) model for CM variables [H.-H. Chen \textit{et al}., \textbf{arXiv:2305.11020} (2023)], in which exchange coupling is effectively transformed into kinetic energy with an exceptionally light effective mass. This model precisely predicts SOT-driven transients, hysteretic excitation process, and the dynamic phase diagram across the full current range. Finally, we account for the theoretical mismatch in the sub-critical current regime by identifying a 'Rigid-Body Breaking' effect: a surge in effective friction caused by the self-resonance of RM variables induced by the translation of the CM variable, as mediated by the in-plane anisotropy (IPA) component of the uniaxial crystalline anisotropy.
\end{abstract}

\maketitle


\section{\label{sec:level1}INTRODUCTION}
In the quest for next-generation spintronic devices, the focus of research has undergone a paradigm shift from traditional ferromagnets (FMs) to anti-ferromagnetic (AFM) and ferri-magnetic (FiM) systems. Unlike FMs, which are limited by gigahertz-scale dynamics and cross-talk issues arising from stray magnetic fields, AFMs and FiMs leverage strong internal exchange interactions to unlock sub-terahertz (sub-THz) and terahertz operational frequencies. These materials offer unique advantages, including ultra-fast spin dynamics, zero or negligible macroscopic magnetization, and high stability against external magnetic perturbations. Recent breakthroughs in collinear AFMs (e.g., CuMnAs, NiO)\cite{Ivanov2014,Cheng2014,Cheng2015,Cheng2016,Khymyn2017,Sulymenko2017,Zheng2024}, and compensated FiMs (e.g., GdFeCo)\cite{Zhao2015,Roschewsky2016,Mishra2017,Kim2017,Zhu2018,Wang2018,Lisenkov2019,SanchezTejerina2024}, non-collinear chiral AFMs (e.g., $\mathrm{Mn_{3}}$\emph{X} (\emph{X} = Sn, Ge, Ir, or Pt) and antiperovskites $\mathrm{Mn_{3}}$\emph{A}N(\emph{A} = Ga, Sn, or Ni))\cite{Nakatsuji2015,Yamane2019,Zhao2021,Shukla2022,Hu2024,Wu2024} have demonstrated their potential across a broad spectrum of applications. These include high-frequency auto-oscillators for telecommunication\cite{Cheng2016,Khymyn2017,Sulymenko2017,Shen2019,Shukla2022,Hu2024}, high-density magnetic random-access memory (MRAM) with zero stray fields\cite{Wadley2016}, and energy-efficient neuromorphic computing based on ultra-fast artificial neurons\cite{Sulymenko2018,Grollier2020}.

While collinear anti-ferromagnets (AFMs) have long been celebrated for their terahertz dynamics, they often suffer from 'read-out' challenges due to their high degree of magnetic symmetry.
In contrast, non-collinear chiral AFMs (NC-AFMs) have emerged as a superior platform, offering a decisive advantage in signal readability. Due to their unique chiral spin texture and the resulting large Berry curvature in momentum space, these materials exhibit the spin Hall effect \cite{Zhang2017,Zhang2018},
anomalous Hall Effect \cite{Chen2014,Nakatsuji2015,Liu2018}, and anomalous Nernst effect\cite{Guo2017,Zhou2020} even at room temperature.

This characteristic enables direct electrical detection of magnetic states and THz-scale oscillations with high sensitivity. Recent experimental milestones, such as the all-antiferromagnetic tunnel junction (AATJ) \cite{Qin2023,Chen2023} composed of NC-AFM/MgO/NC-AFM layers, have successfully demonstrated room-temperature tunneling magnetoresistance (TMR). These breakthroughs, combined with their robust electrical controllability
\cite{Hajiri2019,Zhao2021,Takeuchi2021,Shukla2022,Higo2022,Shukla2023}, efficient spin-current generation
\cite{ifmmodeZelseZfielezny2017,Bai2021,Nan2020,You2021}, and robust topological nature enhanced stability against external perturbations\cite{Wu2024}, underscore the potential of NC-AFMs as a highly readable, ultra-fast platform for next-generation spintronic logic and memory devices\cite{Song2021,ChenMonDec26000000CST2022}.

The theoretical exploration of non-collinear spintronics has evolved through several sophisticated stages, providing profound insights into steady-state behaviors. Early efforts, such as the seminal work by Zhao \textit{et al.} \cite{Zhao2021}, effectively leveraged the $120^{\circ}$ rotational symmetry to reduce the complexity of coupled Landau-Lifshitz-Gilbert (LLG) equations into an analytically tractable form. Subsequent models, notably the pendulum analogy introduced by Shukla and Rakheja \cite{Shukla2022}, offered an intuitive grasp of antiferromagnetic order and helped clarify THz auto-oscillation thresholds. More recently, Hu \textit{et al.} \cite{Hu2024} employed atomic-scale simulations to reveal intricate spin-flop transitions within an AATJ. While these established frameworks are indispensable, they often pre-impose specific geometric constraints or rely on Neel vector perspectives optimized for excitations near the exchange ground state. Consequently, such treatments may bypass the intrinsic, infinitely degenerate manifold of the exchange interaction itself, suggesting that a more generalized exploration of the underlying dynamical symmetries could further enrich our understanding of these systems across a broader range of excitations.

We suggest that a bottom-up approach, prioritizing the collective-mode symmetry of the pure exchange Hamiltonian, is essential before introducing specific anisotropies or torques. In this paper, we employ infinitesimal continuous transformations within a Poisson Bracket framework \cite{goldstein2014}
to systematically examine the system's operational symmetries. This analysis reveals that the exchange interaction inherently supports an infinitely degenerate manifold of rigid-body precession-a fundamental '\textit{collective mode}' that serves as the dynamical foundation for both collinear and NC-AFM even for FM systems. By partitioning the dynamics into \textit{Center-of-Mass} (CM) and \textit{Relative Motion} (RM) coordinates, we bridge the microscopic coupling with a macroscopic particle-like description, reminiscent of the universal modeling approach pioneered by Slavin and others\cite{Slavin2009}. This hierarchical strategy allows us to derive a Terminal Velocity Motion (TVM) model [Chen \textit{et al.}, arXiv:2305.11020] \cite{chen2023a} that remains valid across the entire current range, rather than being confined to the low-energy regime.

Unlike the first-order approximation methods ( i.e., Alder's equation \cite{Bhansali2009} or first-order Kuramoto model\cite{Acebron2005}) commonly employed \cite{Slavin2009}, the TVM model is derived by applying a Legendre transformation to the phase-space motion equations of a two-dimensional autonomous nonlinear frequency shift oscillator. By substituting the canonical momentum, which is equivalent to the variable power $p$ defined in the complex amplitude in the Universal Model\cite{Slavin2009}, with the azimuthal (phase) angular velocity, we obtain a second-order Newton-like equation of motion, which somehow can also be seen as the generalized version of the pendulum-like model
\cite{HaoHsuan2011,HaoHsuan2012,HaoHsuan2015,HaoHsuan2016,HaoHsuan2017,HaoHsuan2018,Chen2021,Markovic2022} or second-order Kuramoto model \cite{Peron2015,Barabash2021}, that preserves the full dynamic information of the original phase-space system (or the Universal Model). Within this framework, the effective mass directly encapsulates the nonlinear frequency-shift coefficient of the oscillator. This allows the TVM model to precisely predict intricate dynamical details across the entire current range-including hysteretic excitations, hysteretic mutual phase-locking, locked amplitude/phase angle, comprehensive dynamical phase diagrams, and even transient processes-which are often beyond the reach of conventional approximations.

Furthermore, we emphasize that in both FM and AFM systems, the dynamical modes driven by exchange interactions represent a fundamental collective mode. The robust exchange coupling imposes internal constraints that effectively reduce the system's dimensionality, allowing for a unified theoretical paradigm. This perspective is inherently consistent with the Universal Model proposed by Slavin and Tiberkevich \cite{Slavin2009}, which treats various spin-wave excitations-whether standing or propagating-as collective auto-oscillations. Within our TVM framework, the primary distinction between FM and AFM collective modes is elegantly captured by their effective mass signatures\cite{chen2023a}: FM systems exhibit a frequency red-shift (characterized by a negative effective mass), whereas AFM systems display a blue-shift (positive effective mass). Importantly, we clarify that the perceived absence of 'inertia' in traditional FM order parameters is typically an artifact of employing a Macrospin approximation that considers the system already locked in its parallel ground state. In contrast, our model recovers the essential dynamical inertia by treating the exchange-coupled multi-spin system as a unified collective entity. This distinction is crucial, as it reveals that 'inertia' is not a property absent in FMs, but rather one that is 'hidden' when the internal degrees of freedom are pre-constrained into a single vector.

The predictive power and practical relevance of such a collective-mode description (TVM model) have been elegantly validated by recent milestones in neuromorphic computing. Notably, the work by Markovi\'{c} \textit{et al.} \cite{Markovic2022}, has successfully demonstrated the use of easy-plane spin Hall nano-oscillators as spiking neurons. Their groundbreaking realization that these oscillators can emulate Josephson-junction-like phase dynamics as artificial neurons perfectly aligns with the analytical foundation laid by our TVM and earlier pendulum-based models. By integrating these pioneering experimental architectures into our theoretical framework, we believe that the TVM model is not only a tool for fundamental stability analysis but also an indispensable platform for designing high-speed, energy-efficient sub-THz spintronic intelligence.

In this work, we aim to develop a unified theoretical framework suitable for analyzing spin torque nano-oscillators (STNOs) based on either FM or AFM systems, including NC-AFMs. The paper is organized as follows: Sec. \ref{sec:2A} introduces the Landau-Lifshitz-Gilbert (LLG) equations with spin-orbit torques (SOTs) for three sub-lattice's spins in an NC-AFM. Sec. \ref{sec:2B} establishes the generalized LLG equations via the Poisson bracket formalism (Sec. \ref{sec:2B1}), identifying exchange couplings as generators of Rigid Body Rotational Transformation (RBRT) in the vector perspective (Sec. \ref{sec:2B2}) and Rigid Body Uniform Translational Transformation (RBUTT) in the particle perspective (Sec. \ref{sec:2B3}). These symmetries lead to an infinite manifold of degenerate states and allow for the decomposition of dynamics into center-of-mass (CM) and relative motion (RM) coordinates. Sec. \ref{sec:2B4} calculates the frequencies of these intrinsic dynamic states from the CM viewpoint. In Sec. \ref{sec:2C}, we apply time-dependent RBRT and RBUTT techniques to solve for stable SOT-driven degenerate states. Building on this, Sec. \ref{sec:2D1} estimates the transient time required for the total momentum to align with the spin polarizer. Sec. \ref{sec:2D2} develops the TVM model for a representative state-characterized by an equal inter-spin angle $120^{\circ}$ configuration-in the absence of uniaxial anisotropy to estimate a characteristic transient interval under such constraints. Sec. \ref{sec:2E} investigates the impact of the out-of-plane anisotropy (OPA) component of the in-plane crystalline uniaxial anisotropy, detailing how it lifts exchange degeneracy (Sec. \ref{sec:2E1}) and induces slow decaying oscillations toward the final steady state \textbf{s} (Sec. \ref{sec:2E2}). Sec. \ref{sec:2E3} develops the TVM model for the state \textbf{s} in the presence of the in-plane uniaxial anisotropy to analytically determine threshold current densities and characterize the intricate hysteretic excitation. Sec. \ref{sec:2F} and Sec. \ref{sec:2G} employ the macrospin and TVM numerical simulations to verify critical currents and display the hysteretic frequency-current density characteristics. Crucially, we provide an in-depth analysis of the breakdown of rigid-body dynamics in the sub-critical current regime, where CM motion through the in-plane anisotropy (IPA) component of the in-plane uniaxial anisotropy drives the RM degrees of freedom into self-resonances. Finally, Sec. \ref{sec:3} concludes the paper by summarizing our core findings and discussing their implications for future spintronic applications.

\section{\label{sec:2}Model and Theory}
\subsection{\label{sec:2A} Landau-Lifshitz-Gilbert Equations for a Non-collinear AFM }
\begin{figure}
\begin{center}
\includegraphics[width=6cm]{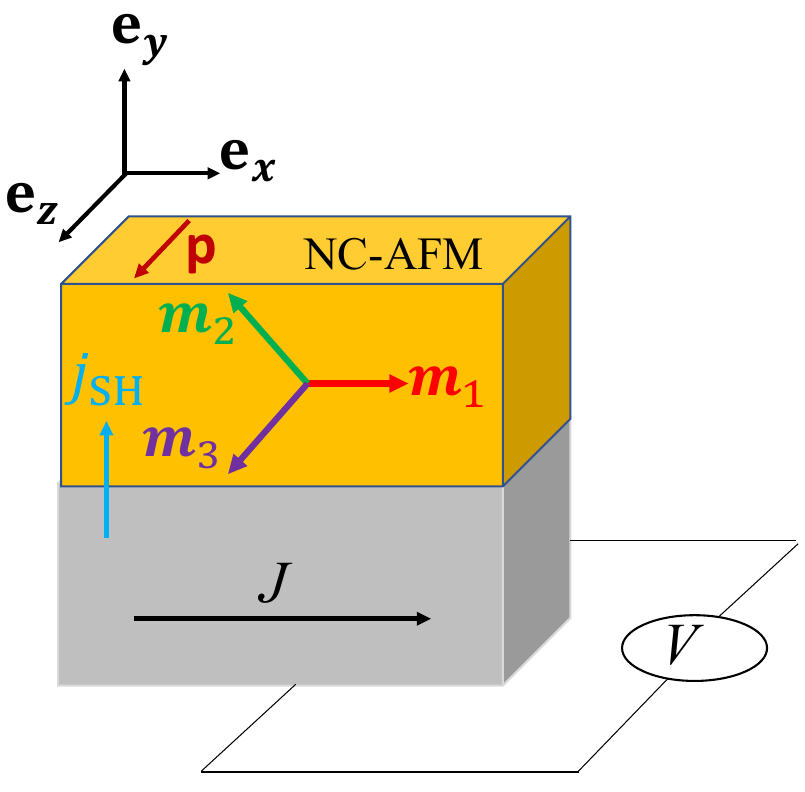}
\end{center}
\caption{(Color online) Schematic illustration of the NC-AFM spin-torque nano-oscillator (STNO) \cite{Zhao2021}. (a) The bilayer setup comprises a heavy metal (HM) base and a NC-AFM free layer. An in-plane charge current $J$ along the $x$-direction generates a vertical spin current via the spin Hall effect, with the spin polarization oriented along the $z$-axis. (b) Configuration of the three-sublattice spins $\mathbf{m}_{1}, \mathbf{m}_{2},$ and $\mathbf{m}_3$. Notably, the three uniaxial crystalline anisotropic symmetric axes for the three sublattices' spins respectively are all defined to lie on the $(x, y)$ plane.}
\label{SH_structure}
\end{figure}

As depicted in Fig. \ref{SH_structure}, the investigated device is a standard spintronic bilayer comprising a heavy metal (HM) and a non-collinear antiferromagnetic (NC-AFM) layer\cite{Zhao2021}. Under an applied in-plane charge current, the HM layer facilitates the injection of spin-polarized electrons into the NC-AFM film. The dynamics of the three coupled sublattice macrospins in an NC-AFM can be assumed to be governed by the Landau-Lifshitz-Gilbert (LLG) equation with the spin-orbit torques (SOTs)\cite{Zhao2021}:
\begin{eqnarray}
\frac{d \mathbf{m}_{i}}{d\tau}&=&-\left(\nabla_{\mathrm{m}_{i}} E\right) \times \mathbf{m}_{i}+\alpha \mathbf{m}_{i} \times\left(\frac{d \mathbf{m}_{i}}{d \tau}\right)\nonumber\\
&&+a_{J_{i}}\left(m_{zi}\right)\left[\mathbf{m}_{i} \times\left(\mathbf{m}_{i} \times \mathbf{p}\right)\right],
\label{LLGinLab}
\end{eqnarray}
where $\textbf{m}_{i}=\textbf{M}_{i}/M_{s}$  is the scaled unit vector of the sublattice magnetization and $M_{s}$ is the saturation magnetization. $\alpha$ is the Gilbert damping constant. $\tau$ is the scaled time $\tau=((\omega_{\mathrm{ex}}/2\pi)(2\pi))t$, where $\omega_{\mathrm{ex}} =\gamma A'_{\mathrm{ex}}/(\mu_{0}M_{s})$ with the gyromagnetic ratio $\gamma=1.76\times10^{11}$ $\textrm{s}^{-1}\textrm{T}^{-1}$, $A'_{\mathrm{ex}}$ the
 inter-sublattice exchange constant, $\mu_{0}$ the vacuum permeability, and $M_{s}$ the saturation magnetization of each sublattice.

 The last term on the right-hand side of Eq. (\ref{LLGinLab}) is the SOT term, where $\mathbf{p}=\mathbf{e}_{z}$ is the unit vector of the spin polarization. $a_{Ji}(\textbf{m}_{i})=a_{Ji0}$ is the scaled SOT strength, and $a_{Ji0}=(u/d)/\omega_{\mathrm{ex}}$ with $d$ being the thickness of AFM layer. $u$ has
 the dimension of velocity, $u=\mu_{\mathrm{B}}/(eM_{\mathrm{s}})\xi J$ with $\mu_{\mathrm{B}}$ being the Bohr magneton, $e$ being the element charge, and $J$ the current density. $\xi$ is the SOT efficiency which equals $T_{\mathrm{int}}\theta_{\mathrm{sH}}$
 \cite{Zhu2018a}, with $\theta_{\mathrm{sH}}$ being the spin Hall angle, and $T_{\mathrm{int}}$ the spin
 transparency of the interface \cite{Wang2020}.\\

 In this work, we choose to take the magnetic material parameters of a typical NC-AFM $\mathrm{M}\mathrm{n}_{\mathrm{3}}\mathrm{S}\mathrm{n}$ to do macrospin and simulations as well as theoretical analyses\cite{Zhao2021}, including  $A'_{\mathrm{ex}}=1\times10^{8}$ $\mathrm{Joul}/\mathrm{m}^{3}$, anisotropic strength coefficient $K=1.6\times10^{6}$ $\mathrm{Joul}/\mathrm{m}^{3}$, and $\mu_{0}M_{\mathrm{s}}=1.26$ T. Then, we have
 $\omega_{\mathrm{ex}}/(2\pi)=2.79$ THz and $\omega_{k}/(2\pi)=44.67$ GHz. Besides, other experimentally reasonable parameters are taken as $d=2$ nm, $\xi=0.32$, and $\alpha=0.01$.

Here, the total Hamiltonian $H$ (Energy $E$) scaled by the exchange coupling strength reads
\setlength\abovedisplayskip{6pt}
\setlength\belowdisplayskip{6pt}
\begin{eqnarray}
H(p_{i},\phi_{i},t)&=&H_{\mathrm{\mathrm{ex}}}+H_{\mathrm{u}},\nonumber\\
&=&\frac{A_{\mathrm{ex}}}{2}\sum_{i,j=1(i\neq j)}^{3}\mathbf{m}_{i}\cdot\mathbf{m}_{j}
-\frac{k}{2}\sum_{i=1}^{3}\left(\mathbf{m}_{i}\cdot \mathbf{e}_{\mathrm{u},i}\right)^{2},\nonumber\\
&=&\frac{A_{\mathrm{ex}}}{2}\sum_{i,j=1(i\neq j)}^{3}\bigg[\sqrt{\left(1-p_{i}^{2}\right)\left(1-p_{j}^{2}\right)}
\nonumber\\
&&\times\cos(\phi_{i}-\phi_{j})+p_{i}p_{j}\bigg]-\frac{k}{2}\sum_{i=1}^{3}\bigg[\left(1-p_{i}^{2}\right)\nonumber\\
&&\times\cos^{2}\left(\phi_{i}-(i-1)\frac{2\pi}{3}\right)\bigg],
\label{Hamiltonian}
\end{eqnarray}
with the exchange coupling strength $A_{\mathrm{ex}}=1$ being much larger than the uniaxial anisotropic strength $k=\omega_{k}/\omega_{\mathrm{ex}}\ll 1$. Also, the three unit vectors $\mathbf{e}_{\mathrm{u},i}$ defines the three symmetric axes of the anisotropy.

\subsection{\label{sec:2B} Continuous Symmetry Operation of Exchange Coupling }
\subsubsection{\label{sec:2B1} Generalization of the Conservative Part of the LLG Equation by Poisson Brackets }
In terms of the canonical variables (or cylindrical coordinates) $(\phi_{i},p_{i}\equiv-m_{zi})$, the LLG equation (Eq. (\ref{LLGinLab})) reads as \cite{chen2023a}
\setlength\abovedisplayskip{6pt}
\setlength\belowdisplayskip{6pt}
\begin{eqnarray}
\dot{p}_{i}&=&-\frac{\partial H}{\partial\phi_{i}}-\alpha(1-p_{i}^{2})\dot{\phi}_{i}+(1-p_{i}^{2})
a_{Ji}(-p_{i}),\nonumber\\
\dot{\phi}_{i}&=&\frac{\partial H}{\partial p_{i}}+\frac{\alpha}{(1-p_{i}^{2})}
\dot{p}_{i},
\label{canon_Eq_LLGS}
\end{eqnarray}
with $H\equiv E$.

Equation (\ref{canon_Eq_LLGS}) can be easily derived from the variational principle incorporating a dissipation function\cite{goldstein2014,Chen2021,chen2023a}: Firstly, by taking $\delta I=\delta\int
dtL(p,\dot{p},\phi,\dot{\phi})=0$ where the Lagrangian is defined as $L(p,\dot{p},\phi,\dot{\phi})=\sum_{i=1}^{n}p_{i}\dot{\phi}_{i}-H(p,\phi)$, we obtain the Hamiltonian equations (the conservative part of Eq. (\ref{canon_Eq_LLGS})) through the standard \textit{Euler-Lagrange} equations: $d(\partial L/\partial\dot{p}_{i})/dt-\partial L/\partial p_{i}=0$ as well as $d(\partial L/\partial\dot{\phi}_{i})/dt-\partial L/\partial \phi_{i}=0$; secondly, the full set of Eq. (\ref{canon_Eq_LLGS}) is obtained by utilizing the Euler-Lagrange equations incorporating the dissipative forces:
\setlength\abovedisplayskip{6pt}
\setlength\belowdisplayskip{6pt}
\begin{eqnarray}
\frac{d}{dt}\left(\frac{\partial L}{\partial\dot{p}_{i}}\right)-\frac{\partial L}{\partial p_{i}}+\frac{\partial F_{\mathrm{d}}}{\partial\dot{p}_{i}}=0,\nonumber\\
\frac{d}{dt}\left(\frac{\partial L}{\partial\dot{\phi}_{i}}\right)-\frac{\partial L}{\partial\phi_{i}}+\frac{\partial F_{\mathrm{d}}}{\partial\dot{\phi}_{i}}=0,
\label{Euler_Lag_dissipation}
\end{eqnarray}
with the dissipation function defined as: $F_{\mathrm{d}}=\sum_{i=1}^{n}(\alpha/2)[(1/(1-p_{i}^{2}))\dot{p}_{i}^{2}
+(1-p_{i}^{2})\dot{\phi}_{i}^{2}]-a_{Ji}(-p_{i})(1-p_{i}^{2})\dot{\phi}_{i}$,
Eq. (\ref{canon_Eq_LLGS}) can be obtained completely.

Note that the SOT term here being attributed as a pure dissipative force implies the dynamic states it drives being out-of-plane (OP) precessions that are normal to the spin polarization vector $\mathbf{p}$, where the SOT completely compensates with the damping force\cite{Chen2019a}. Actually, whether the SOT (or STT) can be treated as a damping-like force depends mainly on its net energy injection along the conserved trajectory during one period\cite{GBertotti2009nonlinear,Chen2019a}. The rigorous argument about this crucial point, which self-consistently validates our initial treatment, can be given in Sec. \ref{sec:2C}.

\begin{figure}
\begin{center}
\includegraphics[width=8.7cm]{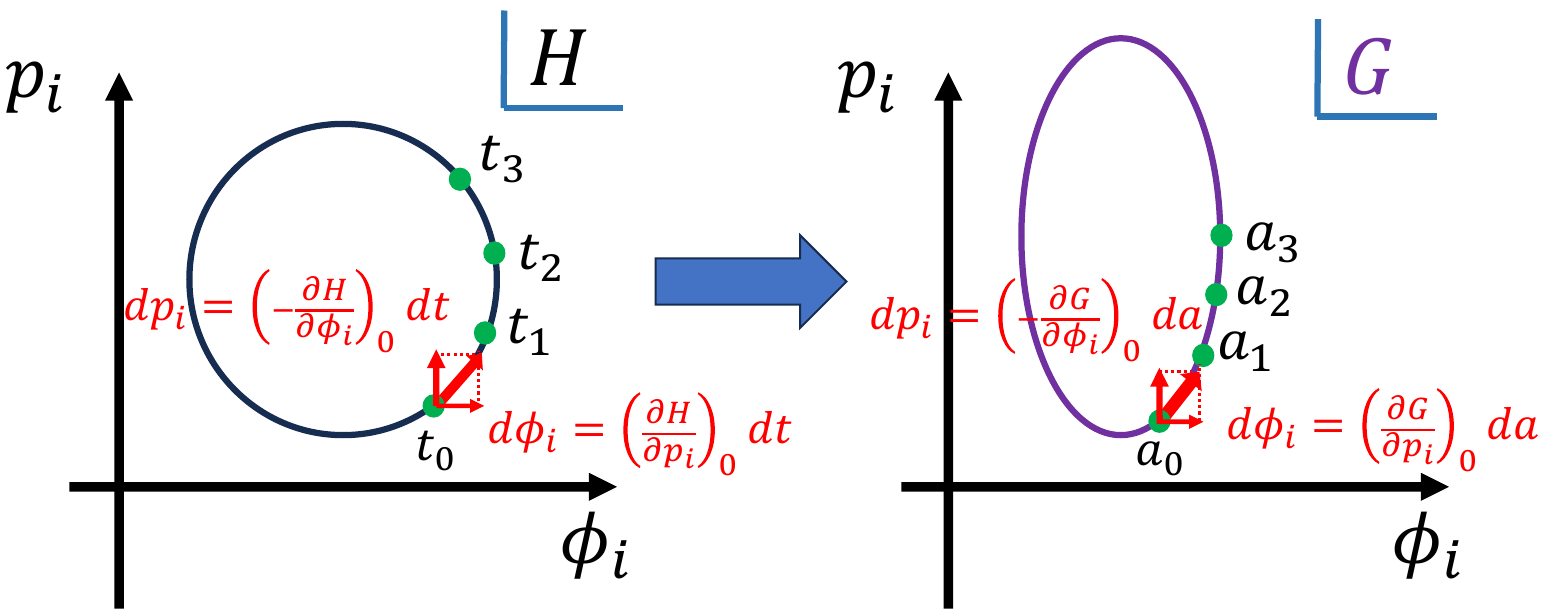}
\end{center}
\caption{(Color online) Schematics of a continuous transformative viewpoint of the Hamiltonian formulation (left panel) and its generalization to any other physical quantities (right panel). The black and purple curves denote the transformed trajectories driven by the generators $H$ and $G$, respectively. The green circles mark the state points on the two curves, which are designated by the parameters time $t$ and $a$, respectively. The bold red arrows indicate the infinitesimal displacements taken at the moment $t_0$ or at $a_0$ on the trajectories, respectively. Their respective projections $(d\phi_{i},dp_{i})$, shown by the thin red arrows, are $((\partial H/\partial p_{i})_{0}dt,(-\partial H/\partial\phi_{i})_{0}dt)$ and $((\partial G/\partial p_{i})_{0}da,(-\partial G/\partial\phi_{i})_{0}da)$, respectively.}
\label{sysmoper_of_H}
\end{figure}

 An interesting viewpoint full of inspiration can be provided here for the conservative part of Eq. (\ref{canon_Eq_LLGS}) as follows. If the time $t$ is replaced with a typical parameter $a$ to designate the state points on the $a$(time)-evolving trajectory $(\phi_{i}(a),p_{i}(a))$, as schematically displayed in Fig. \ref{sysmoper_of_H}, then the role the Hamiltonian plays is to give an infinitesimal displacement of the system on this path. Specifically, when $a$ changing from $a_{0}$ to $a_{0}+da$, the displacement is: $(d\phi_{i},dp_{i})=((\partial H/\partial p_{i})da,(-\partial H/\partial\phi_{i})da)$. More importantly, $H$ here can be generalized to any other physical quantities termed $G$ to drive the system to transform on various types of paths. These physical quantities are thereby treated as the \textit{generators} of \textit{continuous} (\textit{canonical}) \textit{transformations}\cite{goldstein2014}.

\begin{figure*}
\begin{center}
\includegraphics[width=15.8cm]{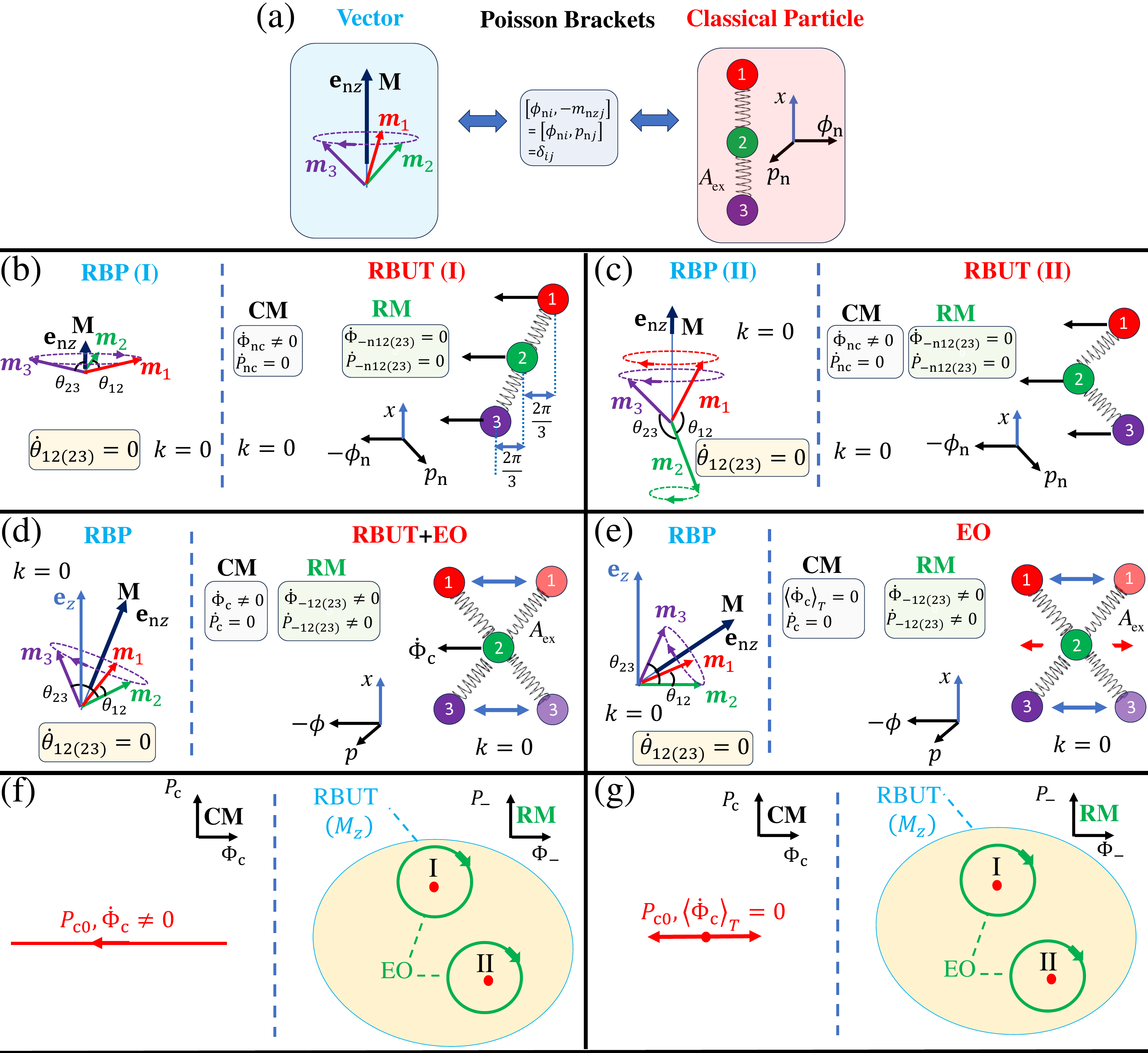}
\end{center}
\caption{(Color online) Two types of perspectives in spin dynamics are presented by the Poisson brackets based continuous transformation (Panel(a)). The \textbf{first (left panel)} is the vector (geometry) perspective, where $\mathbf{M}$ is the total magnetic moment of the three exchange-coupled spins $\mathbf{m}_{i}$ ($i=1,2,3$) and $\mathbf{e}_{\mathrm{n}z}$ is its unit vector. The \textbf{second (right panel)} is the classical particle perspective, with the $x$ axis indicating the arrangement direction of the three non-linear strings coupled particles (all having the same elastic coefficient $A_{\mathrm{ex}}$) and the phase plane $\phi_{\mathrm{n}}-p_{\mathrm{n}}$ (normal to the $x$ axis) being used to designate the states of the particles. Panels (b) and (c) schematically present two examples of degenerate RBP (RBUT) states under the two perspectives, specifically for the absence of anisotropy ($k=0$). Both states share the same constant non-zero total moment $M_{\mathrm{n}z}$ (where $\dot{P}_{\mathrm{n}c}=0$ and $\dot{\Phi}_{\mathrm{n}c}$ is a non-zero constant), and $M_{\mathrm{n}x(y)}= 0$. However, they feature different spin configurations: \textbf{State I} in Panel (b) has identical $z$ components of the spins and equal phase angle differences ($\Phi_{-\mathrm{n}12}=\Phi_{-\mathrm{n}23}=2\pi/3$). \textbf{State II} in Panel (c) has non-identical $z$ components and non-equal phase angle differences ($\Phi_{-\mathrm{n}12}\neq\Phi_{-\mathrm{n}23}$). The terms RBP and RBUT signify a \textbf{rigid body} motion (\textbf{precession}/\textbf{translation}) where the angles between spins are static ($\dot{\theta}_{12(23)}=0$), and the RM variables are static ($\dot{\Phi}_{-\mathrm{n}12(23)}=0$ and $\dot{P}_{-\mathrm{n}12(23)}=0$). Panels (d) and (e) display two instances of $M_{\mathrm{n}z}$-degenerate RBP states where the total moment is tilted, resulting in $M_{x(y)}\neq0$. The difference between them lies in the spin trajectories: Panel (d) shows a state where at least one of the spins' trajectories \textbf{encloses} \textbf{the} $z$ \textbf{axis}. This motion is a mixed state, comprising the RBUT and the RM's EO around this RBUT state, characterized by dynamic RM variables ($\dot{\Phi}_{-12(23)}\neq0$ and $\dot{P}_{-12(23)}\neq0$); while, Panel (e) shows a state where \textbf{no} spin's trajectory encloses the $z$ axis. This is a pure RM's EO around one of the stationary RBUT states, characterized by a static CM momentum ($\dot{P}_{\mathrm{c}}=0$) and zero time-averaged angular velocity ($\langle\dot{\Phi}_{\mathrm{c}}\rangle_{T}=0$). Notably, the red double arrow in the right sub-panel of (e) signifies a tiny oscillatory amplitude of $\Phi_\mathrm{c}$ around some point, which results in $\langle\dot{\Phi}_\mathrm{c}\rangle_{T} = 0$. Schematics (f) and (g) present the phase portraits in the CM and RM phase spaces for the two cases introduced in Panels (d) and (e), respectively. In the \textbf{CM space}, the red horizontal line indicates the CM's uniform translation, while the red double arrows signify its oscillation around an equilibrium point along the $\Phi_{\mathrm{c}}$ axis. In the \textbf{RM space}, the light transparent yellow area marks the set of RBUT states (or RBP states with $M_{x(y)}= 0$) that are degenerate to $M_{z}$ ($P_{\mathrm{c}}$). Furthermore, the green circle enclosing one of these RBUT states (State I or II) indicates the trajectory of the RM's elastic oscillation (EO).}
\label{two_prespectives}
\end{figure*}

 As suggested by \textit{classical mechanics}\cite{goldstein2014}, \textit{Poisson brackets} can be used to formulate the conserved spin dynamics from the viewpoint of continuous transformation. Furthermore, based on the Poisson brackets, two distinct perspectives are offered here to gain insight into the spin dynamics (as illustrated in Fig. \ref{two_prespectives}(a)): one is \textit{geometric} (\textit{vector}) perspective exemplified by the LLG equations; the other is \textit{classical} particles dynamics perspective, which treats the spins as independent classical particles described by the canonical variables $(\phi_{i}, p_{i})$.

Utilizing the concept of \textit{Poisson brackets}, Eq. (\ref{canon_Eq_LLGS}) can be generalized to the time derivative of any physical quantity $A$, which is a function of the spin variables $(\phi_{i}, p_{i})$ and time $t$, as follows:
\setlength\abovedisplayskip{6pt}
\setlength\belowdisplayskip{6pt}
\begin{eqnarray}
\dot{A}=[A,H]+\frac{\partial A}{\partial t}+\sum_{i=1}^{3}\bigg[\frac{\partial A}{\partial p_{i}}\bigg(-\frac{\partial F_{\mathrm{d}}}{\partial\dot{\phi}_{i}}\bigg)+\frac{\partial A}{\partial \phi_{i}}\frac{\partial F_{\mathrm{d}}}{\partial\dot{p}_{i}}\bigg],\nonumber\\
\label{any_quantity}
\end{eqnarray}
where the first term $[A,H]$ rigorously represents the conservative dynamics. The definition of \textit{Poisson brackets} of two functions $f$, $g$ with respect to $(\phi_{i},p_{i})$ being given to be
\setlength\abovedisplayskip{6pt}
\setlength\belowdisplayskip{6pt}
\begin{eqnarray}
[f,g]_{\phi,p}=\sum_{i=1}^{3}\frac{\partial f}{\partial\phi_{i}}\frac{\partial g}{\partial p_{i}}-\frac{\partial f}{\partial p_{i}}\frac{\partial g}{\partial\phi_{i}}.
\label{Poissonbrackets}
\end{eqnarray}

If these two functions $f$, $g$ are taken out of the set of canonical variables $(\phi_{i},p_{i})$ themselves, the brackets have the values
\setlength\abovedisplayskip{6pt}
\setlength\belowdisplayskip{6pt}
\begin{eqnarray}
[\phi_{i},\phi_{j}]_{\phi,p}=0=[p_{i},p_{j}]_{\phi,p},\nonumber
\label{}
\end{eqnarray}
and
\setlength\abovedisplayskip{6pt}
\setlength\belowdisplayskip{6pt}
\begin{eqnarray}
[\phi_{i},p_{j}]_{\phi,p}=\delta_{ij}=-[p_{j},\phi_{i}]_{\phi,p}.\nonumber
\label{}
\end{eqnarray}
where $i,j=1,2,3$  are used to designate different magnetic moments, and $\delta_{ij}$ is Kronecker delta.

As suggested by classical mechanics \cite{goldstein2014}, the Poisson bracket can be used to construct a certain finite (canonical) transformation by integrating a series of infinitesimal transformations. This mapping can be expressed as a Taylor expansion (or exponential mapping): \cite{goldstein2014}:
\setlength\abovedisplayskip{6pt}
\setlength\belowdisplayskip{6pt}
\begin{eqnarray}
f(a)&=&f_{0}+a[f,G]_{0}+\frac{a^2}{2!}[[f,G],G]_{0}\nonumber\\
&&+\frac{a^3}{3!}[[[f,G],G],G]_{0}+....
\label{Finite_Canotransform}
\end{eqnarray}
Here, $a$ is the parameter used to designate a specific trajectory (curve) in phase space (see Fig. \ref{sysmoper_of_H}). $f=f(\phi,p)$ is a continuous function of the system configuration, and $f(a)$ is its value along the trajectory, with an initial value $f_{0}=f(0)$. The first derivative of $f$ with respect to $a$, evaluated at $a=0$, is given by the Poisson bracket: $[f,G]_{0}=(df/da)_{0}$, where $G=G(\phi,p)$ is the generator of the transformation. Subsequently, the higher-order derivatives are represented by recursive brackets: $[[f,G],G]_{0} = (d^{2}f/da^{2})_{0}$ and  $[[[f,G],G],G]_{0}=(d^{3}f/da^{3})_{0}$, respectively. If $f$ is chosen to be the canonical variables $(\phi_{i}, p_{i})$, with $f_{0}$ being the starting set $(\phi_{i0}, p_{i0})$, then Eq. (\ref{Finite_Canotransform}) provides the formal description of the finite (canonical) transformation generated by $G$.

Obviously, $\phi_{i}$ and $p_{i}$ are each other's generators of \textit{infinitesimal} (\textit{canonical}) \textit{transformations}. Applying the fundamental Poisson bracket relations, we immediately confirm:$d\phi_{i}=da[\phi_{i},p_{i}]_{\phi,p}=da$ and $dp_{i}=da[p_{i},\phi_{i}]_{\phi,p}=-da$. As we know, for a single spin, whose length is kept constant (i.e., $m_{xi}^2+m_{yi}^2+m_{zi}^2=1$), there exists a corresponding \textit{continuous symmetry operation} termed the \textit{rotational transformation group}. The generators of this group are the spin components themselves: $(m_{xi},m_{yi},m_{zi})$. These generators are expressed in terms of the canonical variables as: $(m_{xi},m_{yi},m_{zi}) =(\sqrt{1-p_{i}^{2}}\cos\phi_{i}, \sqrt{1-p_{i}^{2}}\sin\phi_{i},-p_{i})$. We can demonstrate how the canonical variables $(\phi_{i}, p_{i})$ are transformed under this group. Taking the generator $m_{zj}$ as an instance, the Poisson brackets are $[\phi_{i},m_{zj}]=[\phi_{i},-p_{j}]=-\delta_{ij}$ and $[p_{i},m_{zj}] = 0$. According to the infinitesimal transformation (Eq. (\ref{Finite_Canotransform}), first order), the transformed variables are $(\phi'_{i}, p'_{i})=(\phi_{i0}-a\delta_{ij},p_{i0})$. This transformation corresponds to a rotation around the $z$-axis and, critically, preserves every spin's length.

Using Eq. (\ref{Poissonbrackets}), their commutator relations (the \textit{Lie algebra}) read
\setlength\abovedisplayskip{6pt}
\setlength\belowdisplayskip{6pt}
\begin{eqnarray}
[m_{bi},m_{cj}]=-\delta_{ij}\epsilon_{bcd}m_{di},
\label{Poisonbracformoment}
\end{eqnarray}
with $b$, $c$, $d$ being $x$, $y$, $z$, respectively, and $\epsilon_{bcd}$ being a permutation symbol. It should be stressed here that due to the poisson brackets being canonical invariant\cite{goldstein2014} the subscript used to indicate the set of canonical variables in terms of which the brackets are defined has been dropped out (see also Eq. (\ref{any_quantity})).

In addition to the single-spin rotational transformation, there still exists another similar one termed \textit{Rigid Body Rotational Transformation} (RBRT), which preserves the angles between spins $\theta_{ij}=\cos^{-1}(\mathbf{m}_{i}\cdot \mathbf{m}_{j})$ as well as the length of their total magnetization $\mathbf{M}\equiv\sum_{i=1}^{3}\mathbf{m}_{i}$. One can create one of the generators of this type of transformation by enforcing all spins to revolve synchronously around the $z$-axis. This transformation results in the canonical variables: $(\phi'_{i}, p'_{i}) = (\phi_{i0}-a, p_{i0})$, where the phase angles $\phi_{i}$ of all spins have a synchronized shift by an amount $a$ without changing their canonical momenta $p_{i}$. The corresponding generator for this transformation must be the total $z$-component of magnetization, $M_{z}=\sum_{i=1}^{3}m_{zi}$. This is verified by the Poisson brackets: $[\phi_{i},M_{z}]=[\phi_{i},\sum_{j=1}^{3}
-p_{j}]=-1$ and $[p_{i},M_{z}]=0$. Similar to the single-spin case, the other generators can be constructed to be the total components $M_{x}=\sum_{i=1}^{3}m_{xi}$ and $M_{y}= \sum_{i=1}^{3}m_{yi}$. From Eq. (\ref{Poisonbracformoment}), the commutator relations (Lie algebra) of them are
\setlength\abovedisplayskip{6pt}
\setlength\belowdisplayskip{6pt}
\begin{eqnarray}
[M_{b},M_{c}]=-\epsilon_{bcd}M_{d}.
\label{PoisonbracforTotmoment}
\end{eqnarray}

Implied by Fig. \ref{two_prespectives}(a), since every spin of the system can be viewed as a classical particle described by a pair of canonical variables $(\phi_{i},p_{i})$, the $z$-component of the total magnetic moment $M_{z}=-P_{\mathrm{c}}\equiv-\sum_{i=1}^{3}p_{i}$, can also be treated as a generator of the \textit{Rigid Body Uniform Translational Transformation} (RBUTT). The RBUTT moves the system in the $\phi$ direction as a whole without triggering any change in the relative configuration among particles. This is mathematically verified by the Poisson brackets of the relative coordinates and momenta being zero with respect to $P_{\mathrm{c}}$: $[p_{1}-p_{2},P_{\mathrm{c}}]=[p_{2}-p_{3},P_{\mathrm{c}}]=[\phi_{1}-\phi_{2},P_{\mathrm{c}}]
 =[\phi_{2}-\phi_{3},P_{\mathrm{c}}]=0$.

\subsubsection{\label{sec:2B2} Geometric (Vector) Perspective of the Exchange Couplings as an RBRT Generator: Infinite Number of Degenerate States }
It is known that the total magnetic moment $\mathbf{M}$ must be conserved by the exchange coupling $H_{\mathrm{ex}}$, regardless of whether the exchange coefficient $A_{\mathrm{ex}}$ is positive (AFM) or negative (FM), whether sub-neighboring couplings are taken into consideration, or even for the number of spins being beyond three. This is because $\mathbf{M}$ itself is the generator of the RBRT, a transformation which, by definition, does not alter the relative angles $\theta_{ij}$ between any two spins. In the language of Poisson brackets, this is expressed as the commutation relation $[H_{\mathrm{ex}},\mathbf{M}]=0$. Therefore, in our system, specifically in the absence of anisotropies (i.e., $k=0$), the total magnetic moment $\mathbf{M}$ is conserved, meaning $d\mathbf{M}/dt = [\mathbf{M},H_{\mathrm{ex}}]=0$. This conservation implies the existence of an infinite number of degenerate spin configurations $\{\mathbf{m}_i\}$ that result in the same total magnetic moment $\mathbf{M}$:

\setlength\abovedisplayskip{6pt}
\setlength\belowdisplayskip{6pt}
\begin{eqnarray}
M_{x}&=&\sum_{i=1}^{n}m_{xi}=\sum_{i=1}^{n}
\sqrt{1-p_{i}^{2}}\cos\phi_{i},\nonumber\\
M_{y}&=&\sum_{i=1}^{n}m_{yi}=\sum_{i=1}^{n}
\sqrt{1-p_{i}^{2}}\sin\phi_{i},\nonumber\\
M_{z}&=&\sum_{i=1}^{n}m_{zi}=-\sum_{i=1}^{n}
p_{i}.
\label{degenerate}
\end{eqnarray}

Moreover, for our symmetric system ($n=3$, equal spin length), the exchange Hamiltonian $H_{\mathrm{ex}}$ can be expressed as a function of the total magnetization length $|\mathbf{M}|$ only:
\setlength\abovedisplayskip{6pt}
\setlength\belowdisplayskip{6pt}
\begin{eqnarray}
H_{\mathrm{ex}}&=&\left(\frac{A_{\mathrm{ex}}}{2}\right)\left(|\mathbf{M}|^{2}-3\right),
\label{H_ex_M}
\end{eqnarray}
This relationship implies that $H_{\mathrm{ex}}$ itself, acting as a time-evolution operator, intrinsically features the nature of the RBRT alone. Consequently, $H_{\mathrm{ex}}$ possesses the same group of states that are degenerate with respect to $\mathbf{M}$ as its eigen (dynamic) states. We denote these degenerate energy states as the \textit{Rigid Body Precessional} (RBP) states.

However, if the three identical coupling strengths are replaced by a \textit{non-identical} set of strengths, then $H_{\mathrm{ex}}$ cannot be expressed solely as a function of $|\mathbf{M}|$ at all, unlike in Eq. (\ref{H_ex_M}). This breakdown of the functional relationship leads to the states that were originally degenerate with respect to $\mathbf{M}$ being unable to remain degenerate with respect to $H_{\mathrm{ex}}$ anymore.

\subsubsection{\label{sec:2B3} Classical Particle Perspective of the Exchange Couplings as an RBUTT Generator: Centre of Mass and Relative Motion Coordinates }
When chosen $\mathbf{M}=\mathbf{e}_{z}M_{z}$, in terms of the particle perspective, $H_{\mathrm{ex}}$ for the general cases with multiple spins, as already mentioned in the very beginning of Sec. \ref{sec:2B2} reads $H_{\mathrm{ex}}=H_{\mathrm{CM}}(P_{\mathrm{c}})+H_{\mathrm{RM}}(P_{\mathrm{c}},\Phi_{-},P_{-})$. Here, the subscriptions CM and RM represent the abbreviations for \textit{center of mass} and \textit{relative motion}, respectively. $P_{\mathrm{c}}$ is the CM's momentum, and $\Phi_{-12(23)}\equiv\phi_{1(2)}-\phi_{2(3)}$ as well as $P_{-12(23)}\equiv p_{1(2)}-p_{2(3)}$ are the RM variables. The CM's momentum $P_{\mathrm{c}}$ must be conserved because it is canonically decoupled from the RM variables: $[P_{-},P_{\mathrm{c}}]=[\Phi_{-},P_{\mathrm{c}}]=0$, ensuring $[H_{\mathrm{RM}},P_{\mathrm{c}}]=0$, as already argued in the end of Sec. \ref{sec:2B1}. This universally agrees with $\mathbf{M}$ being a conserved physical quantity for any form of $H_{\mathrm{ex}}$.
Notably, the non-zero $H_{\mathrm{RM}}$ often presents in general multi-spin systems, suggesting \textit{non-identical} exchange coupling strengths between spins.  Due to the non-zero Poisson bracket between the RM variables, $[\Phi_{-12},P_{-23}]=[\Phi_{-23},P_{-12}]\neq0$, the $H_{\mathrm{RM}}$ term may still induce a non-stationary (dynamic) state of the RM variables.

Based on the fact that $P_{\mathrm{c}}$ must always be conserved regardless of whether the RM variables are stationary or in motion, there is an infinite number of RM's stationary/dynamic states degenerate to $P_{\mathrm{c}}$. However, in the general case, these states are not degenerate to $H_{\mathrm{ex}}$ itself, precisely because they are not exactly degenerate to the RM Hamiltonian $H_{\mathrm{RM}}$. Interestingly, in our specific NC-AFM system, the RM Hamiltonian is exactly zero ($H_{\mathrm{RM}}=0$), a consequence of the identical set of exchange coupling coefficients.
Consequently, the total exchange Hamiltonian is entirely determined by the CM part (see also Eq. (\ref{H_ex_M})):
\setlength\abovedisplayskip{6pt}
\setlength\belowdisplayskip{6pt}
\begin{eqnarray}
H_{\mathrm{ex}}&=&\left(\frac{A_{\mathrm{ex}}}{2}\right)\left(P_{\mathrm{c}}^{2}-3\right).
\label{H_ex_Pc}
\end{eqnarray}

Since $H_{\mathrm{ex}}$ is independent of the RM variables, it is unable to drive any motion of the RM variables, leading to $\dot{P}_{-}=[P_{-},H_{\mathrm{ex}}]=0$ and $\dot{\Phi}_{-}=[\Phi_{-},H_{\mathrm{ex}}]=0$. That is, $H_{\mathrm{ex}}$, acting as the time-evolution operator, features an RBUT characteristic alone. Thus, it also inherits the infinite number of degenerate stationary states of RM variables originally degenerate to $P_{\mathrm{c}}$. These states, which are now degenerate to both $P_{\mathrm{c}}$ and $H_{\mathrm{ex}}$, are called \textit{Rigid Body Uniform Translational} ($\text{RBUT}$) states. This key point that $P_{\mathrm{c}}$ and $H_{\mathrm{ex}}$ share the same degenerate dynamic states is very similar to the conclusion derived from the vector perspective (see Sec. \ref{sec:2B2}).

Just as pointed out by \textit{classical mechanics} \cite{goldstein2014}, the conservation of $P_{\mathrm{c}}$ implies the existence of a \textit{cyclic coordinate} in the exchange Hamiltonian $H_{\mathrm{ex}}$, where $H_{\mathrm{ex}}$ does not involve it at all; this variable is precisely the conjugate to $P_{\mathrm{c}}$. Finding this variable would be beneficial for simplifying the analysis of RBP states where the RM variables are stationary. For convenience, we adapt a unique set of Cartesian coordinates whose $z$-axis always points to the total momentum $\mathbf{M}=\mathbf{e}_{\mathrm{n}z}M_{\mathrm{n}z}$, ensuring $M_{\mathrm{n}x}=M_{\mathrm{n}y} = 0$. A new set of canonical cylindrical coordinates $(\phi_{\mathrm{n}i}, p_{\mathrm{n}i})$ attached to these Cartesian ones can then be well-defined.

Given that any set of conjugate variables are the generators of infinitesimal transformations to each other, and that the change the CM's momentum $P_{\mathrm{nc}}$ can induce is just a synchronized shift in $\phi_{\mathrm{n}i}$ without altering $p_{\mathrm{n}i}$ themselves at all, then one can easily find that the quantity $\Phi_{\mathrm{nc}}$ conjugate to $P_{\mathrm{nc}}$ has a non-zero constant $[\Phi_{\mathrm{nc}}, P_{\mathrm{nc}}]$. The transformation $\Phi_{\mathrm{nc}}$ can make is only a phase-locked shift in $p_{\mathrm{n}i}$ without changing $\phi_{\mathrm{n}i}$ themselves at all. Hence, we believe $\Phi_{\mathrm{nc}}$ as a function of $(\phi_{\mathrm{n}i}, p_{\mathrm{n}i})$ should have a similar appearance to $P_{\mathrm{nc}}(p_{\mathrm{n}i})$, simply replacing $p_{\mathrm{n}i}$ in $P_{\mathrm{nc}}$ with $\phi_{\mathrm{n}i}$ in $\Phi_{\mathrm{nc}}$. Consequently, $\Phi_{\mathrm{nc}}$ can be built to be $\Phi_{\mathrm{nc}} = (1/3)\sum_{i=1}^{3}\phi_{\mathrm{n}i}$, which is obviously cyclic since $dH_{\mathrm{ex}} = [H_{\mathrm{ex}}, P_{\mathrm{nc}}]d\Phi_{\text{nc}}=0$ \cite{goldstein2014}.

Due to $(\Phi_{\mathrm{nc}},P_{\mathrm{nc}})$ being the set of total canonical variables of the three spins, they can be used to describe the configuration of the system as a whole and are termed CM (canonical) phase angle/momentum, just as in classical particles' dynamics where the CM variables can only sense external forces instead of internal forces. In this system, the internal forces partially responsible for phase-locking spins are apparently induced by the exchange couplings, which can be seen from their even functions of $\phi_{\mathrm{n}i}-\phi_{\mathrm{n}j}$ (see Eq. (2)). This mimics central forces in classical particle dynamics, leaving the total (linear) momentum $P_{\mathrm{nc}}$ conserved, or rather, owing to $[P_{\mathrm{nc}},\phi_{\mathrm{n}i}-\phi_{\mathrm{n}j}] = 0$. In addition, since the CM momentum is not cyclic ($\Phi_{\mathrm{nc}}$ is cyclic), i.e. $dH_{\mathrm{ex}}=[H_{\mathrm{ex}},\Phi_{\mathrm{nc}}]dP_{\mathrm{nc}}\neq0$, the exchange couplings will thus have a non-zero contribution to the time rate of the CM phase angle, namely, $\dot{\Phi}_{\mathrm{nc}}=[\Phi_{\mathrm{nc}},H_{\mathrm{ex}}]$.

The CM variables $(\Phi_{\mathrm{nc}},P_{\mathrm{nc}})$ alone, undoubtedly, fail to give a complete description for these three coupled spins, as the system's dimensionality is six. Thus, we must introduce four new extra variables to describe the relative motion (RM) between spins. A change of variables is given below to define the CM as well as RM coordinates simultaneously:

\setlength\abovedisplayskip{6pt}
\setlength\belowdisplayskip{6pt}
\begin{eqnarray}
\left(
     \begin{matrix}
     \Phi_{\mathrm{n}\mathrm{c}}\\
     \Phi_{\mathrm{n}-12}\\
     \Phi_{\mathrm{n}-23}\\
     \end{matrix}
\right)=
\left(
      \begin{matrix}
      \frac{1}{3}&\frac{1}{3}&\frac{1}{3}\\
      1&-1&0\\
      0&1&-1\\
      \end{matrix}
\right)
\left(
      \begin{matrix}
      \phi_{\mathrm{n}1}\\
      \phi_{\mathrm{n}2}\\
      \phi_{\mathrm{n}3}\\
      \end{matrix}
\right),
\label{transphi_to_Phi}
\end{eqnarray}
and
\setlength\abovedisplayskip{6pt}
\setlength\belowdisplayskip{6pt}
\begin{eqnarray}
\left(
     \begin{matrix}
     P_{\mathrm{n}\mathrm{c}}\\
     P_{\mathrm{n}-12}\\
     P_{\mathrm{n}-23}\\
     \end{matrix}
\right)=
\left(
      \begin{matrix}
      \frac{1}{3}&\frac{1}{3}&\frac{1}{3}\\
      1&-1&0\\
      0&1&-1\\
      \end{matrix}
\right)
\left(
      \begin{matrix}
      p_{\mathrm{n}1}\\
      p_{\mathrm{n}2}\\
      p_{\mathrm{n}3}\\
      \end{matrix}
\right),
\label{transp_to_P}
\end{eqnarray}
and their inverses
\setlength\abovedisplayskip{6pt}
\setlength\belowdisplayskip{6pt}
\begin{eqnarray}
\left(
     \begin{matrix}
     \phi_{\mathrm{n}1}\\
     \phi_{\mathrm{n}2}\\
     \phi_{\mathrm{n}3}\\
     \end{matrix}
\right)=
\left(
      \begin{matrix}
      1&\frac{2}{3}&\frac{1}{3}\\
      1&-\frac{1}{3}&\frac{1}{3}\\
      1&-\frac{1}{3}&-\frac{2}{3}\\
      \end{matrix}
\right)
\left(
      \begin{matrix}
      \Phi_{\mathrm{n}\mathrm{c}}\\
      \Phi_{\mathrm{n}-12}\\
      \Phi_{\mathrm{n}-23}\\
      \end{matrix}
\right),
\label{inverssphi_to_Phi}
\end{eqnarray}
and
\setlength\abovedisplayskip{6pt}
\setlength\belowdisplayskip{6pt}
\begin{eqnarray}
\left(
     \begin{matrix}
     p_{\mathrm{n}1}\\
     p_{\mathrm{n}2}\\
     p_{\mathrm{n}3}\\
     \end{matrix}
\right)=
\left(
      \begin{matrix}
      1&\frac{2}{3}&\frac{1}{3}\\
      1&-\frac{1}{3}&\frac{1}{3}\\
      1&-\frac{1}{3}&-\frac{2}{3}\\
      \end{matrix}
\right)
\left(
      \begin{matrix}
      P_{\mathrm{n}\mathrm{c}}\\
      P_{\mathrm{n}-12}\\
      P_{\mathrm{n}-23}\\
      \end{matrix}
\right),
\label{inversep_to_P}
\end{eqnarray}
where $(\Phi_{\mathrm{nc}},P_{\mathrm{nc}})$, $(\Phi_{\mathrm{n}-12} ,P_{\mathrm{n}-12} )$, and $(\Phi_{\mathrm{n}-23},P_{\mathrm{n}-23})$ indicate the CM and RM variables, respectively. In order to make both the transformed forms of $\Phi_{\mathrm{nc}}$ and $P_{\mathrm{nc}}$ look more symmetric, we choose to define $P_{\mathrm{nc}}=(1/3)\sum_{i=1}^{3}p_{\mathrm{n}i}$ rather than the usual $P_{\mathrm{nc}}=\sum_{i=1}^{3}p_{\mathrm{n}i}$. Although this scaling results in a non-unity Poisson bracket relation, $[\Phi_{\mathrm{n}\mathrm{c}},P_{\mathrm{n}\mathrm{c}}]=1/3\neq1$, there is no hindrance of them being each other's conjugate variables.

Note here that the commutator relations between CM and RM variables are all zero: $[\Phi_{\mathrm{nc}},\Phi_{\mathrm{n}-12(23)}]=[\Phi_{\mathrm{nc}},P_{\mathrm{n}-12(23)}]=
[P_{\mathrm{nc}},\Phi_{\mathrm{n}-12(23)}]=[P_{\mathrm{nc}},P_{\mathrm{n}-12(23)}]=0$. This means that any transformation involving CM variables alone causes no change in RM variables at all, and vice versa. This property, which confirms the independence of the CM (macro) and RM (micro) modes, will be essential later for analyzing the system when SOT and damping are present. In addition, notably, the commutator relations between RM variables are not all zero, indicating internal coupling within the RM subspace.

As mentioned previously, in the particle perspective,  the RBP state with $\mathbf{M}=\mathbf{e}_{\mathrm{n}z}M_{\mathrm{n}z}$ corresponds to an RBUT state This RBUT state is fundamentally characterized by the fact that all RM variables are stationary, specifically requiring: $\dot{P}_{\mathrm{n}-12(23)}=[P_{\mathrm{n}-12(23)},H_{\mathrm{ex}}]=0$ and $\dot{\Phi}_{\mathrm{n}-12(23)}=[\Phi_{\mathrm{n}-12(23)},H_{\mathrm{ex}}]=0$ (see Figs. \ref{two_prespectives}(b) and (c)). This yields four quite complicated equalities of constraints to effectively reduce the dimensionality of the spins from six to two. Under these constraints, there is an infinite number of combinations of stationary $(\Phi_{\mathrm{n}-12(23)},P_{\mathrm{n}-12(23)})$, manifesting the existence of an infinite number of dynamic states degenerate to $P_{\mathrm{n}c}$ as well as $H_{\mathrm{ex}}$. Moreover, this can also prove to be completely equivalent to Eq. (\ref{degenerate}).

It shout be stressed here that for the RBP states with their total magnetization $\mathbf{M}$ being not collinear to the z axis, i.e. $\mathbf{e}_{\mathrm{n}z}\neq\mathbf{e}_{z}$, two other types of dynamic states exist in terms of the $(x,y,z)$ Cartesian coordinate (see Fig. \ref{two_prespectives}(d) and (e)):
one is an RBUT state accompanied by an \textit{elastic oscillation} (EO) of RM variables, where at least one of the spin vectors' trajectories encloses around the z axis and the EO is around one of the equilibrium RBUT states satisfying the constraints $M_{x}=M_{y}=0, M_{z}\neq0$; the other is characterized as a pure EO of RM variables with CM variables being almost stationary with $\langle\dot{\Phi}_{\mathrm{c}}\rangle_{T}=0$, with none of the spins' precessional orbits enclosing around the z axis.

\subsubsection{\label{sec:2B4} Angular Frequency of RBP States }
 The angular frequency of the degenerate RBP states can be directly derived form their CM's angular frequency to be
$\omega_{o}=|\dot{\Phi}_{\mathrm{nc}}|=|[\Phi_{\mathrm{nc}},H_{\mathrm{ex}}]|=(1/3)|(\partial H_{\mathrm{ex}}/\partial P_{\mathrm{nc}})|=3A_{\mathrm{ex}}|P_{\mathrm{nc}}|=A_{\mathrm{ex}}|\mathbf{M}|$, where
\setlength\abovedisplayskip{6pt}
\setlength\belowdisplayskip{6pt}
\begin{eqnarray}
H_{\mathrm{ex}}
=\left(\frac{A_{\mathrm{ex}}}{2}\right)\left(9P_{\mathrm{nc}}^{2}-3\right).
\label{H_ex_RBP}
\end{eqnarray}
Thus, the states degenerate to both $\mathbf{M}$ and $H_{\mathrm{ex}}$ are also degenerate with respect to the (angular) frequency $\omega_{0}$. Furthermore, for the case where $\mathbf{M}$ is not collinear with the $\mathbf{e}_{z}$, the angular frequency of the RBUT state as well as the EO of the RM variables are both demonstrably equal to $A_{\mathrm{ex}}|\mathbf{M}|$. This is easily anticipated from the nature of RBRT of $H_{\mathrm{ex}}$, eliminating the need to solve the full equations in the particle perspective.

Interestingly, such an EO can thereby be treated as a \textit{simple harmonic oscillation} (SHO). It is analogous to a light massive particle linked by a mass-free string to the origin that is fixed at one of the RBUT states (RBP states with $\mathbf{e}_{\mathrm{n}z}=\mathbf{e}_{z}$), whose
natural angular frequency is $\omega_{\mathrm{EO}}=A_{\mathrm{ex}}|\mathbf{M}|$.

\subsection{\label{sec:2C} SOT-driven RBP States}
We begin to consider the case for the presence of the non-conservative effects including the damping as well SOT with a spin polarizer vector $\mathbf{p}=\mathbf{e}_{z}$ being collinear with $\mathbf{e}_{\mathrm{n}z}$. There are two equivalent viewpoints on offer here to explore how such an SOT drives stable RBP states, one is the vector perspective, and the other is the particle perspective.

By inspecting the dissipation function $F_{\mathrm{d}}$ appearing in Eq. (\ref{any_quantity}), one immediately finds that such an SOT, being a pure \textit{negative} damping force/field \cite{Chen2021} that always points toward the directions of $\phi_{i}$ increase or decrease, can only compensate with the positive damping force in the direction of $\dot{\phi}_{i}$.
In the particle perspective, this means the SOT can only compensate the damping force along the RBUT's CM velocity, $\dot{\Phi}_{\mathrm{c}}$. In other words, only those RBP states with their $\mathbf{M}$ being exactly collinear to $\mathbf{p}$ (i.e., pure RBUT states, see Figs. \ref{two_prespectives}(b) and (c)) can therefore be driven. This completely excludes the possibility of the SOT-driven EOs of the RM variables, due to the fact that the EOs gain no net energy from this SOT during a single oscillating period (as the SOT force direction is fixed, but the EO velocity direction reverses). Note here that such an SOT can effectively be treated as an \textit{effective conserved field} instead for those EOs \cite{HaoHsuan2015,HaoHsuan2017,Chen2021,chen2023a}.

\begin{figure}
\begin{center}
\includegraphics[width=8.0cm]{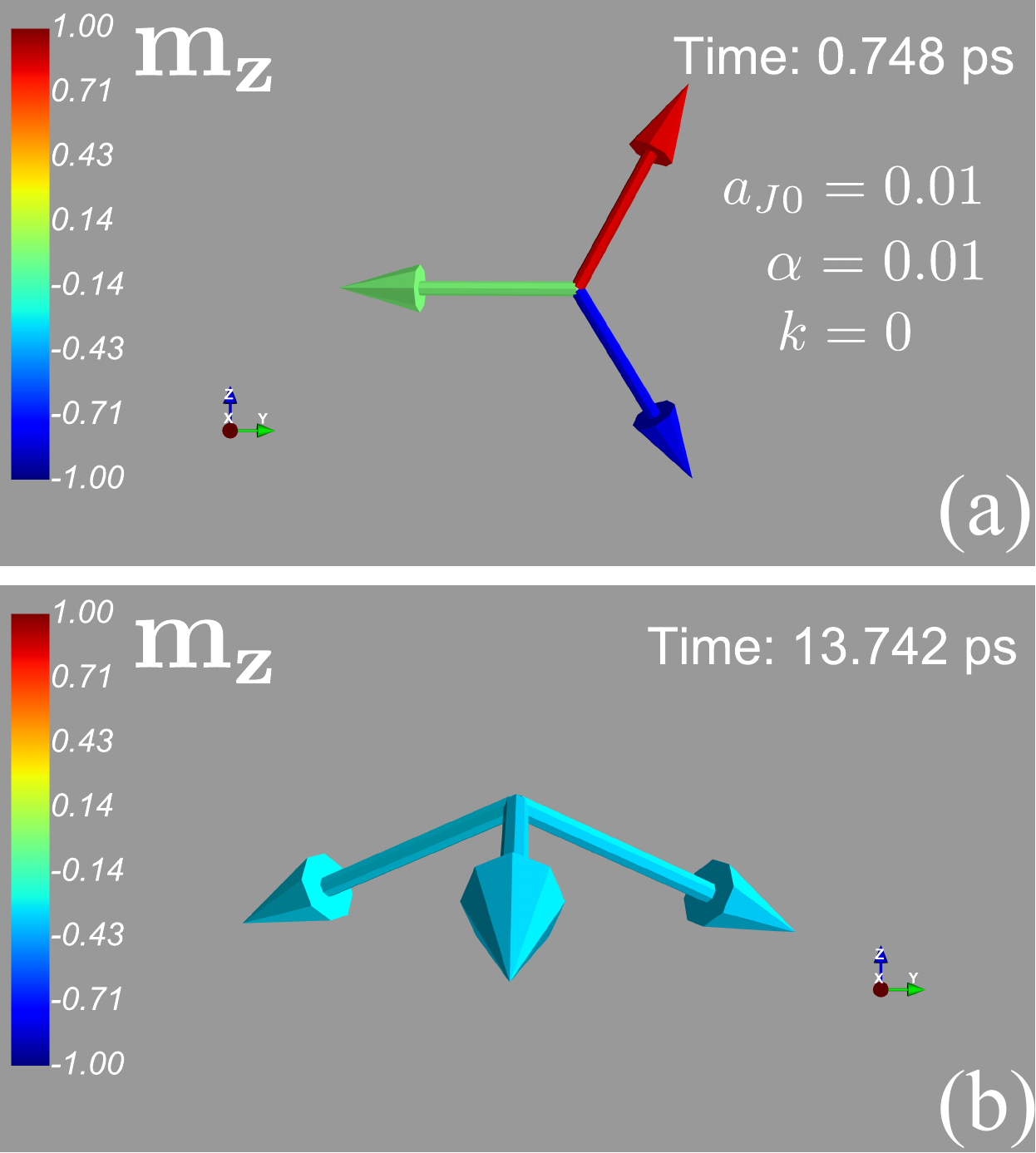}
\end{center}
\caption{(Color online)  RBP state relaxation to stability via SOT and damping cooperation. Macrospin simulated snapshots show the RBP state relaxation to stability, driven by the cooperative action of the SOT and damping.\textbf{ Panel (a) Initial State}: The system starts with the total magnetization direction $\mathbf{e}_{\mathrm{n}z}$ \textbf{misaligned} with the spin polarization vector $\mathbf{p}$ ($\mathbf{e}_{\text{n}z} \neq \mathbf{p}$), also characterized as an RBUT state with an EO. \textbf{Panel (b) Final State}: The system evolves into one of the stable $M_{z}$-degenerate  RBP (RBUT) states with the total momentum's direction \textbf{aligning} with the polarization ($\mathbf{e}_{\text{n}z} = \mathbf{p}$). \textbf{Parameters}: $a_{J0} = 0.01$ ($J =1.8962\times 10^{-9}\text{A/cm}^2$), $\alpha=0.01$, and $k = 0$. More details about such a relaxation process are presented in the supplementary animation in the attachment (see 'SOT-Damp eliminated RM.mp4').}
\label{SOT_DAMP_deletRM}
\end{figure}

\begin{figure}
\begin{center}
\includegraphics[width=8.0cm]{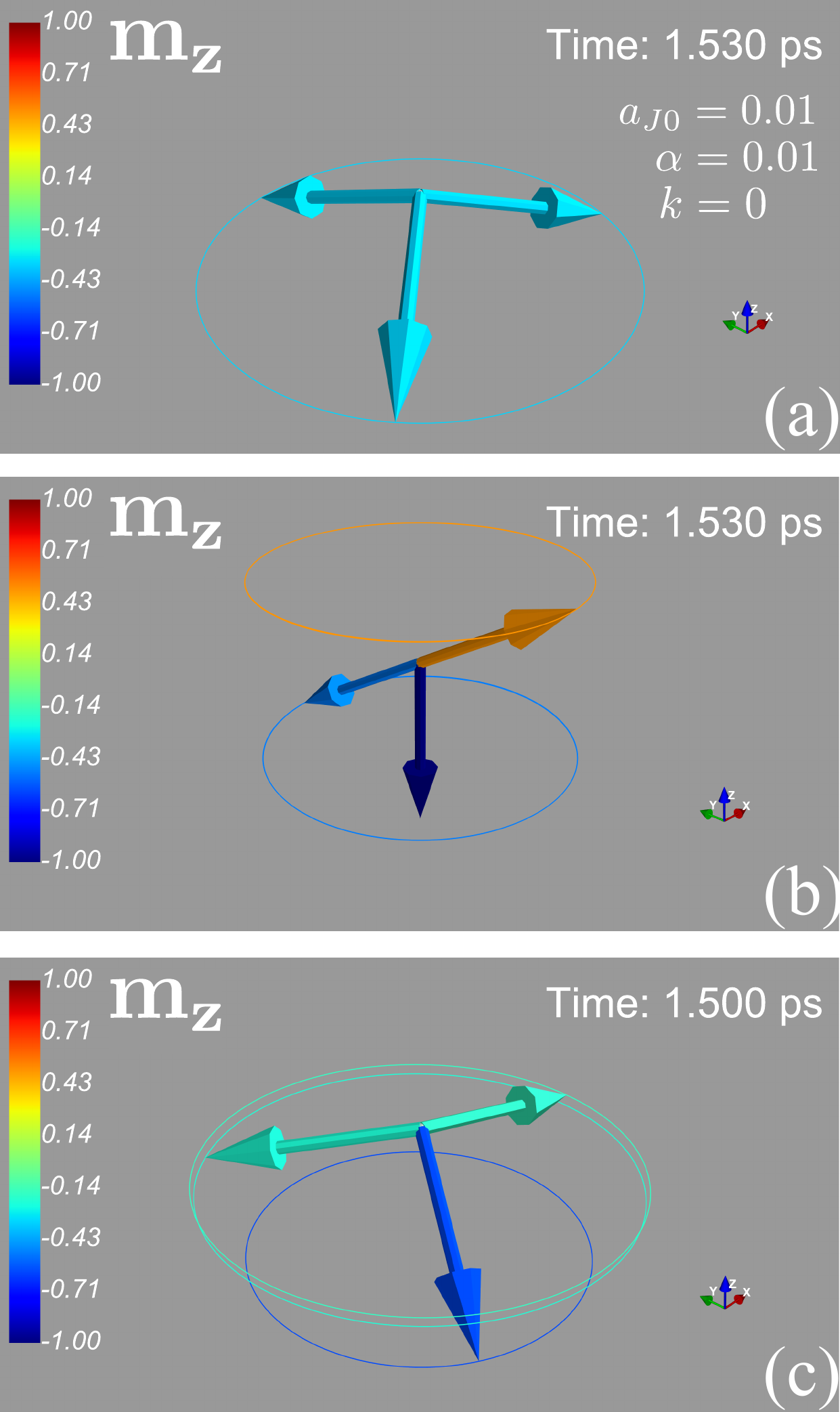}
\end{center}
\caption{(Color online) Snapshots of three selected degenerate SOT-driven RBP (RBUT) states.
These macrospin simulation snapshots show three degenerate RBP (RBUT) states for $a_{J0}=0.01$ ($J=1.8962\times 10^{-9}\mathrm{A}/\mathrm{cm}^2$), $\alpha=0.01$, and $k=0$. The spin-polarization vector $\mathbf{p}$ points toward the $z$ axis, and the colors on the arrows indicate the $m_{z}$ values of the three spins. Panels (a)-(c) illustrate the three distinct states: (a) the state where $m_{zi}=-1/3$ ($\Phi_{-12(23)}=2\pi/3$); (b) the state with $m_{z1}=-1$ and $m_{z2}=-m_{z3}$ ($\Phi_{-23} = \pi$); and (c) the state with $m_{z1}\neq m_{z2}\neq m_{z3}$. More details about the SOT-driven RBP (RBUT) states are presented in the supplementary animation in the attachment (see 'degenerate state 120 degree.mp4', 'degenerate state 180 degree .mp4', and 'degenerate state other.mp4').}
\label{Degenerate_states_snapshots}
\end{figure}

\begin{figure}
\begin{center}
\includegraphics[width=7.8cm]{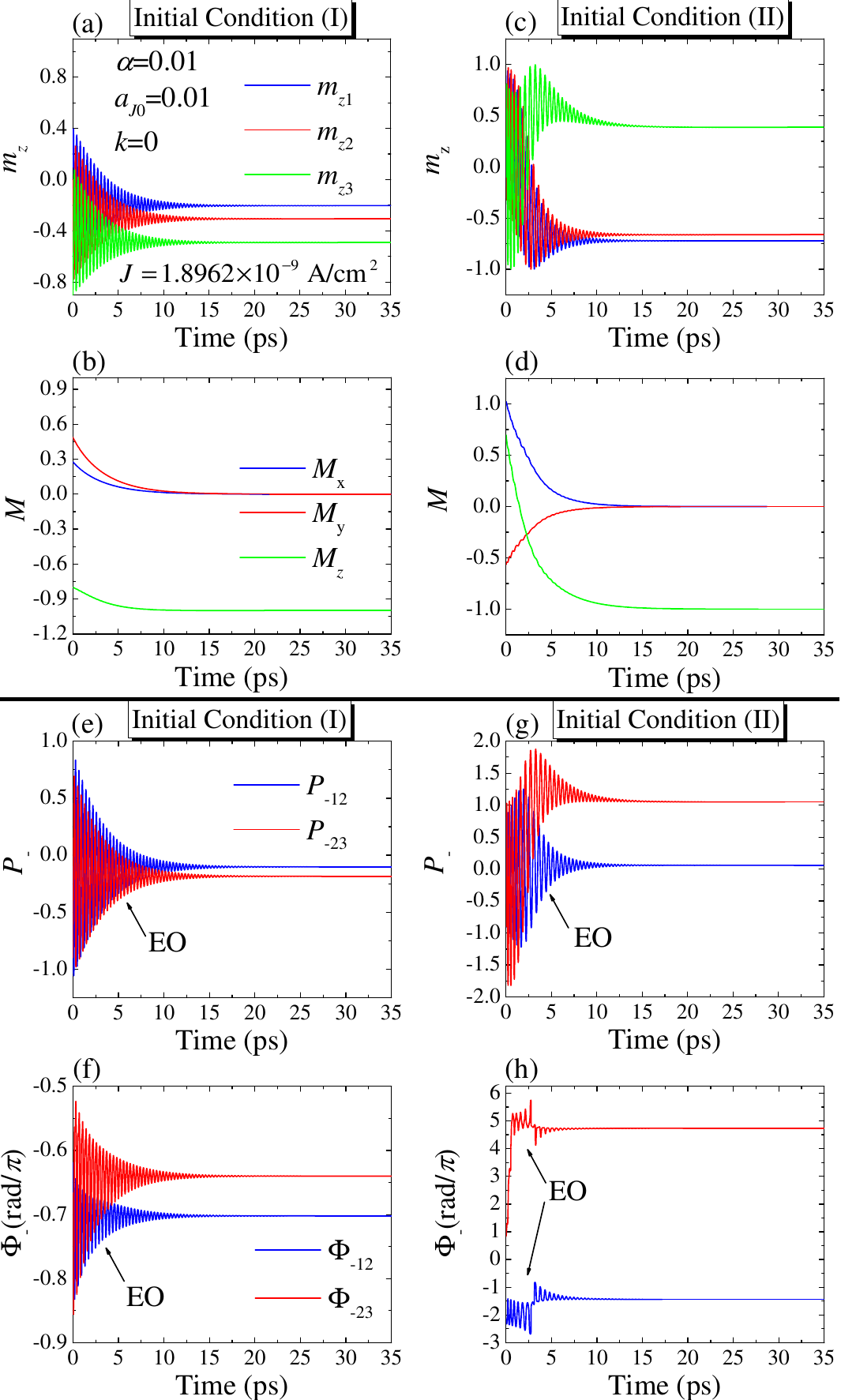}
\end{center}
\caption{(Color online)  Evolution to $M_{z}$-degenerate states from different initial conditions: A comparison of vector and particle perspectives. Macrospin simulations confirm that SOT-driven RBP (RBUT) states evolve into $M_{z}$-degenerate states irrespective of initial conditions. Parameters: $a_{J0}=0.01$ ($J=1.8962\times10^{-9}\mathrm{A}/\mathrm{cm}^{2}$), $\alpha=0.01$, and $k=0$. \textbf{A. Vector Perspective (Upper Panels (a)-(d))} \textbf{Time Traces}: Panels (a) and (c) show the time traces of the three coupled spins' $z$-components ($m_{zi}$), where blue, red, and green lines indicate spins 1, 2, and 3, respectively. \textbf{Total Moment}: Panels (b) and (d) present the time evolutions of the total moment $\mathbf{M}$ (blue: $M_{x}$, red: $M_{y}$, green: $M_{z}$). \textbf{B. Particle Perspective (Lower Panels (e)-(h))} \textbf{RM Momenta}: Panels (e) and (g) show the time traces of the relative motion momenta, $P_{-12}$ (blue line) and $P_{-23}$ (red line). \textbf{RM Coordinates (Phase Angular Differences)}: Panels (f) and (h) show the time evolutions of $\Phi_{-12}$ (red curve) and $\Phi_{-23}$.}
\label{Degenerate}
\end{figure}

\subsubsection{\label{sec:2C1} Vector Perspective: Frame of Reference Synchronous to the RBP States}
Based on the finding that only the RBP states with $\mathbf{e}_{\mathrm{n}z} =\mathbf{p}=\mathbf{e}_{z} $ are driven, we introduce a particular frame of reference that is synchronous to these states. The purpose of this transformation is to render the precessing states stationary. In this synchronous rotating frame, it is straightforward to directly identify the stable equilibrium states by analytically solving for the local minima of the effective Hamiltonian (Energy) \cite{Chen2019a,Chen2021}. This approach effectively transforms the dynamic non-conservative problem into a static stability analysis.

Evidently, exploiting the RBRT characteristic of these RBP states, a coordinate transformation termed a \textit{time-dependent} RBRT is applied to render them stationary: $(\phi'_{i},p'_{i})=(\phi_{i}-\omega(M_{z},J)t,p_{i})$ where $\omega(M_{z},J)$ is the angular velocity of the RBP states. The generator of this RBRT is necessarily a function of the total $z$-component magnetization, $M_{z}\equiv-3P_{\mathrm{c}}$, and the injected current density, $J$. Based on the discussion in Sec. \ref{sec:2B1} regarding the Hamiltonian as a time-evolution generator, this transformation indicates that an extra term must be produced in the effective Hamiltonian $H'$ (effective energy $E'$) in the rotating frame. This term is associated with $M_{z}$ alone and is analogous to the Zeeman energy.

 Moreover, the Poisson brackets for the new variables are invariant $[\phi'_{i},\phi'_{j}]=0=[p'_{i},p'_{j}]$ as well as $[\phi'_{i},p'_{j}]=\delta_{ij}=-[p'_{j},
 \phi'_{i}]$, saying that such a transformation is also \textit{canonical}\cite{goldstein2014}. Notably, just as mentioned in Sec. \ref{sec:2B3},  the dynamics of the RM variables are unchanged at all under such a transformation. Thus, in the new frame of reference, Eq. (\ref{any_quantity}) becomes
\setlength\abovedisplayskip{6pt}
\setlength\belowdisplayskip{6pt}
\begin{eqnarray}
\dot{A}'=[A',H']+\sum_{i=1}^{3}\bigg[\frac{\partial A'}{\partial p'_{i}}\bigg(-\frac{\partial F'_{\mathrm{d}}}{\partial\dot{\phi}'_{i}}\bigg)+\frac{\partial A'}{\partial \phi'_{i}}\frac{\partial F'_{\mathrm{d}}}{\partial\dot{p}'_{i}}\bigg],\nonumber\\
\label{any_quantity_new_frame}
\end{eqnarray}
with $A'=A'(\phi'_{i},p'_{i})$, $H'=H'_{\mathrm{ex}}(|\mathbf{M}'|^{2})+H'_{\mathrm{zee}}(M'_{z},J)$
with $H'_{\mathrm{zee}}(M'_{z},J)=\int^{M'_{z}}dM''_{z}
\omega(M''_{z},J)$ (see also \ref{sec:2B4} for $\omega=\partial H'_{\mathrm{zee}}/\partial M'_{z}$), and the damping forces
\setlength\abovedisplayskip{6pt}
\setlength\belowdisplayskip{6pt}
\begin{eqnarray}
-\frac{\partial F'_{\mathrm{d}}}{\partial\dot{\phi}'_{i}}&=&-\alpha\left[1-(p'_{i})^{2}\right]
\bigg[\dot{\phi}'_{i}+\omega(M'_{z},J)\nonumber\\
&&-\frac{a_{Ji}(-p'_{i})}{\alpha}\bigg],\nonumber\\
\frac{\partial F'_{\mathrm{d}}}{\partial\dot{p}'_{i}}&=&\frac{\alpha}{\left[1-(p'_{i})^{2}\right]}
\dot{p}'_{i}.\nonumber
\end{eqnarray}

Interestingly, if $\omega$ is taken to be $a_{Ji}/\alpha=a_{J0}/\alpha$, then the new Hamiltonian $H'$ and damping force $-\partial F'_{\mathrm{d}}/\partial\dot{\phi}'_{i}$ become $H'=(A_{\mathrm{ex}}/2)(|\mathbf{M}'|^{2}-3)
+(a_{J0}/\alpha)M'_{z}=[(A_{\mathrm{ex}}/2)(M'_{z})^{2}
+(a_{J0}/\alpha)M'_{z}]+[(A_{\mathrm{ex}}/2)((M'_{x})^{2}
+(M'_{y})^{2}-3)]$ as well as $-\partial F'_{\mathrm{d}}/\partial\dot{\phi}'_{i}=-\alpha
\left[1-(p'_{i})^{2}\right]\dot{\phi}'_{i}$, respectively. This result demonstrates that the SOT force has been completely offset by the effective damping force $-\alpha[1-(p'_{i})^2]\omega(M'_{z},J)$ produced by the transformation in the rotating frame \cite{Chen2021,Chen2019a}. Note here that $|\mathbf{M}'|=|\mathbf{M}|$, and $\mathbf{M}'_{z}=\mathbf{M}_{z}$. Apparently, in the new frame, the symmetry degree of the Hamiltonian has been reduced by the SOT-induced Zeeman-like energy term. The symmetry changes from its original invariance under an RBRT around an arbitrary direction (i.e., $[H_{\mathrm{ex}},\mathbf{M}]=0$) to one under the RBRT only around the spin-polarizer vector $\mathbf{p}$ (i.e., $[H',M'_{z}] = 0$). If the non-conservative effect is not present, $H'$ will conserve $M'_{z}$ rather than the full $\mathbf{M}'$, i.e., $[M'_{z},H'] = 0$.

 It follows that for the dynamic states with non-zero $M_{x(y)}$ in the laboratory (lab) frame, due to $[M'_{x(y)},H']=(a_{J0}/\alpha)[M'_{x(y)},M'_{z}]\neq0$,
$\mathbf{M}'$ will be precessing around $\mathbf{p}$ in the rotating frame. Then, thanks to the damping effect (see Eq. (\ref{any_quantity_new_frame})), the system starting with any one of them will eventually be damped out into one of the stationary states synchronous to this frame with zero $M_{x(y)}$, or rather, one of the RBP states of $H_{\mathrm{ex}}$ around $\mathbf{p}$ with the minimum value of the total energy $E'$. This exactly agrees the previously established conclusion that only the RBP states with $\mathbf{M}$ being collinear to $\mathbf{p}$ can be driven by the SOT in the lab frame, which is perfectly evidenced by the macrospin simulation (see Fig. \ref{SOT_DAMP_deletRM}(a) to (b)).

 Therefore, for the RBP states with $M_{x(y)}=0$, the Hamiltonian reduces to a function of $M'_{z}$ only: $H'_{\mathbf{M}}(M'_{z})=(A_{\mathrm{ex}}/2)(M'_{z})^{2}
+(a_{J0}/\alpha)M'_{z}$. By requiring the first derivative to be zero (equilibrium condition) and the second derivative to be positive (stability condition): $(\partial H'_{\mathbf{M}}/\partial M'_{z})_{M'_{z0}}=0$ and $[\partial^{2} H'_{\mathbf{M}}/\partial (M'_{z})^{2}]_{M'_{z0}}>0$. We solve for the degenerate $M'_{z0}$ value of the stable SOT-driven RBP states, which is $M'_{z0}=-a_{J0}/(A_{\mathrm{ex}}\alpha)$. Their shared stability (the curvature of $H'_{\mathbf{M}}$) is confirmed by $[\partial^{2} H'_{\mathbf{M}}/\partial (M'_{z})^{2}]_{M'_{z0}}=A_{\mathrm{ex}}>0$ (since $A_{\mathrm{ex}}$ is defined as positive for AFM exchange coupling). As well evidenced by the macrospin simulations (see Figs. \ref{Degenerate_states_snapshots} and \ref{Degenerate} (a) and \ref{Degenerate} (c)), in the absence of anisotropy ($k=0$), there exist stable degenerate SOT-driven RBP states with different combinations of $m_{zi}$ through setting different initial conditions, where $\mathbf{p} = \mathbf{e}_z$. Additionally, Fig. \ref{Degenerate} (b) and (d)  evidences that non-zero $M_{x}$ and $M_{y}$ components are indeed eliminated by the damping effect.

As a comparison, by the way, for an FM ($A_{\mathrm{ex}}<0$), due to $A_{\mathrm{ex}}<0$, the SOT-driven RBP states are obviously unstable, that is, the $H'_{\mathbf{M}}$ minimum will be at $P_{\mathrm{c}0}=-M_{z}=\pm1$, meaning that FM exchange couplings still favor parallel alignments of spins under the driving of the SOT. This point echoes the argument over why an AFM's collective modes (i.e., its internal degrees of freedom) are pre-constrained into a single vector under the excitation of SOTs (STTs)  (see Sec. \ref{sec:level1}).

In addition, what we would like to stress here is that Hu \textit{et al.} \cite{Hu2024} successfully applied a so-called coordinate transformation (a rotating frame technique), based on their previous works \cite{ZHANG2018458} as well as those of Chen \textit{et al.} \cite{Chen2019a,Chen2021}, to analytically solve the STT-driven dynamic states. However, in that work, there seems to be no clear reference on how to reduce these three spins to a single one by some constraints of spin variables in terms of the LLG equations. Furthermore, no proof was provided for the stability of those driven dynamic states using their technique. So, on the face of it, despite a certain degree of similarity with what we have done here, it would be beneficial for readers to fully understand their analytical results if the theoretical details about these two critical points were both explicitly offered.

\subsubsection{\label{sec:2C2} Particle Perspective: Frame of Reference Synchronous to the RBUT States}
Now that only the RBUT states can be driven by the SOT, thus similar to the vector perspective, exploiting the RBUTT nature of those RBUT states, a CM velocity boosting (or rather, a time-dependent RBUTT) can be performed to render them static: $(\phi'_{i},p'_{i})=(\phi_{i}-v(P_{\mathrm{c}},J)t,p_{i})$ where $v(P_{\mathrm{c}},J)$ is their CM velocity and $J$ is the injected current (density). The generator of this transformation is the CM momentum, $P_{\mathrm{c}}$. The resulting analytical formulation is consequently very similar to that in the vector perspective, specifically matching the generalized equation of motion (Eq. (\ref{any_quantity_new_frame})). The effective Hamiltonian in this CM velocity-boosted frame reads $H' = H'_{\mathrm{ex}}(|\mathbf{M}'|)+ H'_{\mathrm{boost}}(P'_{\mathrm{c}}, J)$, with $H'_{\mathrm{boost}}(P'_{\mathrm{c}}, J)=-3 \int^{P'_{\mathrm{c}}}v(P''_{\mathrm{c}},J) dP''_{\mathrm{c}}=-3\left(a_{J0}/\alpha\right)P'_{\mathrm{c}}$ where we have taken the CM velocity $v(P'_{\mathrm{c}},J) = a_{J0}/\alpha$ (the condition necessary for SOT cancellation). Here, $|\mathbf{M}'|$ has been expressed by the CM and RM variables, namely $(\Phi'_{\mathrm{c}}, P'_{\mathrm{c}})$ and $(\Phi'_{-12(23)}, P'_{-12(23)})$, according to Eqs. (\ref{inverssphi_to_Phi})-(\ref{inversep_to_P}).

Interestingly, different from the vector perspective, due to zero commutator relationships between CM and RM variables, the EO of RM variables is not affected by the CM velocity boosting at all. This suggests that such an EO is still present in the new frame, but will ultimately be eliminated by energy dissipation. It follows that the only stable states in this new frame are those pure RBUT degenerate states, sustained by the cooperation of the SOT and positive damping forces. For these stable states, the infinite number of combinations of stationary RM variables $(\Phi'_{-12(23)},P'_{-12(23)})$ must satisfy the equilibrium conditions: $\dot{P}'_{-12(23)}=0 \quad \text{and} \quad \dot{\Phi}'_{-12(23)}=0$, which is equivalent to satisfying Eq. (\ref{degenerate}) under the constraint that $M_{x}=0$ and $M_{y}=0$.

Therefore, for the RBUT states with $M_{x}=0$ and $M_{y}=0$, the Hamiltonian reduces to $H'_{\mathbf{P}}$, a function of $P'_{\mathrm{c}}$ only: $H'_{\mathbf{P}}(P'_{\mathrm{c}})=(9A_{\mathrm{ex}}/2) (P'_{\mathrm{c}})^2 - \left(3a_{J0}/\alpha\right) P'_{\mathrm{c}}$. By requiring $(\partial H'_{\mathbf{P}}/\partial P'_{\mathrm{c}})_{P'_{\mathrm{c}0}}=0$ and $[\partial^{2} H'_{\mathbf{P}}/\partial (P'_{\mathrm{c}})^{2}]_{P'_{\mathrm{c}0}}>0$, we solve for the degenerate $P'_{\mathrm{c}0}$ value of the stable SOT-driven RBUT states: $P'_{\mathrm{c}0}=a_{J0}/(3A_{\mathrm{ex}}\alpha)$. The shared stability is confirmed by the curvature: $[\partial^{2} H'_{\mathbf{P}}/\partial (P'_{\mathrm{c}})^{2}]_{P'_{\mathrm{c}0}}=9A_{\mathrm{ex}}>0$. This result is perfectly consistent with the vector perspective solution $M'_{z0}=-a_{J0}/(A_{\mathrm{ex}}\alpha)$ through the relation $M_{z} =-3P_{\mathrm{c}}$.

Receiving great evidence from the macrospin simulations (see Fig. \ref{Degenerate} (e)-(h)), in the absence of anisotropy ($k=0$), there exist stable degenerate SOT-driven RBUT states with different combinations of static M variables $(\Phi'_{-12(23)},P'_{-12(23)})$, evolving from different initial states. The simulations also evidence that the initial EO of RM variables will eventually be erased by the dissipation effect, confirming that the SOT only drives the overall CM motion.

With Eq. (\ref{any_quantity}), the equations of motion of CM variables for the RBUT states with the reduced Hamiltonian $H_{\mathrm{ex,\mathbf{P}}}=(9A_{\mathrm{ex}}/2)P_{\mathrm{c}}^{2}$ read
\setlength\abovedisplayskip{6pt}
\setlength\belowdisplayskip{6pt}
\begin{eqnarray}
\dot{P}_{\mathrm{c}}&=&[P_{\mathrm{c}},H_{\mathrm{ex,\mathbf{P}}}]+\left(\frac{1}{3}\right)
\sum_{i=1}^{3}\bigg(-\frac{\partial F_{\mathrm{d}}}{\partial\dot{\phi}_{i}}\bigg),\nonumber\\
&=&\left(\frac{1}{3}\right)\sum_{i=1}^{3}-\alpha(1-p_{i}^{2})\dot{\phi}_{i}+(1-p_{i}^{2})
a_{Ji},\nonumber\\
&=&\left[\left(\frac{1}{3}\right)\sum_{i=1}^{3}(1-p_{i}^{2})\right]\left(-\alpha\dot{\Phi}_{\mathrm{c}}
+a_{J0}\right),\nonumber\\
\dot{\Phi}_{\mathrm{c}}&=&[\Phi_{\mathrm{c}},H_{\mathrm{ex,\mathbf{P}}}]+\left(\frac{1}{3}\right)
\sum_{i=1}^{3}\frac{\partial F_{\mathrm{d}}}{\partial\dot{p}_{i}},\nonumber\\
&=&3A_{\mathrm{ex}}P_{\mathrm{c}}+\left(\frac{1}{3}\right)\sum_{i=1}^{3}
\frac{\alpha}{(1-p_{i}^{2})}\dot{p}_{i},\nonumber\\
&\approx&3A_{\mathrm{ex}}P_{\mathrm{c}},
\label{CM_timerate_with_STT}
\end{eqnarray}
with $|\dot{p}_{i}|$ being much smaller than $|\dot{\Phi}_{\mathrm{c}}|$ to be neglected.
 Notably, $p_{i}$ satisfying RBUT states' constraints (Eq. (\ref{degenerate}) with $M_{x(y)}=0$) shows that for different combinations of $p_{i}$ there will be different values of positive/negative damping forces due to them sharing a common factor $(1/3)\sum_{i=1}^{3}1-p_{i}^{2}$, but, failing to affect the SOT-generated frequency $\dot{\Phi}_{\mathrm{c}}=a_{J0}/\alpha$ at all for a stable SOT-driven RBP state with $\dot{P}_{\mathrm{c}}=0$. Hence, it is impossible to differentiate these driven degenerate RBP (RBUT) states solely through their generated frequency.

 On the face of it, the above analytical technique (a time-dependent RBRT or RBUTT) seems to be only applicable to the simple case with a uniform SOT(current density) injection into spins, i.e. $a_{Ji}=a_{J0}$. However, it can actually also be used to the cases more complicated with a non-uniform SOT (current density) injection. Hence, a generalization of this technique combining both vector and particle perspectives will be developed in Appendix \ref{appa} to address those more challenging issues.

\subsection{\label{sec:2D} Terminal Velocity Motion Model in the Absence of Uniaxial Anisotropy }
\subsubsection{\label{sec:2D1} SOT-driven Transient Evolution from the RBP state with $M_{x(y)}\neq0$ to that with $M_{x(y)}=0$ }
Having known that if the initial state is not at the one of the RBP states with $\mathbf{M}$ being collinear to $\mathbf{p}$ or RBUT ones, then the non-zero $M_{x(y)}$ or the EO of RM variables must be damped out immediately into disappearance. Thus, the complete SOT-driven evolution should contain this decaying process whose transient time could be estimated as follows. Given that taking the particle perspective to address this issue would be a tough job in mathematics, we here can only give a quite rough description about the relaxation of $\delta M_{x}$ or $\delta M_{y}$, which is a very small deviation away from $M_{x(y)}=0$, by using the energy balance technique.

\begin{figure}
\begin{center}
\includegraphics[width=7.0cm]{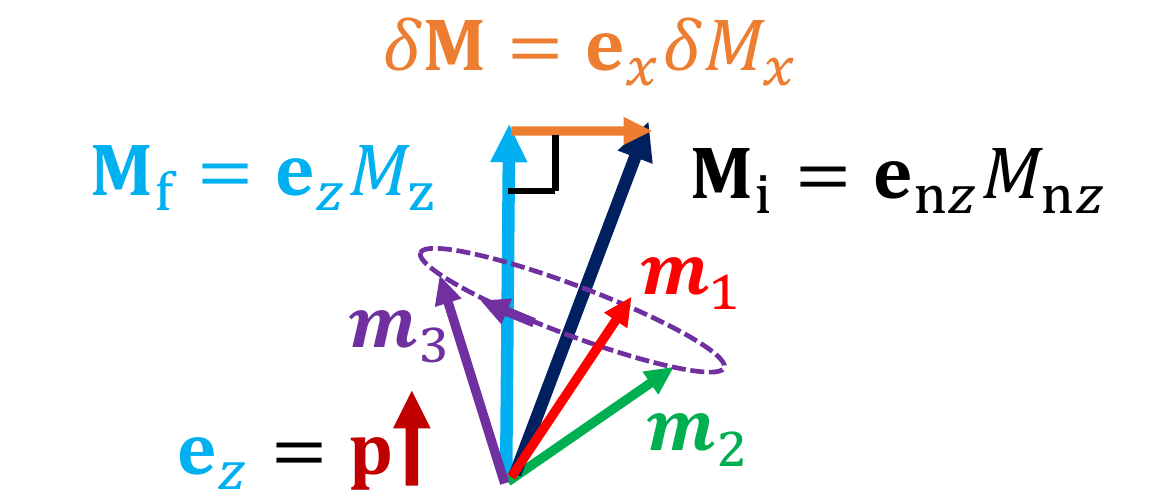}
\end{center}
\caption{(Color online)  Schematic of the transient evolution of the total magnetic moment $\mathbf{M}$. The figure illustrates the evolution from the initial RBP state $\mathbf{M}_{i}=\mathbf{e}_{\mathrm{n}z} M_{\mathrm{n}z}$ (marked by the black arrow) to the final stable SOT-driven state $\mathbf{M}_{f} = \mathbf{e}_{z}M_{z}$ (designated by the blue arrow). The spin-polarization vector $\mathbf{p}$ is indicated by the brown arrow. The earthy yellow arrow indicates the small difference $\delta\mathbf{M}=\mathbf{e}_{x}\delta M_{x}$ between $\mathbf{M}_{i}$ and $\mathbf{M}_{f}$, representing the components out of $\mathbf{p}$ eliminated during the transient process.}
\label{trainsien_Mx}
\end{figure}

As schematically shown in Fig. \ref{trainsien_Mx}, if the initial state has $\mathbf{M}_{\mathrm{i}}=M_{z}\mathbf{e}_{z}=M_{z}\mathbf{e}_{z}+\delta M_{x}\mathbf{e}_{x}$, then under the SOT only being able to drive those RBP states with $\mathbf{e}_{\mathrm{n}z}=\mathbf{p}=\mathbf{e}_{z}$ it is fairly reasonable to assume that the final stable state can have $\mathbf{M}_{\mathrm{f}}=M_{z}\mathbf{e}_{z}$ with $M_{z}=\sqrt{M_{\mathrm{n}z}^{2}-\delta M_{x}^{2}}$. Thus, the difference of the reduced Hamiltonians between the initial and final RBP states, respectively, which is also the energy of the RM's EO, is $\Delta H_{\mathrm{ex}}\equiv H_{\mathrm{ex,\mathbf{M}},\mathrm{f}}-H_{\mathrm{ex,\mathbf{M}},\mathrm{i}}=(A_{\mathrm{ex}}/2)\delta M_{x}^{2}$, according to Eq. (\ref{H_ex_M}).

As we know, $\Delta H_{\mathrm{ex}}$ should be damped out by the net energy dissipation that is given by the difference between the positive energy dissipation rate and the SOT-induced negative one. For simplicity, due to $\delta M_{x}$ being small enough, the SOT-induced energy injection rate for the initial state can roughly be represented by that for the final state, in which, as previously mentioned, the energy injection rate can be completely compensated by the dissipation rate. Thereby, $\Delta H_{\mathrm{ex}}$'s dissipation rate reads
\setlength\abovedisplayskip{6pt}
\setlength\belowdisplayskip{6pt}
\begin{eqnarray}
\frac{d\Delta H_{\mathrm{ex}}}{dt}&\approx&\frac{dH_{\mathrm{ex,\mathbf{M}},\mathrm{i}}}{dt}-
\frac{dH_{\mathrm{ex,\mathbf{M}},\mathrm{f}}}{dt},\nonumber\\
&=-\alpha&\sum_{i=1}^{3}\left[(1-p_{\mathrm{n}i}^{2})\dot{\Phi}_{\mathrm{nc}}^{2}
-(1-p_{i}^{2})\dot{\Phi}_{\mathrm{c}}^{2}\right],\nonumber\\
&\approx&-\alpha\left[\sum_{i=1}^{3}(1-p_{i}^{2})\right]\left(
\dot{\Phi}_{\mathrm{nc}}^{2}-\dot{\Phi}_{\mathrm{c}}^{2}\right),\nonumber\\
&\approx&-\alpha A_{\mathrm{ex}}^{2}\left[\sum_{i=1}^{3}(1-p_{i}^{2})\right]\delta M_{x}^{2},\nonumber
\label{}
\end{eqnarray}
with the use of the reasonable assumption of $p_{\mathrm{n}i}\approx p_{i}$ during the whole relaxation process and the use of $|\dot{\Phi}_{\mathrm{(n)c}}|=A _{\mathrm{ex}}|M_{\mathrm{(n)}z}|$.

 Combining $\Delta H_{\mathrm{ex}}$ as a function of $\delta M_{x}$, one immediately gets the rough relaxation equation for $\delta M_{x}$ as follows:
\setlength\abovedisplayskip{6pt}
\setlength\belowdisplayskip{6pt}
\begin{eqnarray}
\delta\dot{M}_{x}&\approx&-\alpha A_{\mathrm{ex}}\left[\sum_{i=1}^{3}(1-p_{i}^{2})\right]\delta M_{x}.
\label{Mnx_relax}
\end{eqnarray}
By requiring $\exp[-(\alpha A_{\mathrm{ex}})(\sum_{i=1}^{3}(1-p_{i}^{2}))]\sim\exp[-(3\alpha A_{\mathrm{ex}})(1-p_{0}^{2})]\sim0.01$, the estimated transient time is $t_{s}=-\ln0.01/[(3\alpha A_{\mathrm{ex}})(1-p_{0}^{2})]\sim9.8513$ $\mathrm{ps}$ for $J=1.8962\times10^{9}\mathrm{A}/\mathrm{cm}^{2}$ ($a_{J0}=0.01$), which is very close to the simulated result (see Fig. \ref{Degenerate}). 

\subsubsection{\label{sec:2D2} Terminal Velocity Motion Model for the SOT-driven Representative RBP State with the RM's Constraints $(\Phi_{-12(23)},P_{-12(23)})=(\pm2\pi/3,0)$ }

Theoretically, we fails to solve which degenerate state with $M_{x(y)}=0$ would become the final state under the SOT driving even in the absence of $\delta M_{x(y)}$, because it obviously depends on the initial condition. However, one can still obtain the representative dynamic behavior of such a system through the derivation of its CM variables's equations of motion under a particular set of stationary RM variables, such as $(\Phi_{-12(23)},P_{-12(23)})=(\pm2\pi/3,0)$, which is also one of the degenerate states in line with $M_{x(y)}=0$. Note that, the reason we choose such a combination of RM variables is that it is the only one that can hold the RM variables unchanged during the whole SOT-driven process.

It is fairly reasonable that if the initial states are very close to one of the ground states with $(\Phi_{\mathrm{c}},P_{\mathrm{c}})=(\Phi_{\mathrm{c}},0)$ and $(\Phi_{-12(23)},P_{-12(23)})=(\pm2\pi/3,0)$, then in the whole SOT-driven evolution the RM variables would be always kept at around $(\Phi_{-12(23)},P_{-12(23)})=(\pm2\pi/3,0)$, that is, $p_{i}\approx P_{\mathrm{c}}$. Therefore, in the particle perspective, with Eq. (\ref{CM_timerate_with_STT}),
one gets the equation for the second time derivative of $\Phi_{\mathrm{c}}$ as (see also Appendix \ref{appb})
\setlength\abovedisplayskip{6pt}
\setlength\belowdisplayskip{6pt}
\begin{eqnarray}
\ddot{\Phi}_{\mathrm{c}}&=&[[\Phi_{\mathrm{c}},H_{\mathrm{ex,\mathbf{P}}}],H_{\mathrm{ex,\mathbf{P}}}]
+A_{\mathrm{ex}}\sum_{i=1}^{3}\left(-\frac{\partial F_{\mathrm{d}}}{\partial\dot{\phi}_{i}}\right),\nonumber\\
&=&(3A_{\mathrm{ex}})\left[\left(\frac{1}{3}\right)
\sum_{i=1}^{3}(1-p_{i}^{2})\right]\left(-\alpha\dot{\Phi}_{\mathrm{c}}
+a_{J0}\right)\nonumber\\
&=&(3A_{\mathrm{ex}})\left[-\alpha S(P_{\mathrm{c}})\dot{\Phi}_{\mathrm{c}}+\beta(P_{\mathrm{c}},J)\right],\nonumber
\end{eqnarray}
with $S(P_{\mathrm{c}})=(1-P_{\mathrm{n}}^{2})$ and $\beta(P_{\mathrm{c}},J)=a_{J0}S(P_{\mathrm{c}})$.

Comparing to a Newton-like particle, we immediately obtain the definition of the effective mass of inertia, namely, $m_{\mathrm{eff}}(P_{\mathrm{c}})\equiv(1/3)(\partial^{2}H_{\mathrm{ex},\mathbf{P}}/\partial P_{\mathrm{c}}^{2})=(3A_{\mathrm{ex}})^{-1}$, which is directly related to the \textit{non-linear frequency shift coefficient}\cite{Chen2021,chen2023a}. Moreover, replacing the CM's momentum with the CM's angular velocity, i.e. $P_{\mathrm{c}}=[1/(3A_{\mathrm{ex}})]\dot{\Phi}_{\mathrm{c}}$, the CM's dynamic equation can be completely expressed in the configuration space $(\Phi_{\mathrm{c}},\dot{\Phi}_{\mathrm{c}})$ to be
\setlength\abovedisplayskip{6pt}
\setlength\belowdisplayskip{6pt}
\begin{eqnarray}
\ddot{\Phi}_{\mathrm{c}}
&=&\frac{1}{m_{\mathrm{eff}}(\dot{\Phi}_{\mathrm{c}})}\left[-\alpha S(\dot{\Phi}_{\mathrm{c}})\dot{\Phi}_{\mathrm{c}}+\beta(\dot{\Phi}_{\mathrm{c}},I)
\right],
\label{TVM_for_zerok}
\end{eqnarray}
with $S(\dot{\Phi}_{\mathrm{c}})=1-[\dot{\Phi}_{\mathrm{c}}/(3A_{\mathrm{ex}})]^{2}$ and $\beta(\dot{\Phi}_{\mathrm{c}},J)=a_{J0}S(\dot{\Phi}_{\mathrm{c}})$. This equation is termed the \textit{terminal velocity motion} (TVM) model, which is exactly the same as that derived through the \textit{Legendre} transformation introduced by Ref. \cite{chen2023a}.

As suggested by Ref. \cite{chen2023a}, the dynamical detail the TVM model has is analytically equal to that of the CM's equations expressed in the phase space. More importantly, Eq. (\ref{TVM_for_zerok}) clearly shows that the SOT-driven auto-oscillation of an NC-AFM has the four fundamental ingredients pressed by a linear TVM particle: \textbf{a}. effective mass of
inertia ($m_{\mathrm{eff}} = (3A_{\mathrm{ex}})^{-1}$), which is related to the non-linearity of the dynamic state energy (or rather, a blue shift); \textbf{b}. positive damping force ($-\alpha S(\dot{\Phi}_{\mathrm{c}})/m_{\mathrm{eff}}>0$); \textbf{c}. constant negative (anti-)damping force
$F_{\mathrm{dc}}=a_{J0}$ taken at $\dot{\Phi}_{\mathrm{c}}=0$; \textbf{d}. stability of terminal velocity $\dot{\Phi}_{\mathrm{c},T}=F_{\mathrm{c}}/\alpha$, which can be given as
follows: Linearizing Eq. (\ref{TVM_for_zerok}) around the terminal velocity $\dot{\Phi}_{\mathrm{c},T}$, one gets  $\delta\ddot{\Phi}_{\mathrm{c}}+\kappa\delta
\dot{\Phi}_{\mathrm{c}}=0$ with $\kappa=\alpha S(\dot{\Phi}_{\mathrm{c},T})/m_{\mathrm{eff}}>0$, implying that any small deviation in $\dot{\Phi}_{\mathrm{c}}$ that is away from $\dot{\Phi}_{\mathrm{c},T}$ will be dragged back to it by the velocity-restoring force, or rather, seen from its solution: $\delta\dot{\Phi}_{\mathrm{c}}(t)=C_{0}e^{-\kappa t}$. By requiring $e^{-\kappa t_{r}}\sim0.01$, the transient time is estimated as $t_{r}=-\ln(0.01)/\kappa\sim9.8513$ $\mathrm{ps}$ for $J=1.8962\times10^{9}\mathrm{A}/\mathrm{cm}^{2}$ ($a_{J0}=0.01$), which is in a perfect agreement with the simulation.

\subsection{\label{sec:2E} Terminal Velocity Motion Model in the Presence of the Uniaxial Anisotropy}
Taking a careful look at the uniaxial anisotropic energy expressed in terms of the canonical coordinates(see Eq. (\ref{Hamiltonian})), one finds such a minor energy is formed by the product of two types of factors: one can be treated as an in-plane anisotropy (IPA) depending solely on the phase angle $\phi_{i}$; the other can be viewed as an out-of-plane anisotropy (OPA) depending only on the z-component of the spins $m_{zi}$ (or $p_{i}$).

For an RBP (RBUT) state whose $\mathbf{M}$ is normal to the film plane the three symmetric axes of the anisotropy share, the IPA-related factor obviously becomes an ultra fast-oscillating term due to $\phi_{i}$ being an ultra fast-growing quantity with the velocity $\dot{\phi}_{i}=\dot{\Phi}_{\mathrm{c}}=A_{\mathrm{ex}}M_{z}$.
This, given that the IPA's oscillating period is much shorter than the characteristic time of the SOT-driven transient evolution (around $10$ ps, see Sec. \ref{sec:2D}), results in a time-averaged IPA-related factor  $\langle\cos^{2}(\phi_{i}-(i-1)(2\pi/3))\rangle_{T}\sim1/2$. It follows that there is in effect only the OPA-related factor left in the uniaxial anisotropy, i.e. $\langle H_{\mathrm{u}}(p)\rangle_{T}=(k/4)\sum_{i=1}^{3}p_{i}^{2}$.

\subsubsection{\label{sec:2E1} OPA-Induced Lifting of Energy Degeneracy of the SOT-Driven RBP (RBUT) States}
 Under the OPA-related energy term, i.e. $\langle H_{\mathrm{u}}\rangle_{T}$, the energy (Hamiltonian) degeneracy of the RBP states with $\mathbf{M}=\mathbf{e}_{z}M_{z}$ will be lifted, which can be easily seen from the fact that different combination of $m_{zi}$ (or $p_{i}$) meeting with the constraints $M_{z0}=\sum_{i=1}^{3}m_{zi}$ (or $P_{\mathrm{c}0}=(1/3)\sum_{i=1}^{3}p_{i}$) must have different $\langle H_{\mathrm{u}}\rangle_{T}$.

In the presence of the non-conservative effect, applying the time-dependent RBRT (RBUTT) to the system like we did in Sec. \ref{sec:2C}, we have the reduced total Hamiltonian in the rotating frame (or angular velocity boosting frame) as $H'_{\mathbf{M}}=H'_{\mathrm{ex},\mathbf{M}}+H'_{\mathrm{SOT}}+\langle H'_{\mathrm{u}}\rangle_{T}=(A_{\mathrm{ex}}/2)(M'_{z})^{2}+(a_{J0}/\alpha)
M'_{z}+(k/4)\sum_{i=1}^{3}(p'_{i})^{2}$. Notably, the energy terms involving $M'_{x(y)}$ have been damping out to disappear, as already argued in Sec. \ref{sec:2C}. In the following, one can find out the most stable state with the minimum $\langle H'_{\mathrm{u}}\rangle_{T}$ from those RBP states originally degenerate to the energy  $H'_{\mathrm{ex},\mathbf{M}}+H'_{\mathrm{SOT}}$ and $M'_{z0}=-a_{J0}/(A_{\mathrm{ex}}\alpha)$ ($P'_{\mathrm{c}0}=a_{J0}/(3A_{\mathrm{ex}}\alpha)$). Firstly, with the constraint $\sum_{i=1}^{3}p'_{i}=3P'_{\mathrm{c}0}$, it is reasonable to express $p'_{i}$ as $p'_{1}=P'_{\mathrm{c}0}+\delta$, $p'_{2}=P'_{\mathrm{c}0}+\Delta$, and $p'_{2}=P'_{\mathrm{c}0}-(\delta+\Delta)$. Then, we have $\langle H'_{\mathrm{u}}\rangle_{T}=(k/4)\sum_{i=1}^{3}(p'_{i})^{2}=(k/4)[3
(P'_{\mathrm{c}0})^{2}+\delta^{2}+\Delta^{2}+(\delta+\Delta)^{2}]$, which obviously has the minimum value $(3k/4)(P'_{\mathrm{c}0})^{2}$ for the most stable state, which is termed the state \textbf{s}, with $p'_{i}=P'_{\mathrm{c}0}$, or rather, with the constraints  $(\Phi'_{\mathrm{c}},P'_{\mathrm{c}})=(\Phi'_{\mathrm{c}},P'_{\mathrm{c}0})$ and $(\Phi'_{-12(23)},P'_{-12(23)})=(\pm2\pi/3,0)$.

 Notably, if the three spins' respective uniaxial anisotropic strengths $k_{i}$ are not identical, then following the above argument over which RBUT state is most stable, it can be concluded that the most stable state with the minimum value $\langle H'_{\mathrm{u}}\rangle_{T}$ would certainly not be the state \textbf{s}. This is quite different from the inference made by De \textit{et al.} \cite{Zhao2021} that the system's SOT-driven magnetic configuration choice is affected by the ($120^{\circ}$) rotational symmetry of such an anisotropy, i.e. IPA, around the z axis.

\subsubsection{\label{sec:2E2} OPA-Related Hamiltonian Driven Slow Decaying Oscillation Going Trough the Degenerate RBP (RBUT) States}
 Under the OPA-related Hamiltonian, as we have known, the system will eventually evolve into the state \textbf{s} by the damping effect no matter which initial condition it begins from. The next question will thus be how the system evolves into such a final state. We would like to adopt the vector as well as particle perspectives, respectively, to address this issue.

In the absence of the non-conservative effect, it can be found, in the vector perspective, that the constraints $M_{x(y)}=0$ will be violated by the OPA-related Hamiltonian $\langle H_{\mathrm{u}}\rangle_{T}$, namely,  $\dot{M}_{x}=[M_{x},\langle H_{\mathrm{u}}\rangle_{T}]
=(k/2)\sum_{i=1}^{3}m_{yi}m_{zi}$ and $\dot{M}_{y}=[M_{y},\langle H_{\mathrm{u}}\rangle_{T}]
=-(k/2)\sum_{i=1}^{3}m_{xi}m_{zi}$, which are almost both non-zero except for the state \textbf{s}, e.g. a randomly picked up SOT-driven RBP state with $(m_{x1},m_{y1},m_{z1})=(0.3263,0.8791,-0.3473)$, $(m_{x2},m_{y2},m_{z2})=(-0.8634,-0.0491,-0.5021)$,
and $(m_{x3},m_{y3},m_{z3})=(0.5370,-0.83,-0.1506)$ obtained from the macrospin simulation having $\dot{M}_{x}=-(k/2)(0.1557)$ and $\dot{M}_{y}=-(k/2)(0.2393)$.

\begin{figure*}
\begin{center}
\includegraphics[width=17cm]{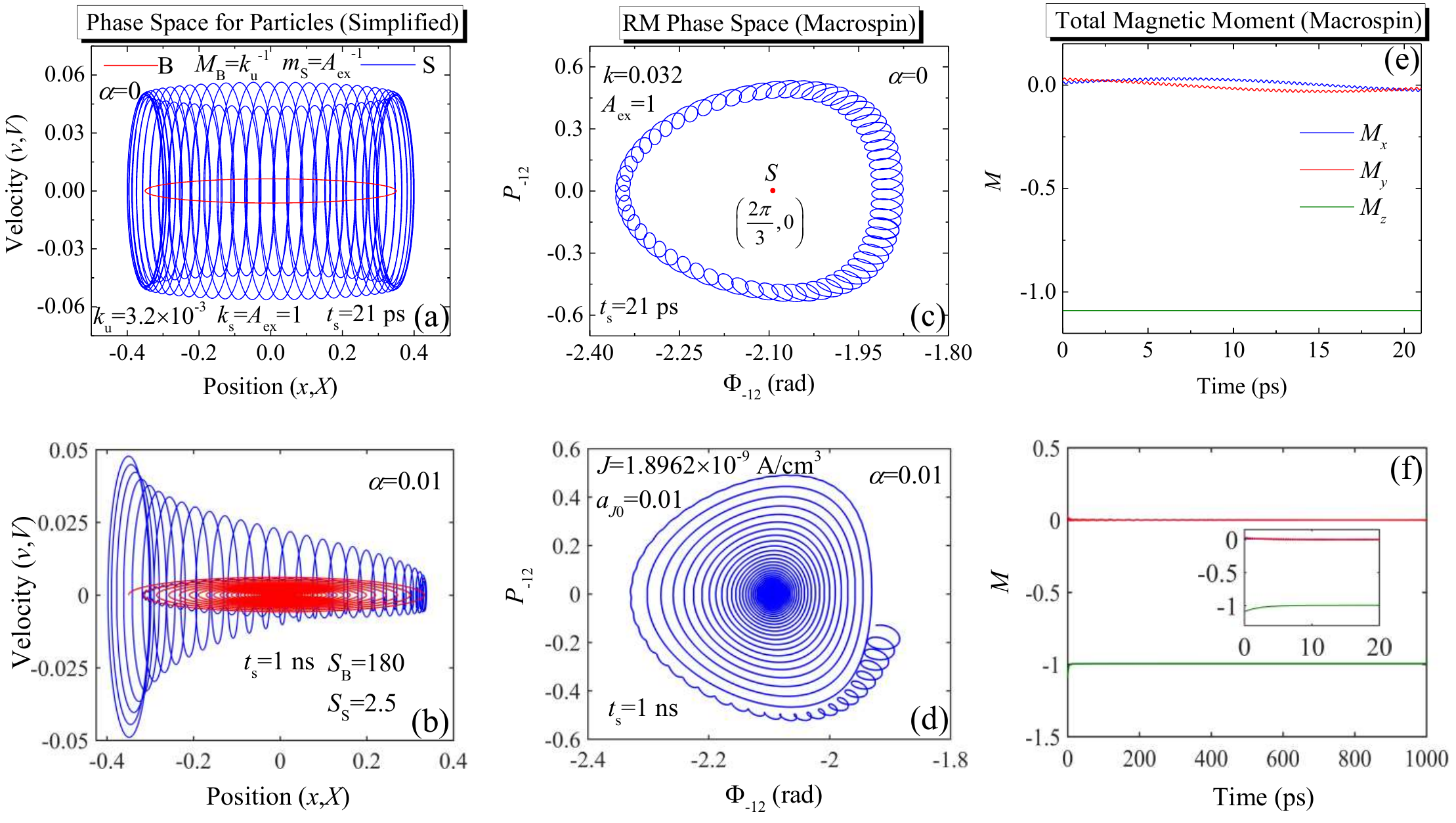}
\end{center}
\caption{(Color online) Analogy of the OPA-driven RM dynamics: Comparison of the simplified model and macrospin simulations. The figure compares the phase space trajectories derived from the simplified model (Panels (a) and (b)) and the macrospin simulation (Panels (c) and (d)) as an analogy for the RM's transient evolution driven by the OPA-related Hamiltonian $\langle H_{\mathrm{ru}} \rangle_{T}$. \textbf{A. Phase Space Comparison} (Panels (a)-(d)): \textbf{Conservative Case} ($\alpha=0$): Panels (a) and (c). \textbf{Non-Conservative Case} ($\alpha=0.01$): Panels (b) and (d). \textbf{Simplified Model Parameters} (Panels (a) and (b)): The trajectories for particles B (red) and S (blue) are shown. Their inertial masses are $M_{\mathrm{B}}=k_{\mathrm{u}}^{-1}=(3.2\times10^{-3})^{-1}$ and $m_{\mathrm{S}}=A_{\mathrm{ex}}^{-1}=1$, respectively, and the elastic coefficient $k_{\mathrm{s}}$ is set equal to the scaled exchange coupling strength $A_{\mathrm{ex}} = 1$.
\textbf{Damping Factors} (Panel (b) only): The effective positive damping factors are $S_{\mathrm{B}}=180\text{ and}$ $S_{\mathrm{S}}=2.5$. \textbf{Simulation Time}: $t_{\mathrm{s}}=21\text{ps}$ for the conservative cases and $t_{\mathrm{s}}=1\text{ns}$ for the non-conservative cases. \textbf{B. Total Magnetic Moment Time Traces} (Panels (e) and (f)): Panels (e) ($\alpha=0$) and (f) ($\alpha=0.01$) show the time traces of the total magnetic moment $\mathbf{M}$ components: $M_{x}$ (blue), $M_{y}$ (red), and $M_{z}$ (olive). \textbf{Zoom-in}: The inset in Panel (f) provides a zoom-in of the simulated results within the first $20 \text{ps}$.}
\label{Longtermdecay}
\end{figure*}

In other words, the minor term $\langle H_{\mathrm{u}}\rangle_{T}$ will draw the system slightly out of one of the degenerate RBP (RBUT) states with $\mathbf{e}_{\mathrm{n}z}=\mathbf{p}=\mathbf{e}_{z}$, moreover, giving rise to a change in the z components of spins, i.e. $\dot{m}_{zi}=[m_{zi},H]=[m_{zi},(A_{\mathrm{ex}}/2)
(M_{x}^{2}+M_{y}^{2})]=A_{\mathrm{ex}}(-m_{yi}
M_{x}+m_{xi}M_{y})$. However, due to $[M_{z},\langle H_{\mathrm{u}}\rangle_{T}]=0$, the whole evolutional
process under the action of $\langle H_{\mathrm{u}}\rangle_{T}$, $M_{z}$ will be kept conserved, as verified by the macrospin simulation in Fig. \ref{Longtermdecay}(e). This also suggests that thanks to the damping effect the transient states that the system goes through, before it going into the state \textbf{s}, should be those RBP states sharing a $M_{z}$ with a quick decaying of an exceedingly tiny $\delta M_{x(y)}$ (around $10$ ps, just as the estimation made in Sec. \ref{sec:2D1}), as verified by the macrospin simulation in Fig. \ref{Longtermdecay}(f) and its panel.

It is very hard to make an insight more deeply into the evolution of $\mathbf{m}_{i}$ with time in the vector perspective, thereby, we here turn to take the particle one. From the RM variables' conservative equations of motion
\setlength\abovedisplayskip{6pt}
\setlength\belowdisplayskip{6pt}
\begin{eqnarray}
\dot{P}_{\mathrm{-12}}&=&[P_{\mathrm{-12}},H]
,\nonumber\\
&=&[P_{\mathrm{-12}},H_{\mathrm{ex}}],\nonumber\\
&=&A_{\mathrm{ex}}\bigg[2\sqrt{\left(1-p_{1}^{2}\right)\left(1-p_{2}^{2}\right)}
\sin\Phi_{-12}\nonumber\\
&&+\sqrt{\left(1-p_{1}^{2}\right)\left(1-p_{3}^{2}\right)}
\sin\left(\Phi_{-12}+\Phi_{-23}\right)\nonumber\\
&&-\sqrt{\left(1-p_{2}^{2}\right)\left(1-p_{3}^{2}\right)}
\sin\Phi_{-23}\bigg],
\label{P_n12}
\end{eqnarray}

\setlength\abovedisplayskip{6pt}
\setlength\belowdisplayskip{6pt}
\begin{eqnarray}
\dot{\Phi}_{-12}&=&[\Phi_{-12},H]
,\nonumber\\
&=&\left(\frac{k}{2}\right)P_{-12}+\left[\Phi_{-12},\left(
\frac{A_{\mathrm{ex}}}{2}\right)\left(M_{x}^{2}+M_{y}^{2}\right)
\right],\nonumber\\
&=&\left(\frac{k}{2}\right)P_{-12}+A_{\mathrm{ex}}\Bigg\{p_{2}
\Bigg[1+\Bigg(\sqrt{\frac{1-p_{1}^{2}}
{1-p_{2}^{2}}}\nonumber\\
&&\times\cos\Phi_{-12}+\sqrt{\frac{1-p_{3}^{2}}{1-p_{2}^{2}}}\cos\Phi_{-23}\Bigg)\Bigg]
-p_{1}\nonumber\\
&&\times\Bigg[1+\Bigg(
\sqrt{\frac{1-p_{2}^{2}}{1-p_{1}^{2}}}\cos\Phi_{-12}+
\sqrt{\frac{1-p_{3}^{2}}{1-p_{1}^{2}}}\nonumber\\
&&\times\cos\left(\Phi_{-12}+\Phi_{-23}\right)\Bigg]\Bigg\},
\label{Phi_n12}
\end{eqnarray}

\setlength\abovedisplayskip{6pt}
\setlength\belowdisplayskip{6pt}
\begin{eqnarray}
\dot{P}_{\mathrm{-23}}&=&[P_{\mathrm{-23}},H]
,\nonumber\\
&=&[P_{\mathrm{-23}},H_{\mathrm{ex}}],\nonumber\\
&=&A_{\mathrm{ex}}\bigg[-\sqrt{\left(1-p_{1}^{2}\right)\left(1-p_{2}^{2}\right)}
\sin\Phi_{-12}\nonumber\\
&&+2\sqrt{\left(1-p_{2}^{2}\right)\left(1-p_{3}^{2}\right)}
\sin\Phi_{-23}\nonumber\\
&&+\sqrt{\left(1-p_{1}^{2}\right)\left(1-p_{3}^{2}\right)}
\sin\left(\Phi_{-12}+\Phi_{-23}\right)\bigg],\nonumber\\
\label{P_n23}
\end{eqnarray}
and
\setlength\abovedisplayskip{6pt}
\setlength\belowdisplayskip{6pt}
\begin{eqnarray}
\dot{\Phi}_{-23}&=&[\Phi_{-23},H],\nonumber\\
&=&\left(\frac{k}{2}\right)P_{-23}+\left[\Phi_{-23},\left(
\frac{A_{\mathrm{ex}}}{2}\right)\left(M_{x}^{2}+M_{y}^{2}\right)\right],\nonumber\\
&=&\left(\frac{k}{2}\right)P_{-23}+A_{\mathrm{ex}}\Bigg\{-p_{2}
\Bigg[1+\Bigg(\sqrt{\frac{1-p_{1}^{2}}{1-p_{2}^{2}}}
\nonumber\\
&&\cos\Phi_{-12}+\sqrt{\frac{1-p_{3}^{2}}{1-p_{2}^{2}}}
\cos\Phi_{-23}\Bigg)\Bigg]\nonumber\\
&&+p_{3}\Bigg[1+\Bigg(
\sqrt{\frac{1-p_{1}^{2}}{1-p_{3}^{2}}}\cos\left(\Phi_{12}
+\Phi_{-23}
\right)\nonumber\\
&&+\sqrt{\frac{1-p_{2}^{2}}{1-p_{3}^{2}}}\cos\Phi_{-23}\Bigg)\Bigg]\Bigg\},\nonumber\\
\label{Phi_n23}
\end{eqnarray}
with $H=H_{\mathrm{ex}}+\langle H_{\mathrm{u}}\rangle_{T}$, one can find for an RBUT state with $[P_{-12(23)},H_{\mathrm{ex}}]=0$ as well as $[\Phi_{-12(23)},(A_{\mathrm{ex}}/2)\left(M_{x}^{2}
+M_{y}^{2}\right)]=0$, the RM's variables are originally stationary
in the absence of $\langle H_{\mathrm{u}}\rangle_{T}$. But, once with $\langle H_{\mathrm{u}}\rangle_{T}$, RM's variables are no longer time-constants except at the state \textbf{s}, which can be unveiled in the following.

The right-hand sides on Eqs. (\ref{P_n12}) and (\ref{P_n23}), as mentioned in Sec. \ref{sec:2B3}, resemble \textit{internal} restoring forces used to hold a rigid body in motion, or rather, to phase-lock the precessing spins. Thus, the terms $(k/2)P_{-12(23)}$ appearing on the right-hand sides of Eqs. (\ref{Phi_n12}) and (\ref{Phi_n23}), apparently, will cause very small (angular) velocity differences $\dot{\Phi}_{-12(23)}$ between particles for those degenerate RBUT states with non-zero $P_{-12(23)}$, namely, the OPA-related Hamiltonian very slightly ruining the original phase-locking between spins, yielding, more specifically, a slight distortion of the rigid body.

\begin{figure}
\begin{center}
\includegraphics[width=7.3cm]{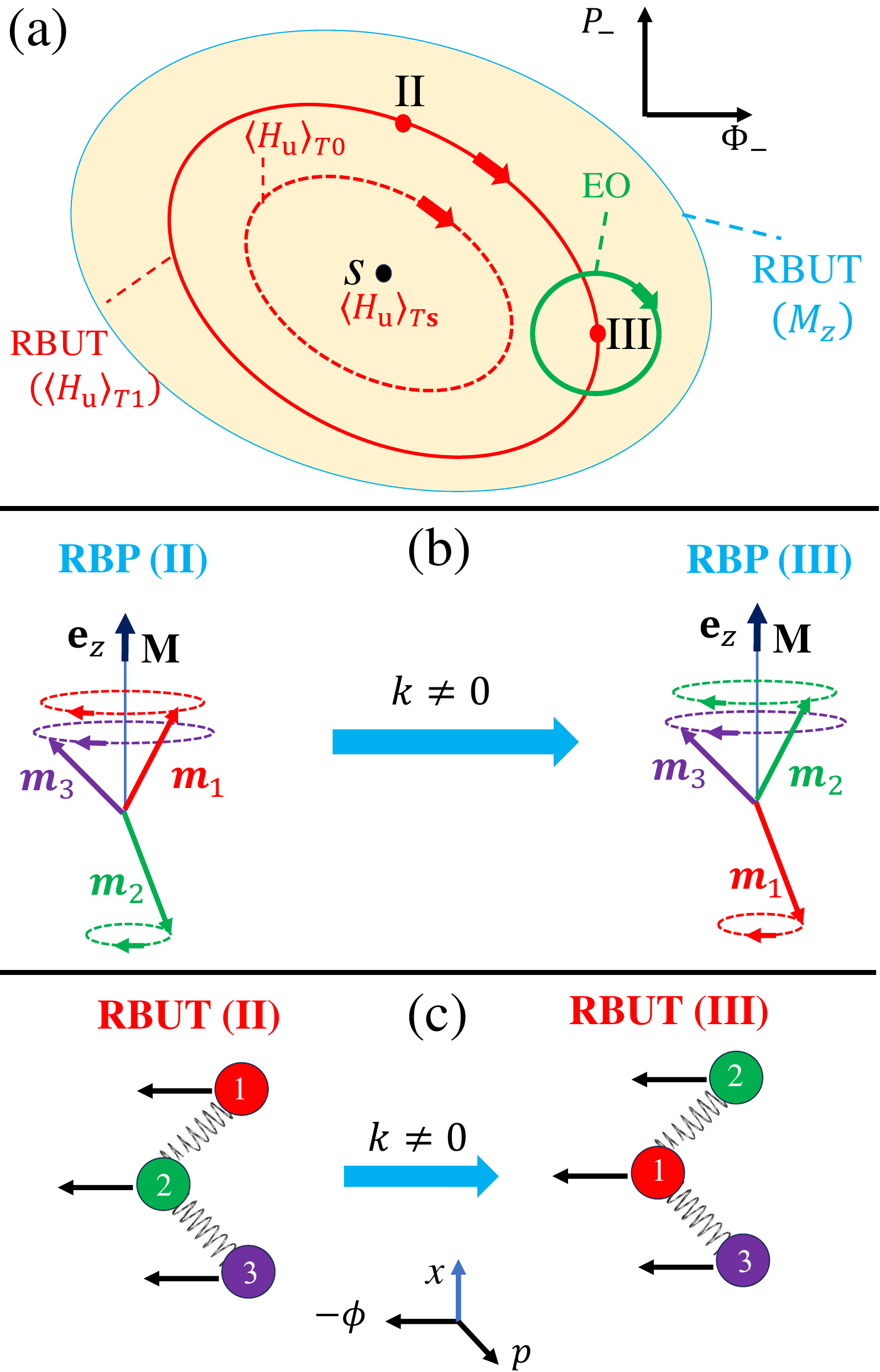}
\end{center}
\caption{(Color online) Schematic of the OPA-related Hamiltonian $\langle H_{u}\rangle_{T}$ driven oscillation in RM phase space. Panel (a) shows the phase space spanned by the RM variables ($\Phi_{-12(23)}, P_{-12(23)}$). The light transparent yellow area is formed by the group of degenerate RBUT states (RBP states with $\mathbf{e}_{\mathrm{n}z}=\mathbf{p}=\mathbf{e}_{z}$) that share the same SOT-driven $M_{z0}$ and exchange energy $H_{\mathrm{ex}}$. The red dashed and solid trajectories, which enclose the equilibrium state \textbf{s}, represent contours of constant $\langle H_{\mathrm{u}} \rangle_{T0}$ and $\langle H_{\mathrm{u}} \rangle_{T1}$, respectively, where $\langle H_{\mathrm{u}} \rangle_{T1} > \langle H_{\mathrm{u}} \rangle_{T0} > \langle H_{\mathrm{u}} \rangle_{T\mathbf{s}}$. The small green trajectory enclosing one of the RBUT states on the red solid trajectory indicates the RM's EO. The evolution from one of the degenerate states II to another state III along the red trajectory can be specifically seen from the vector viewpoint (Panel (b)) and the particle viewpoint (Panel (c)), respectively.}
\label{complex_osci}
\end{figure}

 Such an un phase-locking will further induce two kinds of oscillations:
 one, just as suggested by the above analysis that $\langle H_{\mathrm{u}}\rangle_{T}$ will cause $M_{x(y)}\neq0$, which also actually states $\dot{P}_{-12(23)}\neq0$ and $\dot{\Phi}_{-12(23)}\neq0$, is the EO of RM variables around one of the equilibrium points (see Fig. \ref{complex_osci}(a)) satisfying $\dot{P}_{-12(23)}=0$ and $\dot{\Phi}_{-12(23)}=0$ (RBUT states or RBP states with $\mathbf{e}_{\mathrm{n}z}=\mathbf{p}=\mathbf{e}_{z}$) whose frequency (several hundreds of GHz to several THz) is as high as that of the RBP state (see Secs. \ref{sec:2B3} and \ref{sec:2B4}); the other, when neglecting the terms $[\Phi_{-12(23)},(A_{\mathrm{ex}}/2)\left(M_{x}^{2}
+M_{y}^{2}\right)]$ on the right-hand sides of Eqs. (\ref{Phi_n12}) and (\ref{Phi_n23}) for an RBUT state and thereby the relationship $\dot{\Phi}_{-12(23)}=(k/2)P_{-12(23)}$ being able to be treated as an analogy of $v=p/m$ in terms of the Newton mechanics, is the oscillation around the state \textbf{s} (see Fig. \ref{complex_osci}(a)) with a much lower frequency (several to tens of GHz) than that of the EO of RM variables, which is owing to $k=0.032\ll A_{\mathrm{ex}}=1$, or rather, a huge effective mass of inertia $m=2/k$ only causing tiny responses to the restoring forces appearing in Eqs. (\ref{P_n12}) and (\ref{P_n23}), namely, very small accelerations in the phase-angular differences $\ddot{\Phi}_{-12(23)}$. Note that, during such a unique oscillation, all of the states that the system goes through are certainly very close to the RBUT states (RBP states with $\mathbf{e}_{\mathrm{n}z}=\mathbf{p}=\mathbf{e}_{z}$) that are degenerate to the (SOT-driven) $M_{z}$ and their shared $\langle H_{\mathrm{u}}\rangle_{T}$, which is based on the assumption of $M_{x(y)}=0$, as schematically displayed by Fig. \ref{complex_osci} (b) and (c).

These two kinds of oscillations driven by the OPA-related Hamiltonian can't in fact exist independently, which is because a non-zero $M_{x(y)}$ (or the EO of the RM's variables) will be produced immediately upon the un phase-locking between spins (distortion of the rigid body) occurring. Thus, the complete conservative motion the system would experience in the phase space spanned by the four RM's variables under the influence of the OPA-related Hamiltonian should be the revolution around the state \textbf{s} at a very slow pacing on a trajectory formed by a group of RBUT states sharing the same $\langle H_{\mathrm{u}}\rangle_{T}$ and the SOT-driven $M_{z}$, accompanied by a high frequency tiny amplitude EO of RM's variables about one of those RBUT states (see Fig. \ref{complex_osci}(a)). On the face of it, such an oscillation looks a little bit like the Earth's revolution accompanied by its rotation.

\begin{figure}
\begin{center}
\includegraphics[width=5.5cm]{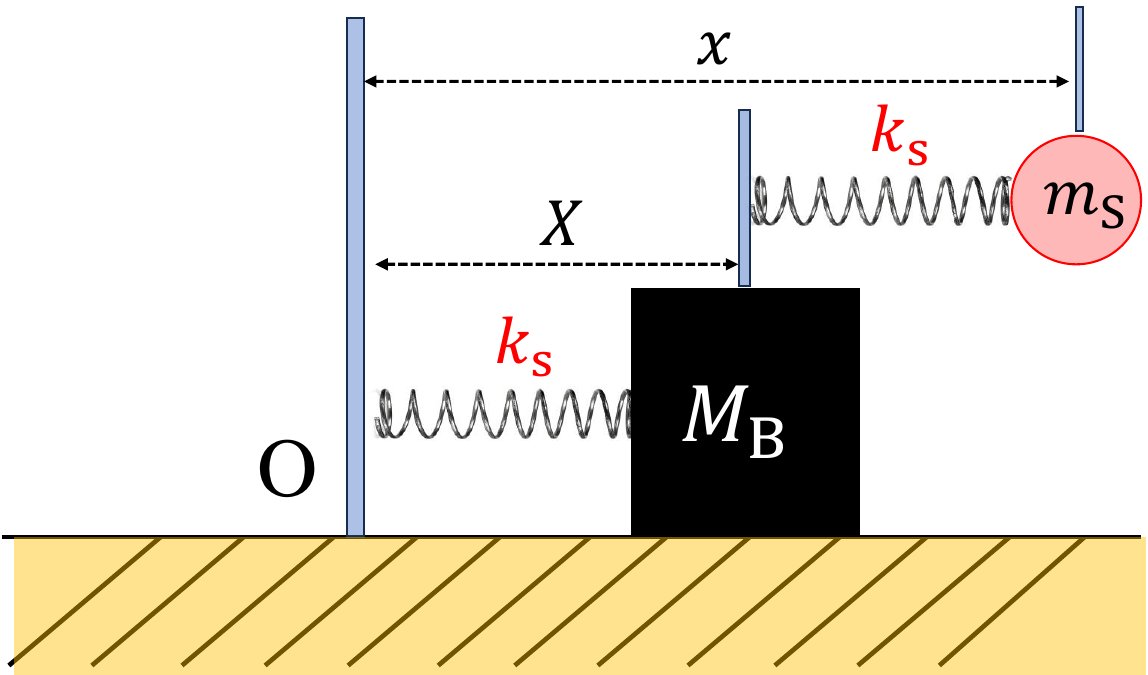}
\end{center}
\caption{(Color online) Schematic of the simplified model for the OPA-related Hamiltonian $\langle H_{\mathrm{u}} \rangle_T$ driven oscillation in terms of the particle perspective. The model consists of two string-coupled massive particles: a black square (\textbf{B}) and a light red circle (\textbf{S}). The parameters are defined as follows: \textbf{Masses of inertia}: $M_{\mathbf{B}}$ (for B) and $m_{\mathbf{S}}$ (for S), with $M_{B}\gg m_{S}$; \textbf{Elastic coefficients}: The coefficients of both the strings (connecting $B$ to $S$ and $B$ to $O$) are both $k_{\mathbf{S}}$; \textbf{Positions}: The instantaneous positions of particles $B$ and $S$ are $X$ and $x$, respectively.}
\label{Simplified_model}
\end{figure}

Though analytically solving such a complicated non-linear oscillation from Eqs. (\ref{P_n12})-(\ref{Phi_n23}) is simply impossible, it is still possible for us to make a simplified model for it. As schematically shown in Fig. \ref{Simplified_model}, this mechanic system is made up with a linear coupled pair of massive particles, where one particle termed B with a big
mass of inertia $M_{\mathrm{B}}$ is tied to the origin "O" by a linear spring with an elastic coefficient $k_{\mathrm{s}}$; the other termed S with a light mass $m_{\mathrm{s}}$ is hooked up to particle B by a linear spring with the same coefficient $k_{\mathrm{s}}$, as shown in Fig. \ref{Simplified_model}.

The reason why such a model's dynamic behavior could resemble the OPA-induced oscillation is that the dimensionality of this unique oscillation is four, including two of them for the fast part of the oscillation, i.e. the EO of RM's variables, which originates from the nature of
RBRT of $H_{\mathrm{ex}}$ and can be modeled by an SHO, as explained in Secs. \ref{sec:2B3} and \ref{sec:2B4}; the other two of them for the slow part of the oscillation, which can be easily seen from the original four-dimensional RM's phase space being reduced to a two-dimensional one by the two constraints of RBUT states (RBP states with $\mathbf{e}_{\mathrm{n}z}=\mathbf{p}=\mathbf{e}_{z}$) $M_{x}=M_{y}=0$. This means that, either qualitatively or quantitatively, using the two coupled massive particles with a very big mass difference between them to roughly unveil this slow decaying oscillatory phenomenon should be fairly appropriate.

Then, the Lagrangian for this system is $L=(1/2)M_{\mathrm{B}}\dot{X}^{2}+(1/2)m_{\mathrm{S}}\dot{x}^{2}-(1/2)
k_{\mathrm{s}}X^{2}-(1/2)k_{\mathrm{s}}(x-X)^{2}$, with $x$ and $X$ being the positions of the particles S and B, respectively. Thus, their equations of motion read
\setlength\abovedisplayskip{6pt}
\setlength\belowdisplayskip{6pt}
\begin{eqnarray}
M_{\mathrm{B}}\ddot{X}&=&-k_{\mathrm{s}}X+k_{\mathrm{s}}(x-X),\nonumber\\
m_{\mathrm{S}}\ddot{x}&=&-k_{\mathrm{s}}(x-X).
\label{sim_model}
\end{eqnarray}
Eq. (\ref{sim_model}) gives us an intuitive inference that once $M_{\mathrm{B}}\gg m_{\mathrm{S}}$ and $|X|$ is significantly bigger than $|x-X|$, then the acceleration $\ddot{X}$ particle B feels, which is mainly caused by the spring linked to the origin O, is certainly much smaller than that particle S experiences, which is induced by the spring tied to B, resulting in a much longer time period of B's oscillation around the point O than that of S's oscillation about B.

\begin{figure*}
\begin{center}
\includegraphics[width=13cm]{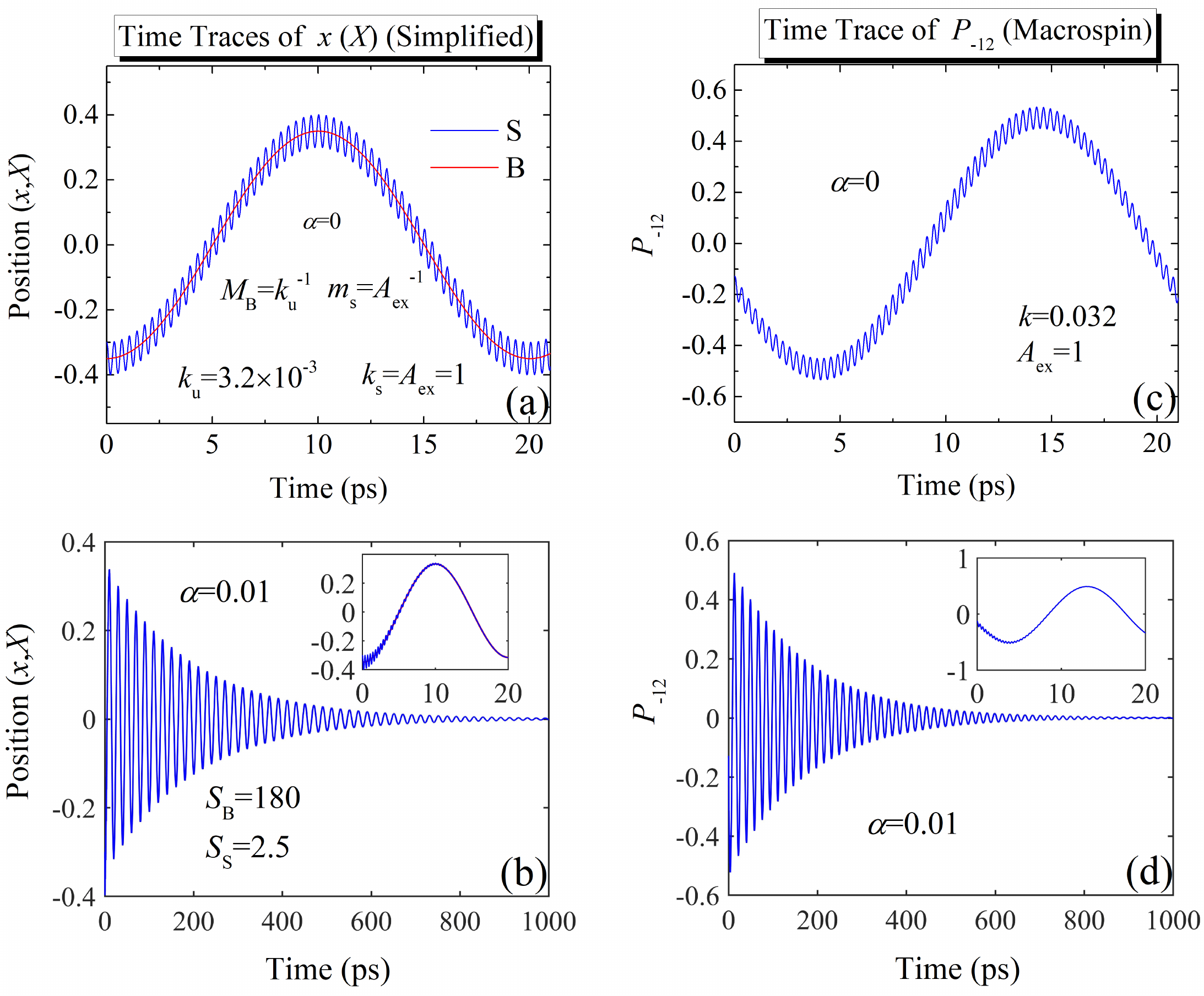}
\end{center}
\caption{(Color online) Time traces comparing simplified model and macrospin simulation. The figure presents time traces of the particles' positions $x(X)$ from the simplified model ((a) and (b)) and the RM variable $P_{-12}$ from the macrospin simulation ((c) and (d)). These are shown as an analogy for the OPA-related Hamiltonian $\langle H_{\mathrm{ru}} \rangle_T$ driven transient evolution of the RM variables. Here are \textbf{key details}. \textbf{Model Correspondence}: Panels (a) and (c) represent \textbf{conservative cases}, while (b) and (d) represent \textbf{non-conservative cases}. \textbf{Simplified Model Variables ((a) and (b))}: Red and blue curves indicate the trajectories for particles \textbf{B} and \textbf{S}, respectively. \textbf{Parameters}: All relative set parameters are identical to those used in Fig. \ref{Longtermdecay}. \textbf{Zoom-ins}: The insets in panels (b) and (d) are zoom-ins focusing on the simulated results within the first $20 \text{ ps}$.}
\label{Longtermdecaya}
\end{figure*}

This point can get verified through conducting the simulation of Eq. (\ref{sim_model}) (see Figs. \ref{Longtermdecay}(a) and \ref{Longtermdecaya}(a)), where the values of all of the relative normalized parameters are set to be $M_{\mathrm{B}}=k_{\mathrm{u}}^{-1}$ with $k_{\mathrm{u}}=3.2\times10^{-3}$, $m_{\mathrm{S}}=A_{\mathrm{ex}}^{-1}$ with $A_{\mathrm{ex}}=1$, $k_{\mathrm{s}}=A_{\mathrm{ex}}$, and $\alpha=0$, so that the simplified model's simulation can be made numerically close enough to the macrospin simulated results, as shown in Figs. \ref{Longtermdecay}(c) and \ref{Longtermdecaya}(c). Notably, the dimensionless position of particle S $x$ or its velocity $v=\dot{x}$ can have a direct analogy with either  $P_{-12(23)}$ or $\Phi_{-12(23)}$, see also the blue curves in Figs. \ref{Longtermdecay}(c) and \ref{Longtermdecaya}(c). The dimensionless position of particle B $X$, as displayed by the red curves in Figs. \ref{Longtermdecay}(a) and \ref{Longtermdecaya}(a), corresponds to the time-averaged trajectory only representing the slow part of the $\langle H_{\mathrm{u}}\rangle_{T}$-driven oscillation in the spin system, as shown in Figs. \ref{Longtermdecay}(c) and \ref{Longtermdecaya}(c). This time-averaged trajectory, as previously discussed, are formed by those degenerate RBUT stats sharing $M_{z}$ and $\langle H_{\mathrm{u}}\rangle_{T}$.

 If $M_{\mathrm{B}}$ goes to infinity or $k_{\mathrm{u}}$ approaches zero, then, evidently, $\ddot{X}\rightarrow0$ (or $X(t)\approx X_{0}+V_{0}t$) with $X_{0}$ and $V_{0}$ being integration constants (initial position and velocity). This means that only the oscillation of particle S around particle B's instantaneous position $X(t)$ remains. This outcome directly corresponds to the case of the absence of uniaxial anisotropy ($k=0$) in the spin system, where only an EO of RM's variables can exist around one of the RBUT states (RBP states with $\mathbf{e}_{\mathrm{n}z}=\mathbf{p}=\mathbf{e}_{z}$) that are degenerate to the SOT-driven $M_{z}$, just as pointed out in Secs. \ref{sec:2B3} and \ref{sec:2B4}.

\setlength\abovedisplayskip{6pt}
\setlength\belowdisplayskip{6pt}
\begin{eqnarray}
M_{\mathrm{B}}\ddot{X}&=&-k_{\mathrm{s}}X+k_{\mathrm{s}}(x-X)
-\alpha S_{\mathrm{B}}\dot{X},\nonumber\\
m_{\mathrm{S}}\ddot{x}&=&-k_{\mathrm{s}}(x-X)-\alpha S_{\mathrm{S}}\dot{x}.
\label{sim_model_damp}
\end{eqnarray}

In the presence of the damping effect ($\alpha=0.01$), as simulated based on Eq. (\ref{sim_model_damp}) with the effective damping factors being $S_{\mathrm{B}}=180$ and $S_{\mathrm{S}}=2.5$, particle S's oscillation will be quickly damped out within approximately $10$ ps (see the panel in Fig. \ref{Longtermdecaya}(b)) due to its very high natural angular frequency ($\omega_{\mathrm{s}}=\sqrt{k_{\mathrm{s}}/m_{\mathrm{s}}}=A_{\mathrm{ex}}$).
In addition, with the energy dissipation at every moment, the complete suppression of particle S's oscillation can be seen almost throughout the evolution.
This means long before particle B's oscillation gradually decaying to the point O, which takes around $1$ ns, particle S has already been phase-locked to particle B, as evidenced by the quick merge of the red and blue trajectories in Figs. \ref{Longtermdecay} (b) and \ref{Longtermdecaya} (b). Thus, with the damping effect, this four dimensional oscillation is in effect reduced to a two dimensional oscillation. Notably, the factors $S_{\mathrm{B}}$ as well as $S_{\mathrm{s}}$ are obtained under the requirement that the simplified simulated results numerically align closely with the macrospin simulation.

Interestingly, the macrospin simulation also displays a similar behavior, as can be seen in Figs. \ref{Longtermdecay}(d) and \ref{Longtermdecaya}(d). The RM's variables' EO is being so quickly dissipated within around $10$ ps (see the panel in Fig. \ref{Longtermdecaya}(d)), which is in good line with the estimation of the time scale on the EO's decaying made in Sec. \ref{sec:2D1}, that the system's following evolution into the state \textbf{s} will completely be restricted on a trajectory formed by the SOT-driven $M_{z}$-degenerate RBUT states (RBP states with $\mathbf{e}_{\mathrm{n}z}=\mathbf{p}=\mathbf{e}_{z}$, see also Fig. \ref{Longtermdecay}(f)), resulting in it being reduced to a two-dimension motion.

\begin{figure*}
\begin{center}
\includegraphics[width=17.7cm]{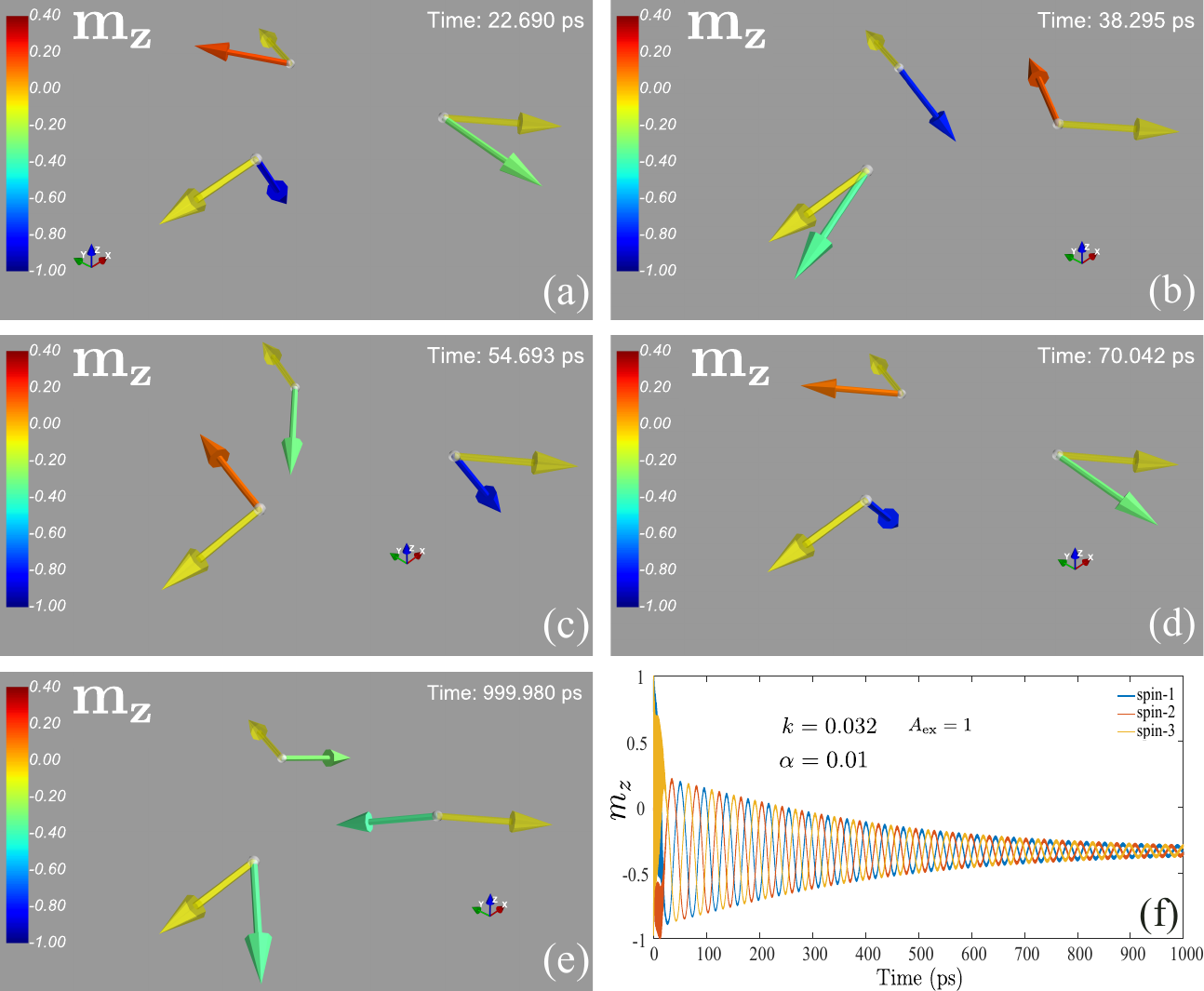}
\end{center}
\caption{(Color online)  Snapshots detailing how the system evolves into the state \textbf{s} from an RBP (RBUT) initial state obtained by the macrospin simulation. Panels (a)-(d) capture the instant spin alignment during one period ($\sim47.35$ ps) of a slow, alternating $z$-component oscillation. This slow period is approximately $142$ times longer than the fast precessional period of the whole system (CM), which is $\sim0.3$ ps. The system exhibits this long-term oscillatory decay before settling into the final stable state \textbf{s} shown in panel (e).  Panel (f) shows the time traces of the individual spin $z$-components ($m_{zi}$) throughout this long-term oscillatory decaying process until they reach the final state $\mathbf{s}$. For a clearer visualization of this dynamic process, please refer to the supplementary animation in the attachment (see 'Long term oscillatory decaying.mp4').}
\label{Long_oscillation_decaying_snapshots}
\end{figure*}

In addition to the evidences made from the particle perspective of the macrospin simulation, the vector perspective of this simulation is also on offer here, as shown in Fig. \ref{Long_oscillation_decaying_snapshots}, where the panels (a)-(d) give the snapshots of the instant spin alignment during one period ($\sim47.35$ ps) of a slow, alternating $z$-component oscillation. Notably, this slow period is approximately $142$ times longer than the fast precessional period of the whole system (CM), which is around $0.3$ ps. After the long term oscillatory decaying process, which takes around $1\mathrm{ns}$, the system eventually evolves into the state \textbf{s}, as displayed by the panel (e). The whole evolution of the individual spin $z$-components ($m_{zi}$) is shown in the panel (f).

Moreover, one can obtain an approximated set of Newton-like equations of motion for this slow oscillatory decaying process, as can be seen in Appendix \ref{appc} for the details. This, in this sense, is a somewhat similar to the phase-locking of a coupled set of FM-based STNOs, which can also be governed by the Newton-like equations\cite{HaoHsuan2011,HaoHsuan2016,HaoHsuan2018,Chen2021,chen2023a}, where their effective masses are also provided by their respective non-linear frequency shifts originated from the shape anisotropy. While, the mass of these FM-based STNOs' phase-locked motion, i.e. CM translation, is also given from their anisotropy, which is completely different from the NC-AFM-based STNOs whose mass of inertia is directly associated to the exchange coupling itself.
At last, since both CM and RM variables being able to be governed by the Newton-like mechanics, the three coupled spins can
 approximately be treated as a coupled set of Newton-like particles described by $(\phi_{i},\dot{\phi}_{i})$ (see Appendix \ref{appc}) under the restriction of $M_{\mathrm{x(y)}}\approx0$, see also Fig. \ref{Pendulum}(b).

\subsubsection{\label{sec:2E3} Hysteretic Excitation and the Driven Frequency for the State \textbf{s} Analyzed by the TVM Model}
Given that in the presence of the uniaxial anisotropy the state \textbf{s} is the most stable, the TVM model for the CM's angular velocity under the constrain of RM' variables at this state is derived in Appendix \ref{appb} as
\setlength\abovedisplayskip{6pt}
\setlength\belowdisplayskip{6pt}
\begin{eqnarray}
\ddot{\Phi}_{\mathrm{c}}&=&\frac{1}{m_{\mathrm{eff,\mathbf{s}}}(\dot{\Phi}_{\mathrm{c}})}\Bigg[-\alpha S(\dot{\Phi}_{\mathrm{c}})\dot{\Phi}_{\mathrm{c}}+\beta(\dot{\Phi}_{\mathrm{c}},J)\nonumber\\
&&-\frac{1}{3}\left(\frac{\partial H_{\mathrm{u},\mathbf{s}}}{\partial\Phi_{\mathrm{c}}}\right)_{P_{\mathrm{c}}=
\frac{\dot{\Phi}_{\mathrm{c}}}{3A_{\mathrm{ex}}}}\Bigg],
\label{TVM}
\end{eqnarray}
with the effective mass of inertia $m_{\mathrm{eff,\mathbf{s}}}=(3A_{\mathrm{ex}})^{-1}$, positive damping function $S(\dot{\Phi}_{\mathrm{c}})=1-[\dot{\Phi}_{\mathrm{c}}/(3A_{\mathrm{ex}})]^{2}$ , negative damping function $\beta(\dot{\Phi}_{\mathrm{c}},J)=a_{J0}S(\dot{\Phi}_{\mathrm{c}})$, and the reduced Hamiltonian $H_{\mathrm{u,\mathbf{s}}}(P_{\mathrm{c}},\Phi_{\mathrm{c}})=-(3k/2)
(1-P_{\mathrm{c}}^{2})\cos^{2}(\Phi_{\mathrm{c}}-3\pi/2)$.

Owing to the TVM model for the state \textbf{s} sharing a
greater resemblance of appearance to that for an FM-based perpendicular-to-plane polarizer (PERP)-STNO \cite{HaoHsuan2015,HaoHsuan2017,Chen2021,chen2023a}, we believe the phenomenon hysteretic excitation appearing in an FM-based PERP-STNO with an in-plane anisotropy and analyzed
by the TVM model would also appear in the NC-AFM we are interested in here.
Thus, adopting the same theoretical technique as used in our previous works
\cite{HaoHsuan2015,HaoHsuan2017,Chen2021,chen2023a}, two distinct threshold currents can be analytically derived in the following sections utilizing the TVM model. Crucially, this hysteretic excitation stems from the second-order dynamics and the effective inertial mass inherent to the TVM formalism.

Firstly, we can reformulate Eq. (\ref{TVM}) into the following, more physically intuitive form, which resembles the equation of motion for a simple pendulum:
\setlength\abovedisplayskip{6pt}
\setlength\belowdisplayskip{6pt}
\begin{eqnarray}
\ddot{\Phi}_{\mathrm{c}}+\alpha_{\mathrm{eff}}\dot{\Phi}_{\mathrm{c}}&=&
\beta_{\mathrm{c}}-g_{\mathrm{u}}\sin(2\Phi_{\mathrm{c}}),
\label{pendulum_like}
\end{eqnarray}
with the effective parameters, $\alpha_{\mathrm{eff}}$, $\beta_{\mathrm{c}}$, and $g_{\mathrm{u}}$, which govern the damping, driving, and restoring (uniaxial anisotropy) forces, respectively:
\setlength\abovedisplayskip{6pt}
\setlength\belowdisplayskip{6pt}
\begin{eqnarray}
\alpha_{\mathrm{eff}}&=&\left(\frac{\alpha}{m_{\mathrm{eff},\mathbf{s}}}\right)\left[1-
\left(\frac{\dot{\Phi}_{\mathrm{c}}}{A''_{\mathrm{ex}}}\right)^{2}\right],\nonumber\\
\beta_{\mathrm{c}}&=&\left(\frac{a_{J0}}{m_{\mathrm{eff},\mathbf{s}}}\right)\left[1-
\left(\frac{\dot{\Phi}_{\mathrm{c}}}{A''_{\mathrm{ex}}}\right)^{2}\right],\nonumber\\
\textrm{and}&&\nonumber\\
g_{\mathrm{u}}&=&\frac{k}{2}\left(\frac{1}{m_{\mathrm{eff},\mathbf{s}}}\right)\left[1-\left(\frac{\dot{\Phi}_{\mathrm{c}}}
{A''_{\mathrm{ex}}}\right)^{2}\right],\nonumber
\label{}
\end{eqnarray}
with $A''_{\mathrm{ex}}\equiv3A_{\mathrm{ex}}$. For analytical convenience, we have redefined the angle $\Phi_{\mathrm{c}}-2\pi/3$ as $\Phi_{\mathrm{c}}$ in Eq. (\ref{pendulum_like}).

Drawing from Ref. \cite{chen2023a}, the \textit{quasi-energy conservation law} for the \textit{local} oscillatory states of the system reads:
\setlength\abovedisplayskip{6pt}
\setlength\belowdisplayskip{6pt}
\begin{eqnarray}
E_{\mathrm{c}}=\frac{1}{2}\dot{\Phi}_{\mathrm{c}}^{2}-g_{\mathrm{u}}(\dot{\Phi}_{\mathrm{c}})
\cos^{2}\Phi_{\mathrm{c}},
\label{quasienergy}
\end{eqnarray}
where the energy $E_{\mathrm{c}}$ is constrained by $-g_{\mathrm{u}}(\dot{\Phi}_{\mathrm{c}}=0)<E_{\mathrm{c}}<0$. Here, $g_{\mathrm{u}}(\dot{\Phi}_{\mathrm{c}}=0)$ represents the potential energy barrier height. Furthermore, the driving force $\beta_{\mathrm{c}}$ is demonstrated to have a conservative-like impact on the nature of these states \cite{HaoHsuan2018,HaoHsuan2017,HaoHsuan2016,HaoHsuan2015,Chen2021,chen2023a}).
This means that $\beta_{\mathrm{c}}$ drive the auto-oscillating state \textbf{s} with a non-zero $\langle M_{\mathrm{z}}\rangle_{T}$ or $\langle P_{\mathrm{c}}\rangle_{T}$ through eliminating all of these local oscillatory states, or rather, equilibrium points of the potential defined by the $\sin(2\Phi_{\mathrm{c}})$ term in Eq. (\ref{pendulum_like}).

Thus, the threshold condition for the onset of auto-oscillation (i.e., the elimination of the potential energy barrier) is attained when the driving force strength $|\beta_{\mathrm{c}}|$ exactly reaches the value of the potential barrier height $g_{\mathrm{u}}(\dot{\Phi}_{\mathrm{c}}=0)$:
\setlength\abovedisplayskip{6pt}
\setlength\belowdisplayskip{6pt}
\begin{eqnarray}
\bigg|\beta_{\mathrm{c}}\left(\dot{\Phi}_{\mathrm{c}}=0\right)\bigg|
&=&g_{\mathrm{u}}\left(\dot{\Phi}_{\mathrm{c}}=0\right),\nonumber
\label{}
\end{eqnarray}
one can solve the critical Spin-Orbit Torque (SOT) or current density $|J_{\mathrm{c}}|$. Remarkably, the resulting critical SOT strength is found to be independent of the damping constant $\alpha$:
\setlength\abovedisplayskip{6pt}
\setlength\belowdisplayskip{6pt}
\begin{eqnarray}
|a_{J0,\mathrm{c}}|=\frac{k}{2}.
\label{TVMforIc}
\end{eqnarray}
Therefore, the value of the threshold current density is $|J_{\mathrm{c}}|=3.0339\times10^{9}\hspace{0.1cm}(\mathrm{A}/\mathrm{cm}^{2})$ ($|a_{J0,\mathrm{c}}|=0.016$) for $k=0.032$.

For the \textit{global} oscillatory states defined by Eq. (\ref{quasienergy}), specifically, the auto-oscillating state \textbf{s} with $E_{\mathrm{c}}>0$, the effect of the driving force $\beta_{\mathrm{c}}$ is necessarily \textit{non-conservative}. Due to this non-conservative nature, a smaller threshold current $|J_{\mathrm{b}}|$ (where $|J_{\mathrm{b}}|<|J_{\mathrm{c}}|$) must exist to excite the non-zero $\langle M_{\mathrm{z}}\rangle_{T}$ or $\langle P_{\mathrm{c}}\rangle_{T}$ state \textbf{s}. The underlying reason is that global oscillatory states possess sufficient kinetic energy to prevent the effective particle from being trapped in the potential well \cite{Chen2021}. This behavior implies the existence of a regime of coexistent states, $\mathrm{S_{\mathbf{s}}/(\mathrm{OP})_{\mathbf{s}}}$, appearing within the current window $|J_{\mathrm{b}}|<|J|<|J_{\mathrm{c}}|$. Here, $\mathrm{S}_{\mathbf{s}}$ denotes the \textit{static} state \textbf{s} with $\dot{\Phi}_{\mathrm{c}}=0$, and $(\mathrm{OP})_{\textbf{s}}$ refers to the \textit{dynamic} auto-oscillating state \textbf{s} characterized by a non-zero time-averaged angular velocity $\langle\dot{\Phi}_{\mathrm{c}}\rangle_{T}$. It is important to note that these threshold currents $|J_{\mathrm{b}}|$ correspond to the current-driven $(\mathrm{OP})_{\textbf{s}}$ states whose energies are slightly above zero. In addition, similar to the PERP-STNO case discussed in Ref. \cite{HaoHsuan2017}, the static state $\mathrm{S_{\mathbf{s}}}$ actually consists of two coexistent static states whose stable in-plane spin alignments are anti-parallel to each other. This is consistent with the angular dependence of the potential energy, i.e.  $V(\Phi_{\mathrm{c}})\propto\cos^{2}\Phi_{\mathrm{c}}$.

Based on the non-conservative effect of $\beta_{\mathrm{c}}$ on the dynamic states, we can determine the sub-critical current $|J_{\mathrm{b}}|$ by calculating the energy balance equation. Firstly, the velocity of the unperturbed trajectories for the dynamic state \textbf{s} is obtained from the quasi-energy conservation law, Eq. (\ref{quasienergy}):

\setlength\abovedisplayskip{6pt}
\setlength\belowdisplayskip{6pt}
\begin{eqnarray}
\dot{\Phi}_{\mathrm{c}}&=&\pm\Bigg[\frac{E_{\mathrm{c}}+
\left(\frac{k}{2m_{\mathrm{eff},\mathbf{s}}}\right)\cos^{2}\Phi_{\mathrm{c}}}{\frac{1}{2}
+\left(\frac{k}{2m_{\mathrm{eff},\mathbf{s}}}\right)\left(\frac{1}{A''^{2}_{\mathrm{ex}}}\right)
\cos^{2}\Phi_{\mathrm{c}}}\Bigg]^{1/2}.
\label{velociunper}
\end{eqnarray}

Secondly, using Eqs. (\ref{velociunper}) and (\ref{pendulum_like}), we calculate the time-averaged energy balance for the unperturbed dynamic state \textbf{s} as follows:

\setlength\abovedisplayskip{6pt}
\setlength\belowdisplayskip{6pt}
\begin{eqnarray}
\langle\dot{E}_{\mathrm{c}}\rangle_{T}&=&\frac{1}{T(E_{\mathrm{c}})}\oint_{C(E_{\mathrm{c}})}
\left(\frac{d\Phi_{\mathrm{c}}}{\dot{\Phi}_{\mathrm{nc}}}\right)
\dot{E}_{\mathrm{c}},\nonumber\\
&=&\frac{1}{T(E_{\mathrm{c}})}\int_{0(2\pi)}^{2\pi(0)}d\Phi_{\mathrm{c}}\left[-\alpha_{\mathrm{eff}}\left(\dot{\Phi}_{\mathrm{c}}\right)\right]
\dot{\Phi}_{\mathrm{c}}\nonumber\\
&&+\frac{1}{T(E_{\mathrm{c}})}\int_{0(2\pi)}^{2\pi(0)}d\Phi_{\mathrm{c}}\beta_{\mathrm{c}}\left(\dot{\Phi}_{\mathrm{c}}\right),
\label{energbalandstate_s}
\end{eqnarray}
with the period of the unperturbed dynamic state \textbf{s} being $T(E_{\mathrm{c}})=\oint_{C(E_{\mathrm{c}})}(d\Phi_{\mathrm{c}}/\dot{\Phi}_{\mathrm{c}})$, and $\Phi_{\mathrm{c}}\in[0(2\pi),2\pi(0)]$ which appears in the upper and lower limits of the integral means that $\dot{\Phi}_{\mathrm{c}}$ is positive or negative, respectively.  Finally, requiring $\langle\dot{E}_{\mathrm{c}}(E_{\mathrm{c}}\rightarrow0))\rangle_{T}=0$, we can solve $|J_{\mathrm{b}}|$, which is proportional to the damping constant $\alpha$:
\setlength\abovedisplayskip{6pt}
\setlength\belowdisplayskip{6pt}
\begin{eqnarray}
a_{J0,\mathrm{b}}=\frac{\int_{0(2\pi)}^{(2\pi)0}d\Phi_{\mathrm{c}}
\alpha_{\mathrm{eff}}\left(\dot{\Phi}_{\mathrm{c}}\right)\dot{\Phi}_{\mathrm{c}}}
{\int_{0(2\pi)}^{(2\pi)0}d\Phi_{\mathrm{c}}\left[1-\left(\frac{\dot{\Phi}_{\mathrm{c}}}
{A''_{\mathrm{ex}}}\right)^{2}\right]}.
\label{energbalandstate_s}
\end{eqnarray}
Thus, for $A_{\mathrm{ex}}=1$, $\alpha=0.01$, and $k=0.032$, the sub-critical current density is $|J_{\mathrm{b}}|=0.3695\times10^{9}\hspace{0.1cm}(\mathrm{A}/\mathrm{cm}^{2})$ ($|a_{J0,\mathrm{b}}|=0.019$).
When $\alpha$ increases to a certain critical value, i.e. $\alpha_{\mathrm{c}}$, making $|J_{\mathrm{c}}|=|J_{\mathrm{b}}|$, then $\mathrm{S}_{\mathbf{s}}/(\mathrm{OP})_{\mathbf{s}}$ states will disappear.

The driven frequency can be given as follows: Using Eq. (\ref{velociunper}), picking up a current value out of its range $|J_{\mathrm{b}}|<|J|<|J_{\mathrm{c}}|$, and
requiring $\langle\dot{E}_{\mathrm{c}}(E_{\mathrm{c}})\rangle_{T}=0$,  the solution $E_{\mathrm{c}}(I)$
used to label the driven $(\mathrm{OP})_{\mathbf{s}}$ state can be obtained.  Then, we have
\setlength\abovedisplayskip{6pt}
\setlength\belowdisplayskip{6pt}
\begin{eqnarray}
f_{(\mathrm{OP})_{\mathbf{s}}}(\mathrm{THz})&=&\left(\frac{\omega_{\mathrm{ex}}}{2\pi}\right)\left(2\pi\right)
\left[\left(\frac{1}{2\pi}\right)\langle\dot{\Phi}_{\mathrm{c}}\rangle_{T}\right],\nonumber\\
&=&\left(\frac{\omega_{\mathrm{ex}}}{2\pi}\right)\left(2\pi\right)\left|\int_{0(2\pi)}^{2\pi(0)}\frac{d\Phi_{\mathrm{c}}}
{\dot{\Phi}_{\mathrm{c}}
(E_{\mathrm{c}})}\right|^{-1}.\nonumber\\
\label{PLfreq}
\end{eqnarray}

\subsection{\label{sec:2F}Threshold and Critical Current Densities Verified by the TVM/Macrospin Simulations}
\begin{figure*}
\begin{center}
\includegraphics[width=15.5cm]{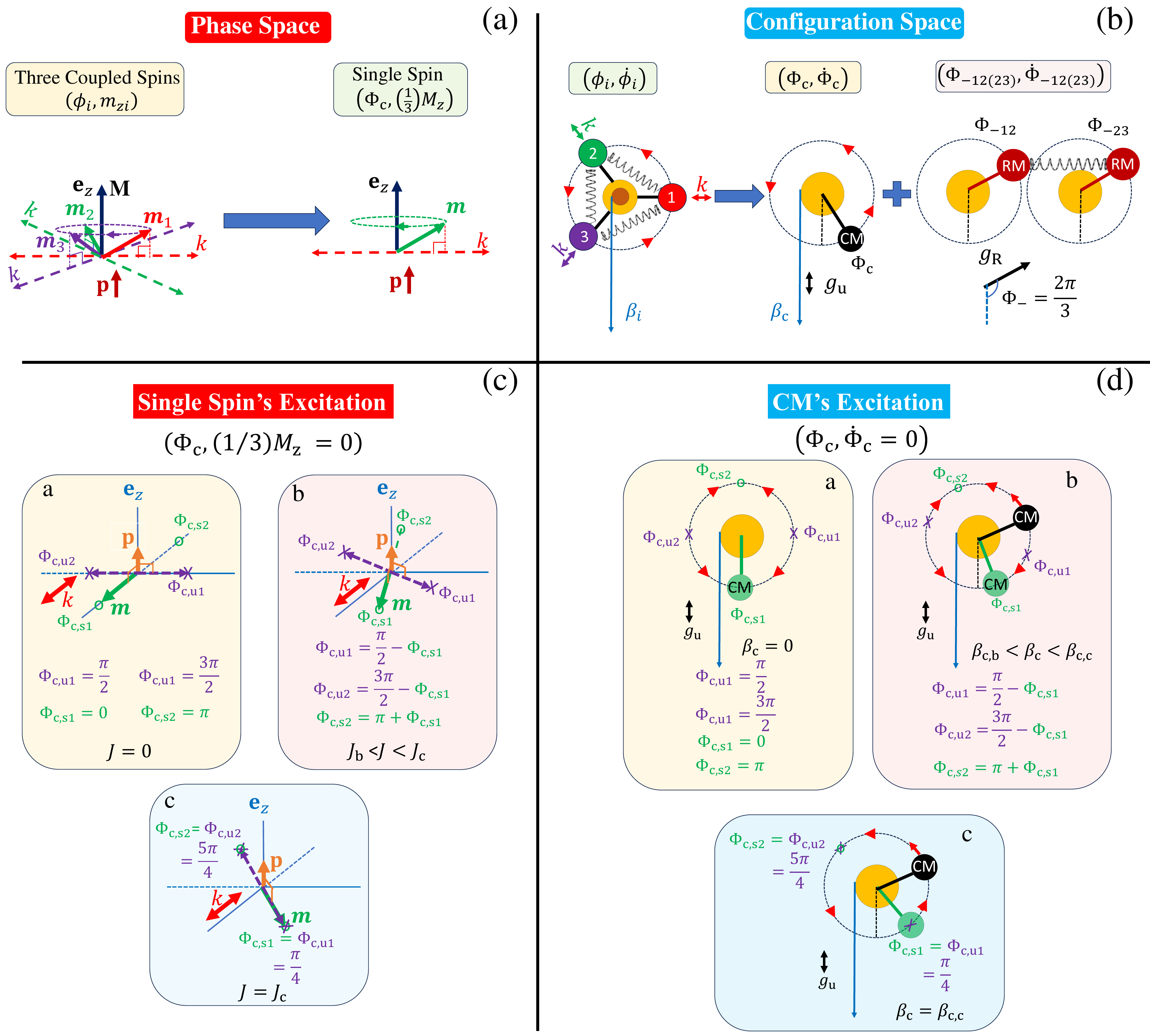}
\end{center}
\caption{(Color Online) Comparison of the threshold excitation of an NC-AFM-based STNO (Left) and a forced pendulum (Right). (a)\textbf{Phase Space [Vector Viewpoint]}. \textbf{Left Sub-Panel}: Schematic of the three SOT-driven coupled spins expressed in the phase space $(\phi_i, m_{zi})$ at the state \textbf{s}. Each spin is subject to an in-plane uniaxial anisotropic energy, where the angle between any two neighboring anisotropic axes is $2\pi/3$. The spin-polarization vector $\mathbf{p}$ is $\mathbf{e}_{z}$.
\textbf{Right Sub-Panel}: Schematic showing the three spins are reduced to a single spin, expressed in the reduced phase space $(\Phi_{\mathrm{c}},(1/3)M_{z})$ at the state \textbf{s}, and subject to an effective anisotropy. (b) \textbf{Configuration Space [Particle Viewpoint]}. \textbf{Left Sub-Panel}: Schematic of a pendulum formed by three massive particles, driven by a tangent force (expressed in $(\phi_{i}, \dot{\phi}_{i})$). All particles revolve synchronously around the pivot with a phase-locked angle of $2\pi/3$ between any two neighboring ones. Each is subject to a uniform gravity with an axial symmetric distribution in angle $\phi$ and a strength coefficient $k$. The angle between any two neighboring gravity symmetric axes is also $2\pi/3$. \textbf{Right Sub-Panel}: Schematic showing the three-particle pendulum is reduced to a single particle, expressed by the CM variables $(\Phi_{\mathrm{c}}, \dot{\Phi}_{\mathrm{c}})$, under the stationary RM conditions: $(\Phi_{-12(23)}, \dot{\Phi}_{-12(23)}) = (\pm2\pi/3,0)$. The original three gravities are reduced to a single one $g_{\mathrm{u}}$. Note that the two sets of RM variables are simplified as two coupled pendulums, both subject to an effective uniform gravity $g_{\mathrm{R}}$ pointing to $\Phi_{-} = 2\pi/3$ (see also Appendix \ref{appc}). \textbf{Panels} (c) and (d) illustrate the threshold current excitation, indicated by sub-panels \textbf{a}, \textbf{b}, and \textbf{c}, respectively: \textbf{a}. $J = 0$ ($\beta_{\mathrm{c}} = 0$), \textbf{b}. $J_{\mathrm{b}} < J < J_{\mathrm{c}}$ ($\beta_{\mathrm{c,b}} < \beta_{\mathrm{c}} < \beta_{\mathrm{c,c}}$), and \textbf{c}. $J = J_{\mathrm{c}}$ ($\beta_{\mathrm{c}} = \beta_{\mathrm{c,c}}$). $J$ and $\beta_{\mathrm{c}}$ denote the injected current and the driving force, respectively. \textbf{Symbols in (c) [Vector Viewpoint]}:  \textbf{Orange arrow}: Spin polarization vector ($\mathbf{p}$). \textbf{Red double-headed arrow}: Anisotropic field. \textbf{Green 'o'}: Stable equilibrium points ($\Phi_{\mathrm{c,s}1(2)}$) of the free layer moment. \textbf{Purple 'X'}: Unstable equilibrium points ($\Phi_{\mathrm{c,u}1(2)}$). \textbf{Symbols in (d) [Particle Viewpoint]}: \textbf{Black double-headed arrow}: Uniform gravitational field. \textbf{Red arrow}: Force exerting on the pendulum (black ball). \textbf{Green ball}: Stable equilibrium angles ($\Phi_{\mathrm{c,s}1(2)}$). \textbf{Purple 'X'}: Unstable equilibrium angles ($\Phi_{\mathrm{c,u}1(2)}$).}
\label{Pendulum}
\end{figure*}

The threshold current densities derived in the preceding sections can be readily verified through the Macrospin or TVM simulations. As demonstrated in our previous works\cite{Chen2021,chen2023a}, the collective dynamics of the three phase-locked spins in the RBP state \textbf{s} can be accurately modeled as a forced pendulum (see Fig. \ref{Pendulum}(a) to (b)). This powerful analogy confirms that the RBP collective motion, treated as a whole, fundamentally behaves like a TVM particle. Therefore, by analyzing the forced pendulum under a uniform gravitational force/field (Eq. (\ref{pendulum_like})), whose anisotropic dependence on the CM phase angle $\Phi_{\mathrm{c}}$ is uniaxial with two local potential minima, we can determine the appropriate initial conditions to find out the critical current densities. Specifically, setting the initial velocity $\dot{\Phi}_{\mathrm{c},i} = 0$ allows us to excite a global oscillation (the RBP state) as successfully depicted in Fig. \ref{Pendulum}(d). For comparison, Fig. \ref{Pendulum}(c) illustrates the corresponding initial conditions in the reduced phase space $(\Phi_{\mathrm{c}},(1/3)M_{z,i}=0)$, mapping the states shown in the forced pendulum analogy. Furthermore, in the phase space representation, the uniform gravitational force/field in configuration space can be replaced by a uniaxial anisotropic field with its easy axis along the $x$-axis.

In the absence of the driving force ($\beta_{\mathrm{c}}=0$ or $J = 0$), as shown in Fig. \ref{Pendulum}(d)\textbf{a}, there are four distinct equilibrium points, defining the system's easy and hard axes, respectively: two of these points are stable equilibrium points (potential minima) at $\Phi_{\mathrm{c,s}1}=0$ and $\Phi_{\mathrm{c,s}2}=\pi$; the other two are unstable equilibrium points (potential maxima or saddle points) at $\Phi_{\mathrm{c,u}1}=\pi/2$ and $\Phi_{\mathrm{c,u}2}=3\pi/2$.

As the positive driving force $\beta_{\mathrm{c}}$ (or current $J > 0$) is gradually increased towards the critical value $\beta_{\mathrm{c,c}}>0$ ($J_{\mathrm{c}}>0$), Figs. \ref{Pendulum}(d) \textbf{b} and \textbf{c} illustrate that the stable point $\Phi_{\mathrm{c,s}1}$ (or $\Phi_{\mathrm{c,s}2}$) on the easy axis will merge with the unstable point $\Phi_{\mathrm{c,u}1}$ (or $\Phi_{\mathrm{c,u}2}$) on the hard axis, coalescing exactly at the phase angle $\pi/4$ (or $5\pi/4$). This critical merging event is consistent with the established angular relationship between the two points: $\Phi_{\mathrm{c,u}1}=\pi/2-\Phi_{\mathrm{c,s}1}$ (and $\Phi_{\mathrm{c,u}2}=3\pi/2-\Phi_{\mathrm{c,s}1}$)\cite{Chen2021,chen2023a}. For negative current driving ($J < 0$), the scenario is similar: the roles are effectively reversed, with $\Phi_{\mathrm{c,s}2}$ (or $\Phi_{\mathrm{c,s}1}$) merging with $\Phi_{\mathrm{c,s}1}$ (or $\Phi_{\mathrm{c,s}2}$) at the phase angles $3\pi/4$ and $7\pi/4$, respectively.

In the coexistent regime ($|J_\mathrm{b}|<|J|<|J_\mathrm{c}|$): If the pendulum is initially placed between any neighboring pair of stable ($\Phi_{\mathrm{c,s}}$) and unstable ($\Phi_{\mathrm{c,s}}$) points, the system will, for most cases, eventually evolve into one of the static states $\Phi_{\mathrm{c,s}1}$ or $\Phi_{\mathrm{c,s}2}$. However, to trigger the global auto-oscillation (the $(\mathrm{OP})_\textbf{s}$ state), a specific initial condition is required. If, for the case of positive (or negative) driving force $\beta_{\mathrm{c}}$, the pendulum is initially positioned close enough to one of the unstable points $\Phi_{\mathrm{c,u}1}$ (or $\Phi_{\mathrm{c,u}2}$) and pushed against the direction of the driving force (from its right side for positive $\beta_{\mathrm{c}}$, or left side for negative $\beta_{\mathrm{c}}$), the pendulum would then gain enough kinetic energy from the driving force (SOT) to escape the potential well and trigger a global oscillation. We term this mechanism a \textit{slingshot effect}.

Finally, at the upper threshold (critical) current, where $|\beta_{\mathrm{c}}|=|\beta_{\mathrm{c,c}}|$ (or $|J|=|J_{\mathrm{c}}|$), the phase space topology simplifies dramatically. The easy and hard axes coalesce at the equilibrium points $\Phi_{\mathrm{c}}=\pi/4$ or $-\pi/4$, corresponding to positive or negative currents, respectively (see Fig. \ref{Pendulum}(d) \textbf{c}). At this point, the stable and unstable points have merged, leaving only a single pair of unstable equilibrium points (\textit{saddle-node bifurcation})\cite{HaoHsuan2017} in the phase space. The dynamic consequence is profound: the pendulum will evolve into a global auto-oscillation (the RBP state) from virtually any initial condition, even a static start with $\dot{\Phi}_{\mathrm{c}}=0$, as presented in Fig. \ref{Pendulum}(d) \textbf{c}. As previously established, there exists a critical damping constant, $\alpha_{c}$, below which the existence of the lower threshold $\beta_{\mathrm{c,b}}$ (or $J_{\mathrm{b}}$) is guaranteed \cite{Chen2021,chen2023a}. Therefore, placing the initial angle at $\Phi_{\mathrm{c}} =\pm\pi/4$ can allows us not only to get $\Phi_{\mathrm{c,s}1(2)}$ to find out $|J_{\mathrm{b}}|$ using the slingshot effect but also to search for $|J_{\mathrm{c}}|$.

\begin{figure*}
\begin{center}
\includegraphics[width=12cm]{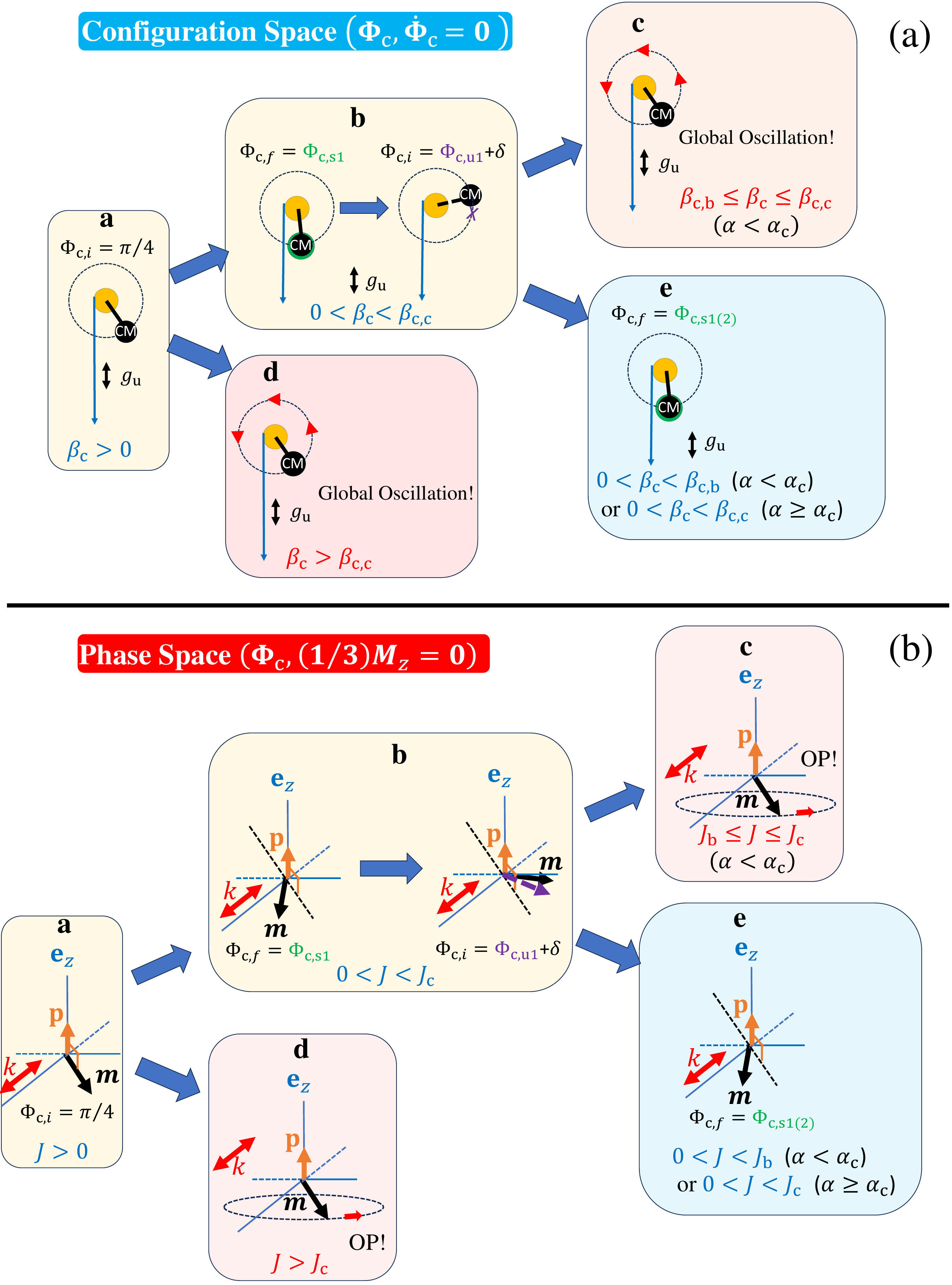}
\end{center}
\caption{(Color Online) Standard operational procedure (SOP) for searching threshold currents $|J_{\mathrm{b(c)}}|$ (driving forces $|\beta_{\mathrm{c,b(c)}}|$ ) via numerical simulation.
The equivalent procedures are demonstrated in the \textbf{Configuration Space (Panel (a))} and \textbf{Phase Space (Panel (b))}. $J$ and $\beta_{\mathrm{c}}$ denote the injected current and the driving force, respectively. The SOP consists of the following steps: \textbf{Step 1 (Initialization - Sub-Panel a)}: The initial phase angle is set at $\Phi_{\mathrm{c},i} = \pi/4$. \textbf{Step 2 (Check for $|J_{\mathrm{b}}|$ (or $|\beta_{\mathrm{c,b}}|$) - Sub-Panels b to d)}: Demonstrates checking for OP state (global oscillation) excitation by first allowing the system to settle into the stable state $\Phi_{\mathrm{c,s}}$, and then using the slingshot effect (where $\delta$ is a small deviation) to trigger the OP precession. \textbf{(Step 3 Check for $|J_{\mathrm{c}}|$ (or $|\beta_{\mathrm{c,c}}|$) - Sub-Panel d)}: The threshold $|J_{\mathrm{c}}|$ (or $|\beta_{\mathrm{c,c}}|$) appears when the stable ($\Phi_{\mathrm{c,s}}$) and unstable ($\Phi_{\mathrm{c,u}}$) states merge and vanish at $\Phi_{\mathrm{c}}=\pi/4$. \textbf{Step 4 (Damping Criterion - Sub-Panel e)}: Emphasizes that whether the slingshot effect can trigger the OP state before $|J_{\mathrm{c}}|$'s appearance determines the relationship between the damping constant and its critical value ($\alpha < \alpha_{\mathrm{c}}$ or $\alpha \geq \alpha_{\mathrm{c}}$).}
\label{initialsetting}
\end{figure*}

Having established the initial conditions for threshold excitation of the NC-AFM (or its single pendulum analogy), we propose a standardized numerical simulation procedure to determine both critical currents, as illustrated in Fig. \ref{initialsetting}: Firstly, as shown in  Figs. \ref{initialsetting}  \textbf{a}, \textbf{b}, and \textbf{d}, the initial state of the moment (pendulum) is first set at $(\Phi_{\mathrm{c},i}=\pm\pi/4,(1/3)M_{z,i}=0)$ (or ($\Phi_{\mathrm{c},i}= \pm\pi/4, \dot{\Phi}_{\mathrm{c},i}= 0$)) to ensure the system first relaxes into the stable static state $(\Phi_{\mathrm{c,s}1},(1/3)M_{z}=0)$ ( or $(\Phi_{\mathrm{c,s}1},\dot{\Phi}_{\mathrm{c}}=0)$). (The $\pm$ signs correspond to the cases of positive or negative driving current $J$ or force $\beta_{\mathrm{c}}$, respectively). Subsequently, utilizing the slingshot effect, the value of $|J_{\mathrm{b}}|$ ($|\beta_{\mathrm{c,b}}|$) is found by gradually increasing the current amplitude (driving force) until the excitation of the RBP state (a global oscillation) is triggered. Secondly, referring to Fig. \ref{initialsetting} \textbf{a} and \textbf{c}, the upper threshold current $|J_{\mathrm{c}}|$ ($|\beta_{\mathrm{c,c}}|$) is determined by placing the initial state at ($\Phi_{\mathrm{c},i}=\pm\pi/4,(1/3)M_{z,i}=0$) (or ($\Phi_{\mathrm{c},i}=\pm\pi/4,\dot{\Phi}_{\mathrm{c},i}= 0$)) and then gradually increasing the current (driving force) amplitude until the triggering of the RBP state (a global oscillation) occurs. Crucially, the search for $|J_\mathrm{c}|$ is performed completely without the help of the slingshot effect.

It is crucial to note here that the successful numerical determination of $|J_{\mathrm{b}}|$ ($|\beta_{\mathrm{c,b}}|$) using the procedure described above automatically implies that the system is operating in the low-damping regime where $\alpha<\alpha_{\mathrm{c}}$. Conversely, if the lower threshold $|J_{\mathrm{b}}|$ fails to be numerically found (i.e., the slingshot effect fails to trigger the auto-oscillation) before the appearance of the upper threshold $|J_{\mathrm{c}}|$ ($|\beta_{\mathrm{c,c}}|$), this signifies that the system is operating at or above the critical damping, meaning $\alpha\geq\alpha_{\mathrm{c}}$. In this high-damping regime, the energy dissipation is too rapid, causing the hysteretic excitation window to vanish (as conceptually illustrated in Fig. \ref{initialsetting} \textbf{a}, \textbf{b}, and \textbf{e}).

\begin{figure*}
\begin{center}
\includegraphics[width=16.5cm]{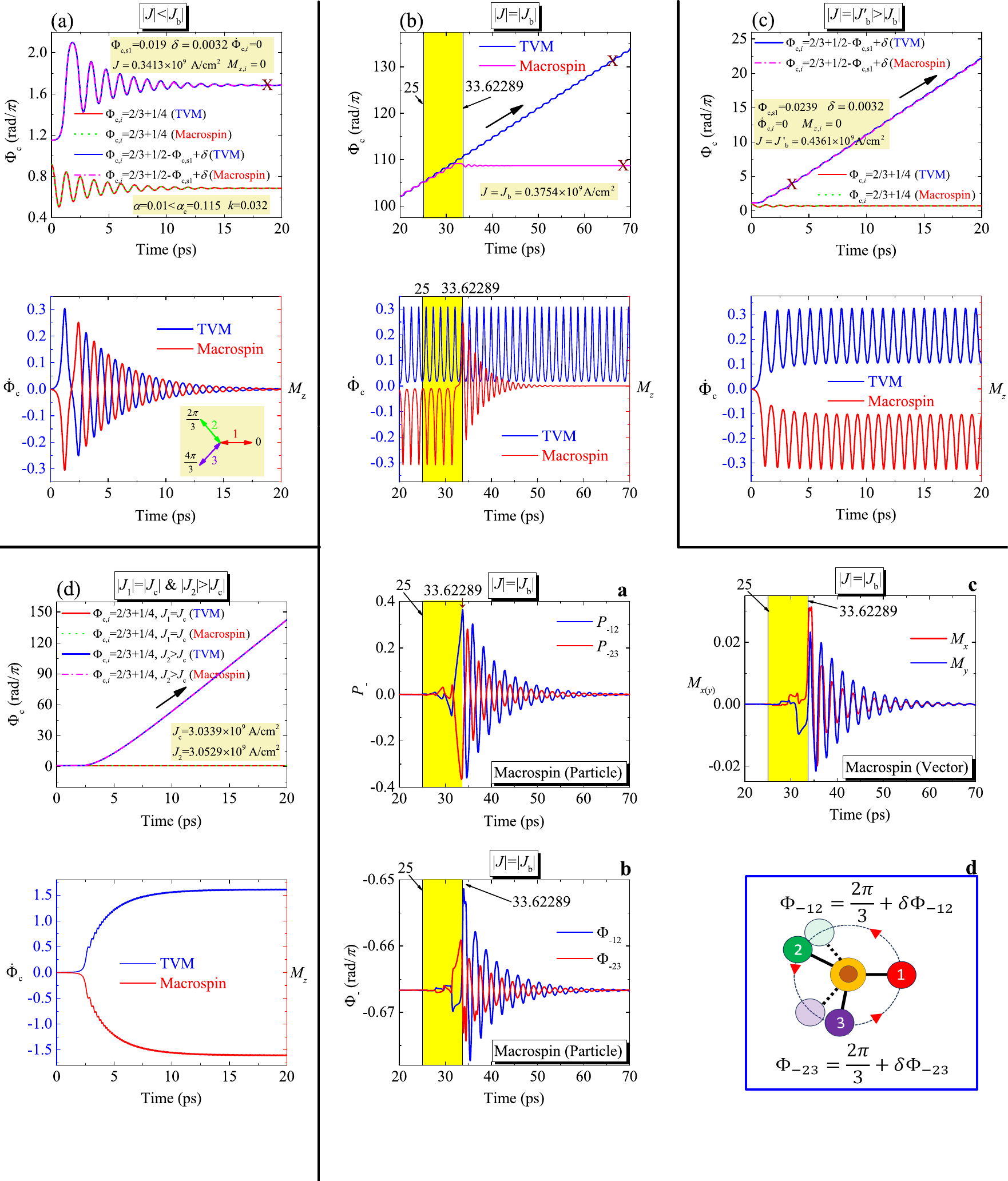}
\end{center}
\caption{(Color Online) Search for threshold currents $|J_{\mathrm{b}}|$ ($|J_{\mathrm{b}}'|$), $|J_{\mathrm{c}}|$ using TVM vs. macrospin simulations and the state \textbf{s} (RBP) rigidity failure analysis. \textbf{I. CM Angle $\Phi_{\mathrm{c}}$ Time Traces (Panels (a)-(d))}. Upper sub-panels in (a) to (d) show the $\Phi_{\mathrm{c}}$ trajectories: \textbf{Red Solid} is the TVM result with the initial state $(\Phi_{\mathrm{c},i}=2\pi/3+\pi/4,\dot{\Phi}_{\mathrm{c},i}=0)$; \textbf{Green Short Dot} is the macrospin result with the initial state $(\Phi_{\mathrm{c},i}=2\pi/3+\pi/4,M_{z,i}=0)$; \textbf{Blue Solid (TVM slingshot)} and \textbf{Magenta Dash-Dot (Macrospin slingshot)} show excitation results after applying the slingshot effect, both marked by "X". Input parameters [$J$, $k$, $\alpha$, $\delta$, $\Phi_{\mathrm{c},\mathrm{s}1}$] are listed in the upper light yellow boxes. Notably, the black arrows indicate the RBP state \textbf{s} driving. \textbf{II. Rigidity Failure Analysis at $|J_{\mathrm{b}}|$}. \textbf{Macrospin Capture}: The yellow box in Panel (b) marks an interval where $\Phi_{\mathrm{c}}$ slows down into a stationary state (only observed in macrospin at $|J| = |J_{\mathrm{b}}|$), confirming RBP rigidity assumption failure. \textbf{Non-Rigid Excitation}: Sub-panels \textbf{a}, \textbf{b}, and \textbf{c} show the self-resonant excitation of RM variables and $M_{x(y)}$ in the macrospin simulation at $|J| = |J_{\mathrm{b}}|$. Sub-panel \textbf{d} is a schematic showing the $\Phi_{-}$ deviation from the RBP state \textbf{s}. \textbf{III. Additional Details}. Panel (c) presents the actual sub-critical current $|J_{\mathrm{b}}'|>|J_{\mathrm{b}}|$ by the macrospin model. Lower sub-panels of (a) to (d) display the $\dot{\Phi}_{\mathrm{c}}$ and $M_{z}$ time traces driven by the slingshot effect. The light yellow box in Panel (a)'s lower sub-panel also shows anisotropy axis distribution of the three sub-lattices.}
\label{time_trace_s}
\end{figure*}

To numerically verify the RBP state excitation, as depicted in Fig. \ref{time_trace_s}, the initial states for the macrospin and TVM model simulations are set as $(\Phi_{\mathrm{c},0}=2\pi/3\pm\pi/4,(1/3)M_{z,0}=0)$ or $(\Phi_{\mathrm{c},0}=2\pi/3\pm\pi/4,\dot{\Phi}_{\mathrm{c},0}=0)$, respectively. Notably, the $\pm$ signs preceding $\pi/4$ depend on the sign of the driving force $\beta_{\mathrm{c}}$ (or current $J$), which can be readily inferred from the RBP precessional directions or the sign of the static phase $\Phi_{\mathrm{c,s}1}$ in direct simulations. Following the initialization, the system first relaxes to the stable static state $(\Phi_{\mathrm{c,s}1},(1/3)M_{z1(2)}=0)$ or $(\Phi_{\mathrm{c,s}1},\dot{\Phi}_{\mathrm{c},0}=0)$. Subsequently, the lower threshold $|J_{\mathrm{b}}|=0.3754\times10^{9}\hspace{0.1cm}(\mathrm{A}/\mathrm{cm}^{2})$ is numerically determined using the slingshot effect (see Fig. \ref{time_trace_s}(a), (b)), which is in perfect line with that of the analytical result $|J_{\mathrm{b}}|=0.3695\times10^{9}\hspace{0.1cm}(\mathrm{A}/\mathrm{cm}^{2})$ (see Eq. (\ref{energbalandstate_s}) in Sec. \ref{sec:2E3}). In addition, when initial conditions for these two models are set as $(\Phi_{\mathrm{c},0}=2\pi/3\pm\pi/4,(1/3)M_{z,0}=0)$ or $(\Phi_{\mathrm{c},0}=2\pi/3\pm\pi/4,\dot{\Phi}_{\mathrm{c},0}=0)$, respectively, the higher threshold $|J_{\mathrm{c}}|=3.0339\times10^{9}\hspace{0.1cm}(\mathrm{A}/\mathrm{cm}^{2})$ is numerically determined (see Fig. \ref{time_trace_s}(d)), which is exactly the same as that of the analyzed result $|J_{\mathrm{c}}|=3.0339\times10^{9}\hspace{0.1cm}(\mathrm{A}/\mathrm{cm}^{2})$ (see Eq. (\ref{TVMforIc}) in Sec. \ref{sec:2E3}).

\begin{figure}
\begin{center}
\includegraphics[width=8.0cm]{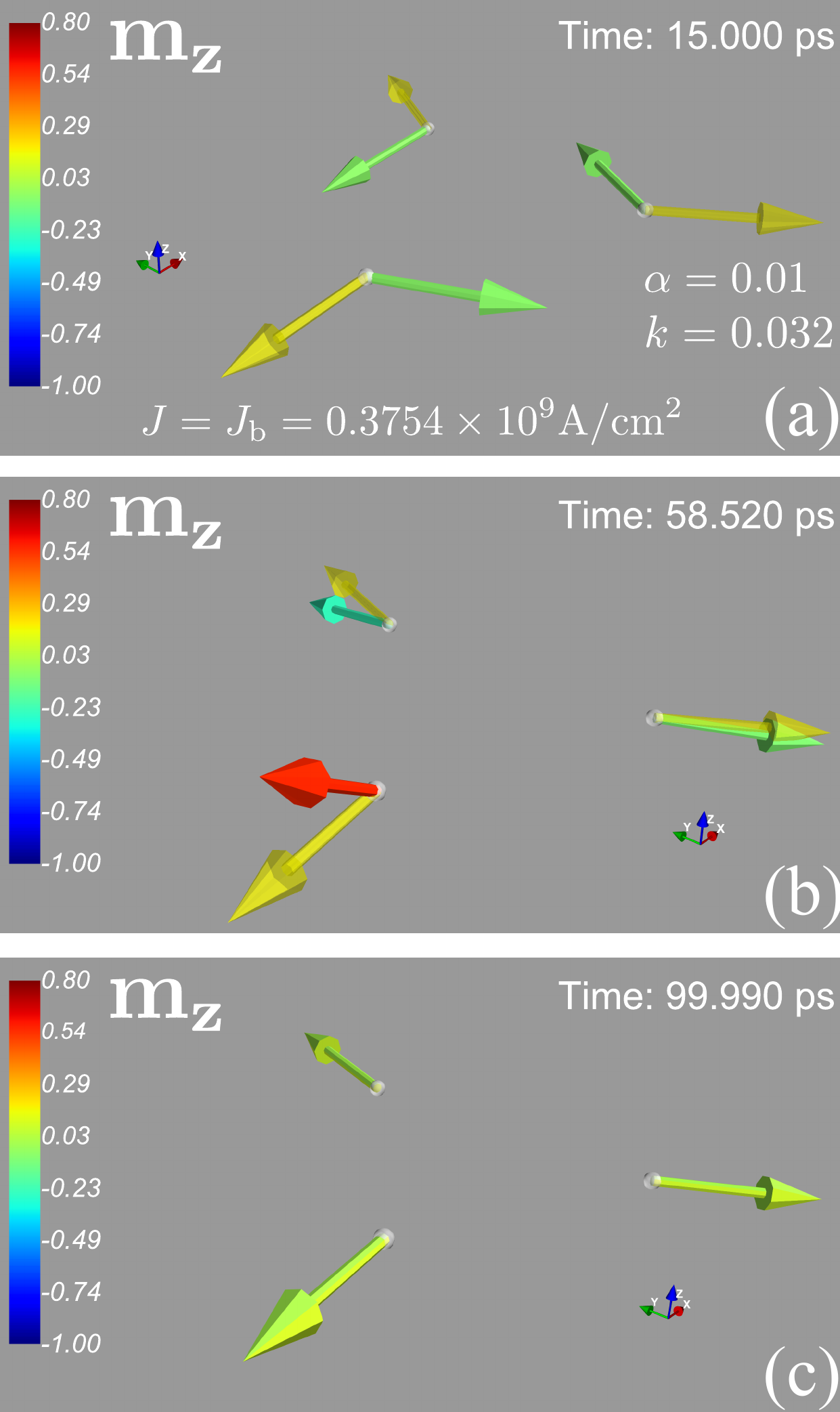}
\end{center}
\caption{(Color Online) Snapshots detailing the \text{RM} burst-induced precessional slowing down of the spins for the smaller threshold current $J=J_{\mathrm{b}}=0.3754\times 10^{9} \mathrm{A}/\mathrm{cm}^2$, with parameters $k=0.032$ and $\alpha=0.01$. These results were obtained using the macrospin model. Figures (a), (b), and (c) illustrate three sequential stages: (a) the initial dynamic state $\mathbf{s}$; (b) the state $\mathbf{s}$ being distorted and the $\text{CM}$'s velocity (or the three spins as a whole precessional frequency) being slowed down by the $\text{RM}$ burst; and (c) the state finally being trapped by the uniaxial anisotropy. The three light, transparent yellow arrows denote the three symmetric axes of the anisotropy. Furthermore, the colors drawn on the arrows indicate the $m_{\mathrm{z}}$ values of the three spins. Notably, the animation provided in the attachment offers a clearer impression of this phenomenon than what is presented here by these static snapshots (see 'RM burst.mp4').}
\label{Snashots_RMburst}
\end{figure}

However, a crucial discrepancy arises between the two models: while the time traces of $\Phi_{\mathrm{c}}$ (as well as $M_{z}$ or $\dot{\Phi}_{\mathrm{c}}$) from the macrospin and TVM simulations show perfect agreement for $|J|<|J_{\mathrm{b}}|$, an obvious disagreement emerges precisely at $|J|=|J_{\mathrm{b}}|$. As indicated by the yellow area in Fig. \ref{time_trace_s} (b), the macrospin simulated angular velocity $\dot{\Phi}_{\mathrm{c}}$ slows down and evolves into a stationary state (see also Fig. \ref{Snashots_RMburst} for the whole slowing down process in the vector viewpoint), rather than successfully evolving into the RBP auto-oscillation state, as the TVM model correctly predicts.

\begin{figure*}
\begin{center}
\includegraphics[width=17cm]{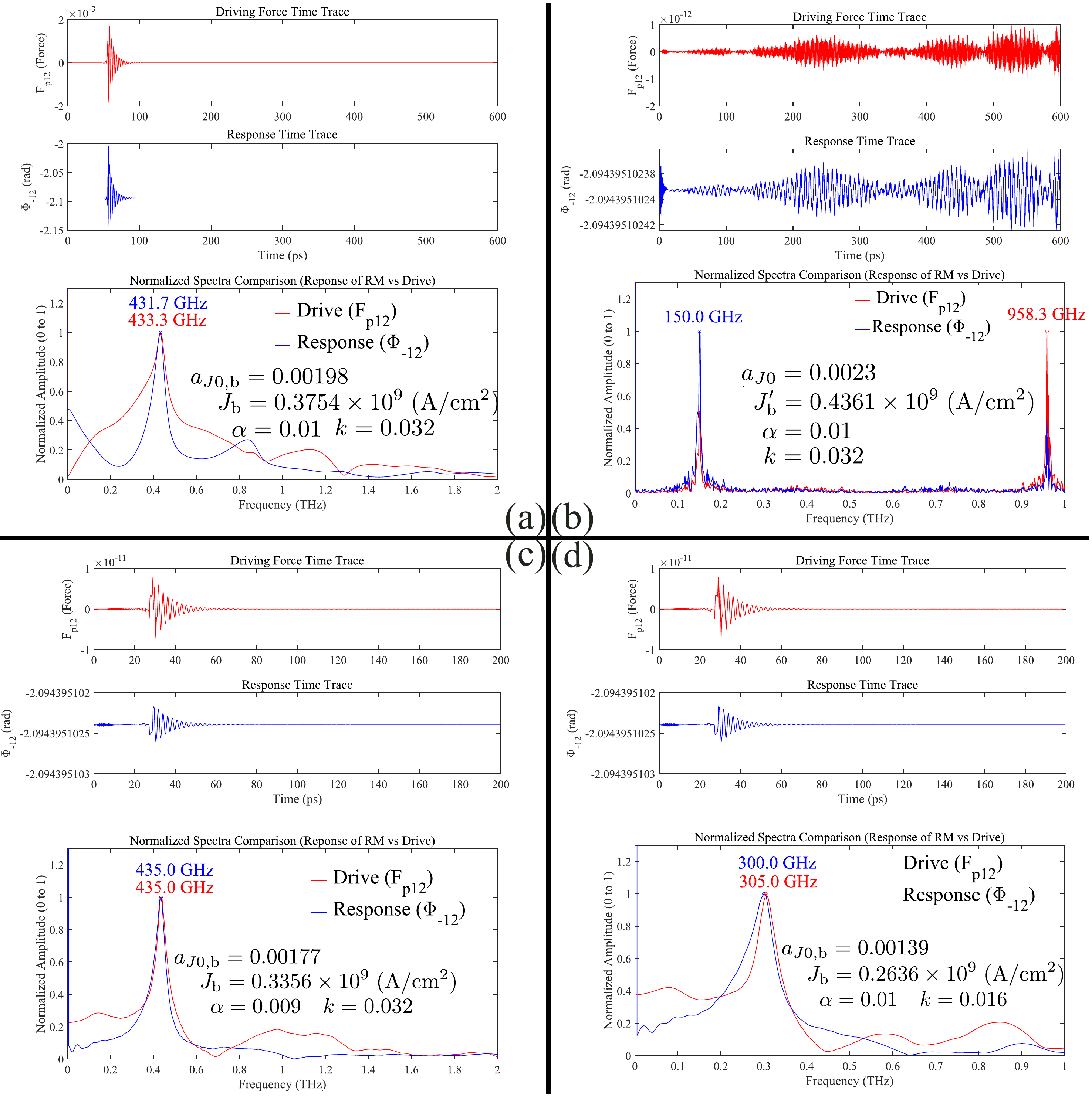}
\end{center}
\caption{(Color Online) Spectral evidence of \textbf{RM burst} (\textbf{RM's} \textbf{self-resonance}) trigged by CM's translation via the IPA component of the in-plane uniaxial anisotropy. Panels (a) and (b) show the system response at injected current densities $J=J_{\mathrm{b}}=0.37549\times10^{9} \hspace{0.1cm}\mathrm{A/cm}^2$ and $J=J_{\mathrm{b}}'=0.43619\times10^{9}\hspace{0.1cm}\mathrm{A/cm}^2$, respectively, with $\alpha=0.01$ and $k=0.032$. Panels (c) and (d) present sub-critical cases with varying parameters: (c) $J=J_{\mathrm{b}}=0.33569\times10^{9}\hspace{0.1cm}\mathrm{A/cm}^{2}$, $\alpha=0.009$, $k=0.032$; (d) $J=J_{\mathrm{b}}=0.26369\times10^{9}\hspace{0.1cm}\mathrm{A/cm}^2$, $\alpha=0.01$, $k=0.016$. In each panel, the top sub-panels display time traces of the driving force $F_{p12}$ and the RM response $\Phi_{-12}$. The bottom sub-panels show their normalized spectra, highlighting the frequency overlap that confirms the resonance-induced RM burst. Notably, all of sub-critical currents $I_{\mathrm{b}}$ for these different cases are calculated by the TVM model.}
\label{RMburst_spectrum}
\end{figure*}

This observed slowing down of $\dot{\Phi}_{\mathrm{c}}$ in the macrospin simulation implies that the system effectively possesses a higher damping constant than the idealized RBP state \textbf{s} assumed by the TVM model. Consequently, the actual lower critical current for the Macrospin must be raised to $|J'_{\mathrm{b}}|>|J_{\mathrm{b}}|$. At this higher threshold $|J'_{\mathrm{b}}|$, the simulated results from both models are again found to be in close agreement, as depicted in Fig. \ref{time_trace_s}(c).

Furthermore, the model comparison provides insight into the oscillation dynamics near the threshold. At the theoretically predicted threshold $|J| = |J_{\mathrm{b}}|$, the TVM model's time trace of $\dot{\Phi}_{\mathrm{c}}$ exhibits a characteristic \textit{comb-like} shape, signifying a strongly non-harmonic oscillation (see the lower panel in Fig. \ref{time_trace_s}(b)). This non-harmonicity occurs because the particle's energy is extremely close to the potential maximum of the state \textbf{s} ($E_{\mathrm{c}}=0$) \cite{chen2023a}, forcing the particle to traverse the potential valley at a velocity much faster than the velocity at the potential peak.

In contrast, at the macrospin's higher threshold $|J|=|J'_{\mathrm{b}}|$, the particle's energy is sufficiently far from the potential maximum. Therefore, the time traces of the stable $\dot{\Phi}_{\mathrm{c}}$ (or $M_{z}$) are closer to a \textit{sinusoidal} (\textit{harmonic}) shape, as illustrated by the lower panel in Fig. \ref{time_trace_s}(c). This indicates that the particle's velocities when passing through the potential valley and crossing over the potential peak are now comparable.

 Evidently, in the current density range $|J_{\mathrm{b}}|<|J|<|J'_{\mathrm{b}}|$, extra \textit{degrees of freedom} must be excited within the macrospin system, which leads to a substantial enhancement of the overall energy consumption. Direct observation of the time traces for the RM variables in the macrospin simulation (as displayed in Fig. \ref{time_trace_s} \textbf{a} and \textbf{b}) provides the physical evidence: A sudden surge of the RM variables, shifting significantly away from their original equilibrium values, occurs within a very short time interval around $8$ picoseconds. This surge is highlighted by the yellow window, which exactly overlaps the time period when the angular velocity $\dot{\Phi}_{\mathrm{c}}$ is observed to be slowing down.

 Therefore, we conclude that within this critical time interval, the entire system temporarily deviates from the ideal Rigid Body Precession (RBP, or state \textbf{s}) regime, exhibiting a non-zero $M_{x(y)}$ component due to the RM excitation (as shown in Fig. \ref{time_trace_s} \textbf{c}). It is this RM surge that we hypothesize causes the effective damping constant of the CM particle to suddenly increase, resulting in the observed failure of the subcritical excitation at $|J| = |J_{\mathrm{b}}|$ in the macrospin simulation.

Building on the previous inference, we investigate the physical mechanism triggering the RM burst within the current density range $|J_{\mathrm{b}}|<|J|<|J'_{\mathrm{b}}|$. We propose that the IPA component of the anisotropy (as defined in Sec. \ref{sec:2E}) mediates the coupling between the CM and RM variables. The argument is as follows: Under the assumption of small $M_{x(y)}$, the CM dynamics is governed by $\dot{\Phi}_{\mathrm{c}}=[\Phi_{\mathrm{c}},H]\approx3A_{\mathrm{ex}}P_{\mathrm{c}}$,
$\dot{P}_{\mathrm{c}}=[P_{\mathrm{c}},H_{\mathrm{u}}]
=(1/3)(k/2)\sum_{i=1}^{3}
[(1-p_{i}^{2})]\sin[2(\phi_{i}-(i-1)(2\pi/3))]$. If the RM variables within $(\phi_{i}, p_{i})$ remain constant, the anisotropic force acting on the CM particle remains strictly conservative, entailing no additional energy dissipation for the CM motion-consistent with a perfect rigid body (TVM particle). However, the RM equations of motion, $\dot{\Phi}_{-12(23)}=[\Phi_{-12(23)},H_{\mathrm{u}}]\approx
(k/2)P_{-12(23)}$ and $\dot{P}_{-12(23)}=[P_{-12(23)},(A_{\mathrm{ex}}/2)
(\delta M_{x}^{2}+\delta M_{y}^{2})]+[P_{-12(23)},H_{\mathrm{u}}]$, suggest a different regime when the system deviates from the state \textbf{s} (see Fig. \ref{time_trace_s}(b)\textbf{d}). Here, the terms $[P_{-12(23)}, (A_{\mathrm{ex}}/2)(\delta M_{x}^2 + \delta M_{y}^2)]$ are given by Eqs. (22) and (24), while the IPA-related periodic forces are defined as $F_{\mathrm{p}12} = [P_{-12}, H_{\mathrm{u}}] = -(k/2) [(1-p_{1}^2)\sin(2\phi_{1}) - (1-p_{2}^2)\sin(2\phi_{2}-2\pi/3)]$ (and similarly for $F_{\mathrm{p}23}$). Since $\Phi_{\mathrm{c}}$ (embedded in $\phi_{i}$) increases with time, the CM translation serves as a periodic driver for RM oscillations via the IPA-related forces. Once the RM frequency aligns with these periodic forces-a condition we term \textit{self-resonance}-the RM variables are significantly excited. Consequently, the anisotropic force acting on the CM particle becomes \textit{non-conservative}, leading to a surge in energy dissipation. Notably, when the CM energy $E_{\mathrm{c}}$ is close to the potential barrier's peak (as shown in Fig. \ref{time_trace_s}(b)), the velocity $\dot{\Phi}_{\mathrm{c}}(t)$ exhibits a highly non-uniform, comb-like profile rather than a constant one. This non-uniformity implies that the IPA-related forces possess a \textit{multi-harmonic} periodicity, which significantly complicates the theoretical analysis of the triggered RM dynamics.

To provide evidence for the existence of the aforementioned self-resonance, we analyze the spectra of the driving force $F_{\mathrm{p}12}$ and the RM response $\Phi_{-12}$ as shown in Fig. \ref{RMburst_spectrum}. By examining whether their spectral peaks overlap, we can verify the resonance condition. Panels (a) and (b) illustrate the system response at different injected currents, $J=J_{\mathrm{b}}$ and $J>J_{\mathrm{b}}$, respectively, with a fixed damping constant ($\alpha = 0.01$) and anisotropy strength ($k = 0.032$). These results suggest that the self-resonance occurs predominantly at the sub-critical current density $J=J_{\mathrm{b}}$.

Furthermore, in panels (c) and (d), we examine cases with a smaller damping constant ($\alpha = 0.009$, panel (c)) and a weaker anisotropy strength ($k = 0.016$, panel (d)) at their respective sub-critical currents. This investigation confirms that self-resonance persists at $J=J_{\mathrm{b}}$ across various parameter regimes. Our findings indicate that the occurrence of self-resonance is robust against changes in intrinsic parameters, whereas the frequency of the RM response's main peak is primarily dictated by the anisotropy strength $k$.

Regarding the RM response frequency, a reasonable inference can be drawn: as suggested in Appendix \ref{appc}, the RM dynamics near the state \textbf{s} can be approximated as a simple harmonic oscillator with a natural angular frequency $\omega_{\mathrm{RM}}=\sqrt{A_{\mathrm{ex}}/M_{\mathrm{eff},\mathbf{s}}}\approx\sqrt{k/2}$. Consequently, the estimated natural frequencies for $k = 0.032$ and $k = 0.016$ are $f_{\mathrm{RM}} \approx 352.91$ THz and $249.5$ THz, respectively.

Although these estimations are simplified, they adequately explain why the RM response's main peak shifts with $k$. Moreover, the downward shift in the peak frequency of RM response observed in Fig. \ref{RMburst_spectrum}(b) at higher injected currents can be attributed to the increase in the average $P_{\mathrm{c}}$. This increase attenuates the effective phase-locking restoring forces, as indicated by the $p_{i}$-dependent pre-factors in Eqs.  (\ref{P_n12}) and (\ref{P_n23})).

Consequently, this attenuation shifts the RM frequency away from the resonance condition, which is the primary reason why the rigid-body constraint (representing the steady state \textbf{s}) underlying the TVM model remains robust-and even more stable-at higher current densities. Additionally, the main peak frequency in panel (a) ($431.7$ GHz) being slightly lower than that in panel (c) ($435.0$ GHz) clearly reflects a red frequency shift in the RM oscillations. This shift is consistent with the order of magnitude of the RM burst amplitude, as observed in the corresponding upper sub-panels, suggesting that a larger oscillation amplitude leads to a more pronounced reduction in the effective resonance frequency.

\begin{figure}
\begin{center}
\includegraphics[width=6cm]{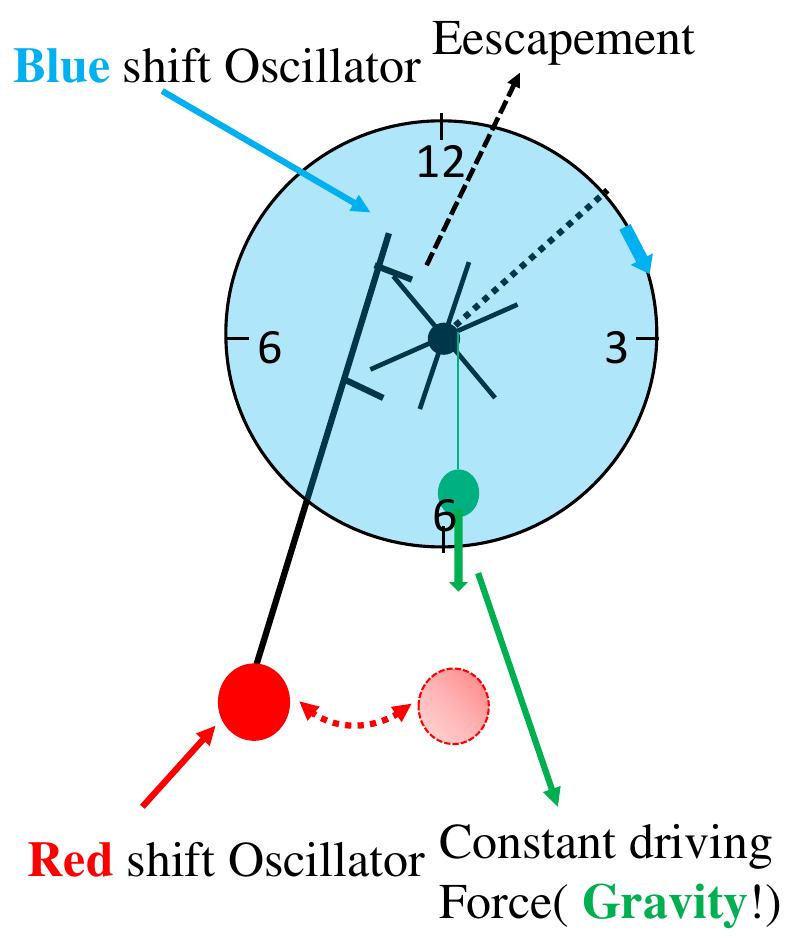}
\end{center}
\caption{(Color Online) Pendulum-Clock Analogy. An interesting and vivid analogy of the self-resonance based on the particle perspective within a pendulum-clock is provided here. The roles of the physical components correspond as follows: \textbf{IPA-Driving Force} $\rightarrow$ \textbf{Escapement}; \textbf{CM Effective Particle} $\rightarrow$ \textbf{Clock Pointer} (a blue shift global oscillator); \textbf{RM Effective Particles} $\rightarrow$ \textbf{Pendulum} (a red shift local oscillator); \textbf{SOT (Energy Provider) } $\rightarrow$ \textbf{Hammer} placed in a uniform gravity.}
\label{pendulm_clock}
\end{figure}

A vivid and interesting analogy for this self-resonance is provided by a uniform gravity-driven pendulum clock, as schematically drawn in Fig. \ref{pendulm_clock}. From the particle perspective, the functional roles of the system components are analogous to those of the clock's mechanism, where:
\begin{itemize}
    \item The $\text{CM}$'s $\text{TVM}$ particle (a \textbf{blue shift global oscillator}) corresponds to the \textbf{pointer}.

    \item The $\text{IPA}$-induced anisotropy corresponds to the \textbf{escapement}.

    \item The $\text{RM}$ particles' slow oscillation (a \textbf{red shift local oscillator}) corresponds to the \textbf{pendulum}.

    \item The $\text{SOT}$ (the energy provider) is analogous to a \textbf{hammer} placed in a uniform gravitational field that exerts a force on the pointer via a rope hanging from its pivot.
\end{itemize}

The pointer's frequency must equal the pendulum's natural frequency, showing that the self-resonance between them not only slows down the pointer's frequency but also \textbf{locks/maintains} its terminal velocity, even when driven by a slightly lighter or heavier hammer.

\vspace{0.5cm}

\subsection{\label{sec:2G} Hysteretic Excitation of the State \textbf{s}}
\begin{figure}
\begin{center}
\includegraphics[width=9cm]{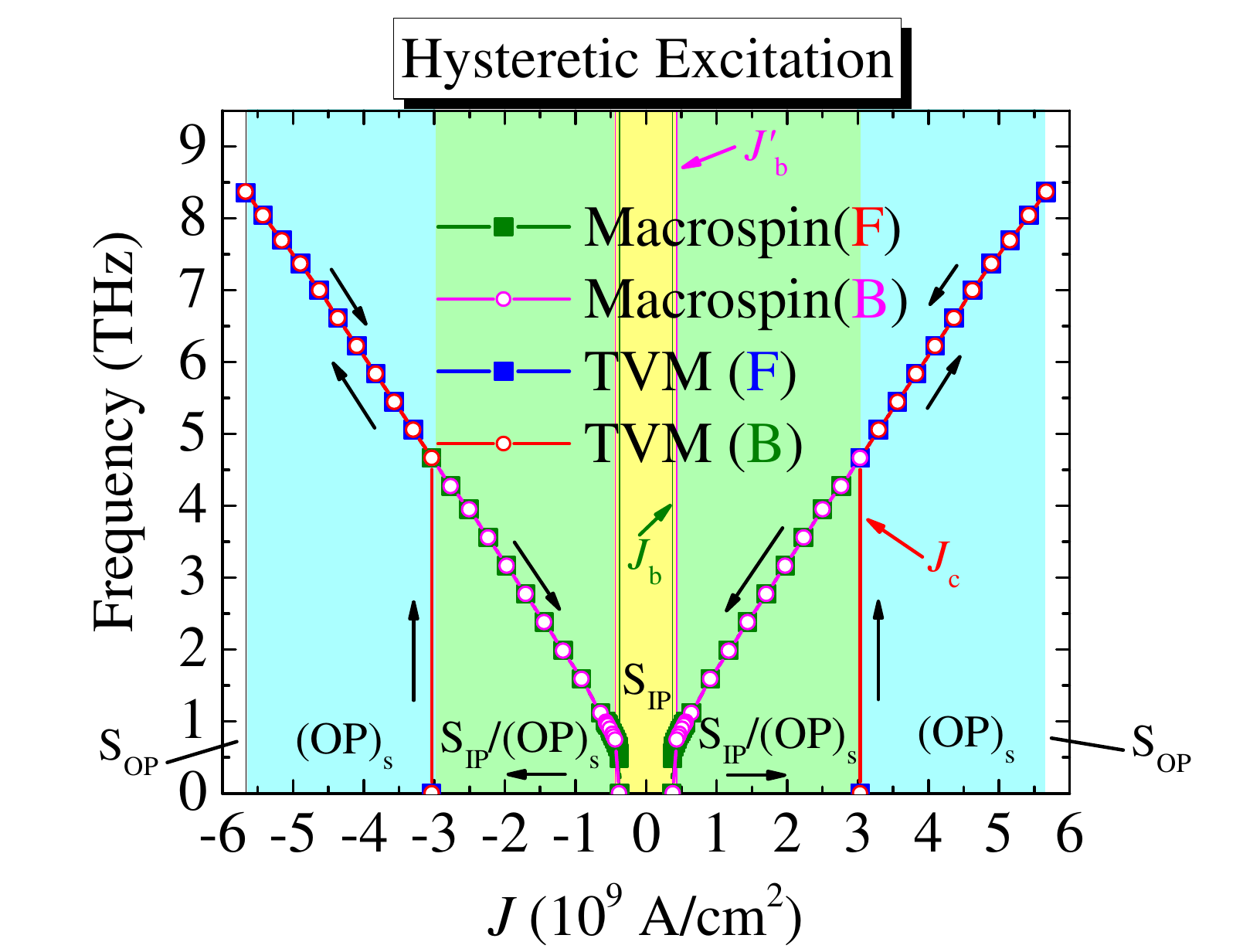}
\end{center}
\caption{(Color online) Hysteric frequency response of the RBP (RBUT) state \textbf{s} against current density. The figure displays the hysteretic frequency response of the RBP state $\mathbf{s}$ versus the injected current density, with parameters $\alpha$ and $k$ set consistent with Fig. \ref{time_trace_s}. The threshold current densities derived from the TVM (theoretical) and macrospin simulations are labeled by vertical lines: $|J_{\mathrm{b}}| = 0.3754 \times 10^{9} \text{A/cm}^2$ marked by the \textbf{dark green vertical lines} (TVM lower bound); $|J'_{\mathrm{b}}| = 0.4361 \times 10^{9} \text{A/cm}^2$ marked by the \textbf{magenta vertical lines} (Macrospin lower bound); $|J_{\mathrm{c}}| = 3.0339\times10^{9}\text{A/cm}^2$ marked by the \textbf{red vertical lines} (upper bound). \textbf{I. Dynamic Regimes Defined by Critical Currents}. These lines divide the current into four distinct dynamic mode regimes, visualized by colored windows. \textbf{Yellow Window ($\mathrm{S}_{\mathrm{IP}}$)}: Static states with two stable equilibriums. \textbf{Green Window ($\mathrm{S}_{\mathrm{IP}}/(\mathrm{OP})_{\mathbf{s}}$)}: Coexisting states with two stable in-plane equilibriums and one stable $\mathbf{s}$ state (RBP). \textbf{Blue Window ($(\mathrm{OP})_{\mathbf{s}}$)}: A single stable $\mathbf{s}$ state (RBP). \textbf{White Window ($\mathrm{S}_{(\mathrm{OP})}$)}: Single domain static state where the three spins point out-of-plane ($|\mathbf{m}_{i}| = 1$). \textbf{II. Data Representation}. \textbf{Macrospin Simulation Results}: Represented by the \textbf{dark green solid squares} and \textbf{magenta hollow circles}. \textbf{TVM (Theoretical) Results}: Represented by the \textbf{blue solid squares} and \textbf{red hollow circles}. \textbf{Hysteretic Process}: \textbf{Black arrows} indicate the hysteretic driving process of state $\mathbf{s}$ by sweeping the current.}
\label{Hys_excitation}
\end{figure}
Given the theoretically calculated or simulated threshold currents, one attends at a dynamic state phase diagram as a function of current (density) alone, as displayed in Fig. \ref{Hys_excitation}. In addition, the generated frequency by the macrospin and TVM model simulations are also given, both of which are apparently in perfect line with each other, except for the disagreement about the lower threshold current, as argued in the previous section. This manifests the TVM model is of an great accuracy in describing the CM's dynamics no matter qualitatively or quantitatively (see also
Ref. \cite{chen2023a}).

More importantly, the coexisting states $\mathrm{S}_{\mathrm{IP}}/(\mathrm{OP})_{\mathrm{s}}$ with two stable in-plane equilibriums and a stable state s (RBP) represents the appearance of a \textit{hysteric} current-driven loop of the state s, as indicated by the black arrows in Fig. \ref{Hys_excitation}, reflecting the existence of the kinetic-like energy that is directly associated with the strong non-linear frequency shift coefficient of an AFM exchange coupling just like a Newton particle such as a pendulum. This point is also shared by most of the other types of STNOs, e.g. a FM-based PERP-STNO \cite{HaoHsuan2017,Chen2021,chen2023a}.

\section{\label{sec:3}Summary and Discussion}
In summary, this work establishes a rigorous analytical framework for NC-AFM STOs. Unlike conventional descriptions bound to geometric torques, we employ Poisson Brackets to inspect the continuous operational symmetries of the system's Hamiltonian, thereby bypassing traditional torque-based limitations. A dual perspective is developed for spin dynamics: the vector perspective and the particle perspective. Within this framework, the intrinsic dynamic modes of the AFM exchange couplings for the three coupled sub-lattices' spins are proven to be energy-degenerate RBP states in the vector perspective. In which, Their complex dynamics are elegantly decomposed into a RBUT mode (represented by CM variables) and an EO mode (represented by RM variables). Furthermore, we show that OPA component of the in-plane uniaxial anisotropy lifts this degeneracy, inducing slow RM oscillations accompanied by a fast EO.

Through time-dependent RBRT/RBUTT techniques, we analytically solve the stable degenerate SOT-driven RBP(RBUT) states in an NC-AFM STO, offering an explanation of why the collective modes of FM exchange couplings fail to be driven by SOTs (STTs) in an FM STO as a comparison. Moreover, our analysis identifies a clear separation of time scales within the system: a rapid SOT-driven transient process ($\sim 10$ ps) followed by a long-term oscillatory decay ($\sim 1$ ns) induced by the OPA component of the in-plane uniaxial anisotropy toward the steady state $\mathbf{s}$. Upon reaching the steady state \textbf{s}, the entire system is accurately modeled as a TVM particle. In this regime, the exchange coupling is effectively mapped onto a kinetic-like energy term with an exceedingly light effective mass of inertia. This TVM model provides precise predictions across the full current range, successfully capturing the stability phase diagram, hysteretic SOT-driven processes, and even the transient processes.

We resolve the theoretical mismatch in the sub-critical current regime by identifying the 'Rigid-Body Breaking' effect. We demonstrate that as the CM energy approaches the potential barrier peak, the velocity profile becomes non-uniform and 'comb-like'. This acts as a multi-harmonic driver that triggers a self-resonant 'RM burst' mediated by the IPA component of the in-plane uniaxial anisotropy. This resonance significantly enhances energy dissipation, thereby preventing the system from reaching the anticipated excited state at the sub-critical current. This discovery not only defines the operational boundaries of the TVM model but also offers profound insights into the nonlinear mode-coupling within topological magnetic systems.

Finally, a pivotal extension of our theory lies in its multi-scale scalability to massive multi-lattice systems. We propose a hierarchical analytical framework that partitions the system's complexity into two self-consistent layers. At the fundamental layer, by applying periodic constraints independently to each sub-lattice species, we demonstrate that even when the $120^{\circ}$ rigid-body configuration is distorted by unconventional SOT polarizations or the absence of anisotropy ($H_{\mathrm{u}i}=0$), the system maintains a robust \textbf{sub-lattice periodic mapping}. This allows each micromagnetic cell to be rigorously represented as a composite of three inter-penetrating CM-blocks. At the macroscopic layer, localized fluctuations, analogous to wave packets, are systematically suppressed by the synergic influence of SOTs and damping. This mode-selection mechanism effectively neutralizes high-frequency relative motion (RM) modes, converging the massive lattice into a coherent assembly of CM units. Crucially, as the Poisson Bracket formalism is intrinsically scale-invariant, this framework enables a seamless transition where long-range textures and collective behaviors can be governed by a unified micromagnetic coordination, providing a versatile foundation for the design of large-scale, non-collinear sub-THz devices.

\section*{Supplementary Material}
See the supplementary material for high-fidelity animations (approximately 380 MB) illustrating the degenerate rigid-body precessional (RBP) states and relative motion (RM) burst processes. These videos provide a direct visualization of the sub-lattice trajectories and the self-resonance mechanism discussed in the Terminal Velocity Motion (TVM) framework.

\section*{Data Availability Statement}
The high-resolution movies supporting the findings of this study are openly available on Figshare at \url{https://doi.org/10.6084/m9.figshare.31153270}. The numerical datasets that support the findings are available from the corresponding author upon reasonable request.

\appendix
\section{\label{appa} Non-uniform SOT Injection into Spins}
Just as with the point raised in Sec. \ref{sec:2E2}, the fact that the SOT, even for only one of the spins receiving a non-zero injection of it such as $a_{J1}=a_{J0}\neq0$ and $a_{J2}=a_{J3}=0$, can only drive the RBP states with $\mathbf{e}_{\mathrm{n}}$ being collinear to $\mathbf{p}=\mathbf{z}$ is still valid. Thus, coming back to Eq. (\ref{any_quantity}) and utilizing the reduced Hamiltonian $H_{\mathrm{ex,\mathbf{P}}}$ for these RBP states (or rather RBUT states) with $M_{x(y)}=0$, i.e. $\dot{p}_{i}=[p_{i},H_{\mathrm{ex,\mathbf{P}}}]=0$ and $\dot{\Phi}_{\mathrm{c}}=[\Phi_{\mathrm{c}},H_{\mathrm{ex,\mathbf{P}}}]$, we get the equations of motion of the variables $(\phi_{i},p_{i})$ as follows:
\setlength\abovedisplayskip{6pt}
\setlength\belowdisplayskip{6pt}
\begin{eqnarray}
\dot{p}_{i}
&=&-\alpha\left(1-p_{i}^{2}\right)\dot{\Phi}_{\mathrm{c}}
+\left(1-p_{i}^{2}\right)a_{Ji}(-p_{i}),\nonumber\\
\dot{\phi}_{i}
&=&\dot{\Phi}_{\mathrm{c}},
\label{LLS_for_non_uniform_SOT}
\end{eqnarray}
with $a_{J1}=a_{J0}\neq0$ and $a_{J2}=a_{J3}=0$.

Requiring the RM variables being stationary, i.e. $\dot{\Phi}_{-12(23)}=0$ as well as $\dot{P}_{-12(23)}=0$, one immediately gets $|p_{2}|=|p_{3}|\neq |p_{1}|$ from
\setlength\abovedisplayskip{6pt}
\setlength\belowdisplayskip{6pt}
\begin{eqnarray}
\dot{P}_{-12}&=&-\alpha
\left[\left(1-p_{1}^{2}\right)-\left(1-p_{2}^{2}\right)\right]
\dot{\Phi}_{\mathrm{c}}+\left(1-p_{1}^{2}\right)\nonumber\\
&&\times a_{J1}(-p_{1})=0,\nonumber\\
\dot{P}_{-23}&=&-\alpha
\left[\left(1-p_{2}^{2}\right)-\left(1-p_{3}^{2}\right)\right]
\dot{\Phi}_{\mathrm{c}}=0,\nonumber
\end{eqnarray}
saying that the $M_{z}$ degeneracy level of those conserved RBP sates meeting with Eq. (\ref{degenerate}) will be greatly reduced by the non-uniform SOT injections. Moreover, the constraints for the SOT-driven RBP states become (see Eq. (\ref{degenerate}))
\setlength\abovedisplayskip{6pt}
\setlength\belowdisplayskip{6pt}
\begin{eqnarray}
\sqrt{1-p_{1}^{2}}\cos\phi_{1}&=&-\sqrt{1-p_{2}^{2}}
\left(\cos\phi_{2}+\cos\phi_{3}\right),\nonumber\\
\sqrt{1-p_{1}^{2}}\sin\phi_{\mathrm{n}1}&=&-\sqrt{1-p_{2}^{2}}
\left(\sin\phi_{2}+\sin\phi_{3}\right),\nonumber\\
P_{\mathrm{c}}&=&\left(\frac{1}{3}\right)\sum_{i=1}^{3}p_{i}.
\label{non_degenerate}
\end{eqnarray}
If $\phi_{1}$ is set to be zero, then we have $\phi_{3}=\phi_{2}+2N\pi$ with $N$ belonging to an integer and $p_{1}=\pm\sqrt{1-4\left(1-p_{2}^{2}\right)}$ with $|p_{2}|>\sqrt{3/4}$. In addition, $P_{\mathrm{c}}=\pm\left(1/3\right)\sqrt{1-4\left(1-p_{2}^{2}\right)}$ with $|p_{2}|>\sqrt{3/4}$ and $|P_{\mathrm{c}}|\in[0,1/3]$, where $p_{3}=-p_{2}$ has been used due to the state for $p_{3}=p_{2}$ having a higher exchange energy, which has a worse stability compared to that of $p_{3}=-p_{2}$. Note here that $p_{(2)3}$ being able to be either positive or negative means that there are still two degenerate RBP states for a given value of $M_{z}$ or angular frequency $\dot{\Phi}_{\mathrm{c}}$.

In terms of the CM variables, the above equations of motion read
\setlength\abovedisplayskip{6pt}
\setlength\belowdisplayskip{6pt}
\begin{eqnarray}
\dot{P}_{\mathrm{c}}&=&\left(\frac{1}{3}\right)\sum_{i=1}^{3}\dot{p}_{i},\nonumber\\
&=&\left(\frac{1}{3}\right)\left[\sum_{i=1}^{3}-\alpha(1-p_{i}^{2})
\dot{\Phi}_{\mathrm{nc}}+(1-p_{1}^{2})
a_{J0}\right],\nonumber\\
&=&-\alpha S(P_{\mathrm{c}})\dot{\Phi}_{\mathrm{c}}+\beta(P_{\mathrm{c}},J),\nonumber\\
\dot{\Phi}_{\mathrm{c}}&=&\left(\frac{1}{3}\right)\frac{\partial H_{\mathrm{ex,\mathbf{p}}}}{\partial P_{\mathrm{c}}}=3A_{\mathrm{ex}}P_{\mathrm{c}},
\label{CM_timerate_with_nonuniform_SOT}
\end{eqnarray}
where $p_{1}(P_{\mathrm{c}})=3P_{\mathrm{c}}$ and $p_{2}(P_{\mathrm{c}})=-p_{3}(P_{\mathrm{c}})=
\pm\sqrt{1-\left(1/4\right)\left[1-\left(3P_{\mathrm{c}}\right)^{2}\right]}$, and where the positive and negative damping factors are $S(P_{\mathrm{c}})=(1/2)\left(1-9P_{\mathrm{c}}^{2}\right)$ and $\beta(P_{\mathrm{c}},J)=\left(1/3\right)\left(1-9P_{\mathrm{c}}^{2}\right)a_{J0}$, respectively\cite{Chen2021}.

Just as we did in Sec. \ref{sec:2E2}, by applying a time-dependent RBRT (RBUTT) to the system\cite{Chen2021}:  $(\phi'_{i},p'_{i})=(\phi_{i}-\omega(M_{z},J)t
 ,p_{i})=(\phi_{i}-v(P_{\mathrm{c}},J)t,p_{i})$, the Hamiltonian in the new frame is $H'=H'_{\mathrm{ex}}(|\mathbf{M}'|^{2})+H'_{\mathrm{boost}}(P'_{\mathrm{c}},J)$ with
 $H'_{\mathrm{boost}}(P'_{\mathrm{c}},J)=-3\int^{P'_{\mathrm{c}}}dP''_{\mathrm{c}}
 v(P''_{\mathrm{c}},J)$ with $v(P'_{\mathrm{c}},J)=\beta(P'_{\mathrm{c}},J)/\alpha S(P'_\mathrm{c})=2a_{J0}/(3\alpha)$.

 Again, $H'$ means that if the initial states have $M'_{x(y)}\neq0$, the whole system must be in motion in the beginning, leading to it being damped out into one of the states with $M'_{x(y)}=0$ eventually. One can therefore focus on the SOT-driven RBP states with $\mathbf{M}'$ being collinear to $\mathbf{p}$, that is, we have the reduced Hamiltonian
 $H'_{\mathbf{P}}(P'_{\mathrm{c}})=(9A_{\mathrm{ex}}/2)(P'_{\mathrm{c}})^{2}
-(2a_{J0}/\alpha)P'_{\mathrm{c}}$.

 By requiring $(\partial H'_{\mathbf{P}}/\partial P'_{\mathrm{c}})_{P'_{\mathrm{c}0}}=0$ as well as $[\partial^{2}H'_{\mathbf{P}}/\partial (P'_{\mathrm{c}})^{2}]_{P'_{\mathrm{c}0}}>0$, we solve the stable driven degenerate RBP states $P'_{\mathrm{c}0}=2a_{J0}/(9A_{\mathrm{ex}}\alpha)$ and their sharing stability (the curvature of $H'_{\mathbf{P}}$): $[\partial^{2}H'_{\mathbf{P}}/\partial (P'_{\mathrm{c}})^{2}]_{P'_{\mathrm{c}0}}=3A_{\mathrm{ex}}>0$, with $0<|a_{J0}|<3A_{\mathrm{ex}}\alpha/2$.

\section{\label{appb} Derivation of the TVM model's Equation of Motion}
Using the Poisson bracket defined in terms of CM's set of variables $(\Phi_{\mathrm{c}},P_{\mathrm{c}})$ and Eq. (\ref{any_quantity}):
\setlength\abovedisplayskip{6pt}
\setlength\belowdisplayskip{6pt}
\begin{eqnarray}
[f,g]_{\Phi_{\mathrm{c}},P_{\mathrm{c}}}=\left(\frac{1}{3}\right)\left(\frac{\partial f}{\partial\Phi_{\mathrm{c}}}\frac{\partial g}{\partial P_{\mathrm{c}}}
-\frac{\partial f}{\partial P_{\mathrm{c}}}\frac{\partial g}{\partial\Phi_{\mathrm{c}}}\right),\nonumber
\label{}
\end{eqnarray}
we have
\setlength\abovedisplayskip{6pt}
\setlength\belowdisplayskip{6pt}
\begin{eqnarray}
\ddot{\Phi}_{\mathrm{c}}&=&[[\Phi_{\mathrm{c}},H_{\mathbf{P},\mathbf{s}}],H_{\mathbf{P},\mathbf{s}}]
+\sum_{i=1}^{3}\bigg[\frac{\partial[\Phi_{\mathrm{c}},H_{\mathbf{P},\mathbf{s}}]}{\partial p_{i}}\left(-\frac{\partial F_{\mathrm{d}}}{\partial\dot{\phi}_{i}}\right)\nonumber\\
&&+\frac{\partial[\Phi_{\mathrm{c}},H_{\mathbf{P},\mathbf{s}}]}
{\partial\phi_{i}}\frac{\partial F_{\mathrm{d}}}{\partial\dot{p}_{i}}\bigg],\nonumber\\
&\approx&\left(\frac{1}{3}\right)\left(\frac{\partial^{2}H_{\mathbf{P},\mathbf{s}}}{\partial P_{\mathrm{c}}^{2}}\right)\left[-\frac{1}{3}\frac{\partial H_{\mathbf{P},\mathbf{s}}}{\partial\Phi_{\mathrm{c}}}+\left(-\frac{\partial F_{\mathrm{d}}}{\partial\dot{\phi}_{i}}\right)_{\mathbf{s}}\right],\nonumber\\
&=&\left(\frac{1}{3}\right)\left(\frac{\partial^{2}H_{\mathrm{o},\mathbf{s}}}{\partial P_{\mathrm{c}}^{2}}\right)\left[-\frac{1}{3}\frac{\partial H_{\mathrm{u},\mathbf{s}}}{\partial\Phi_{\mathrm{c}}}+\left(-\frac{\partial F_{\mathrm{d}}}{\partial\dot{\phi}_{i}}\right)_{\mathbf{s}}\right]\nonumber\\
&&+\left(\frac{1}{3}\right)\frac{d}{dt}\left(\frac{\partial H_{\mathrm{u},\mathbf{s}}}{\partial P_{\mathrm{c}}}\right),\nonumber\\
&\approx&\left[\frac{1}{m_{\mathrm{eff},\mathbf{s}}(P_{\mathrm{c}})}\right]\left[-\frac{1}{3}\frac{\partial H_{\mathrm{u},\mathbf{s}}}{\partial\Phi_{\mathrm{c}}}+\left(-\frac{\partial F_{\mathrm{d}}}{\partial\dot{\phi}_{i}}\right)_{\mathbf{s}}\right],\nonumber
\end{eqnarray}
with the reduced Hamiltonian $H_{\mathbf{P},\mathbf{s}}(P_{\mathrm{c}},\Phi_{\mathrm{c}})=H_{\mathrm{o},\mathbf{s}}(P_{\mathrm{c}})
+H_{\mathrm{u},\mathbf{s}}
(P_{\mathrm{c}},\Phi_{\mathrm{c}})=(9A_{\mathrm{ex}}/2)P_{\mathrm{c}}^{2}-(3k/2)
(1-P_{\mathrm{c}}^{2})\cos^{2}(\Phi_{\mathrm{c}}-2\pi/3)$ for the state \textbf{s}, the effective mass of inertia $m_{\mathrm{eff},\mathbf{s}}(P_{\mathrm{c}})=[(1/3)\partial^{2} H_{\mathrm{o},\mathbf{s}}/\partial P_{\mathrm{c}}^{2}]^{-1}=(3A_{\mathrm{ex}})^{-1}$, $\partial F_{\mathrm{d}}/\partial\dot{p}_{i}$ being neglected due to $|\dot{\phi}_{i}|\gg|\dot{p}_{i}|$, $(1/3)(d/dt)\partial H_{\mathrm{u},\mathbf{s}}/\partial P_{\mathrm{c}}$ being dropped out due to $|H_{\mathrm{o},\mathbf{s}}|\gg|H_{\mathrm{u},\mathbf{s}}|$, and the dissipative force for the state \textbf{s} $(-\partial F_{\mathrm{d}}/\partial\dot{\phi}_{i})_{\mathbf{s}}=-\alpha S(P_{\mathrm{c}})\dot{\Phi}_{\mathrm{c}}+\beta(P_{\mathrm{c}},J)$
with $S(P_{\mathrm{c}})=(1-P_{\mathrm{c}}^{2})$ and $\beta(P_{\mathrm{c}},J)=a_{J0}S(P_{\mathrm{c}})$.

Moreover, using $\dot{\Phi}_{\mathrm{c}}=[\Phi_{\mathrm{c}},H_{\mathbf{P},\mathbf{s}}]=\omega_{\mathrm{o}}(P_{\mathrm{c}})
+(1/3)\partial H_{\mathrm{u},\mathbf{s}}/\partial P_{\mathrm{c}}
=3A_{\mathrm{ex}}P_{\mathrm{c}}
+kP_{\mathrm{c}}\cos^{2}(\Phi_{\mathrm{c}}-2\pi/3)$ and approximately expressing the momentum $P_{\mathrm{c}}$ as a function of the angular velocity $\dot{\Phi}_{\mathrm{c}}$ as (see Ref. \cite{chen2023a}
for the details)

\setlength\abovedisplayskip{6pt}
\setlength\belowdisplayskip{6pt}
\begin{eqnarray}
P_{\mathrm{c}}&=&\omega_{0}^{-1}\left(\dot{\Phi}_{\mathrm{c}}-kP_{\mathrm{c}}
\cos^{2}(\Phi_{\mathrm{c}}-2\pi/3)\right),\nonumber\\
&\approx&\omega_{\mathrm{o}}^{-1}\left(\dot{\Phi}_{\mathrm{c}}\right)
-\left(\frac{\partial\omega_{\mathrm{o}}}{\partial P_{\mathrm{c}}}\right)_{\omega_{\mathrm{o}}^{-1}\left(\dot{\Phi}_{\mathrm{c}}
\right)}^{-1}k\omega_{\mathrm{o}}^{-1}\left(\dot{\Phi}_{\mathrm{c}}\right)\nonumber\\
&&\times\cos^{2}(\Phi_{\mathrm{c}}-2\pi/3),\nonumber\\
&=&\left(\frac{\dot{\Phi}_{\mathrm{c}}}{3A_{\mathrm{ex}}}\right)-\frac{k}
{(3A_{\mathrm{ex}})^{2}}\dot{\Phi}_{\mathrm{c}}\cos^{2}(\Phi_{\mathrm{c}}-2\pi/3),\nonumber\\
&\approx&\frac{\dot{\Phi}_{\mathrm{c}}}{3A_{\mathrm{ex}}}.
\label{P_replaced_by_dotPhi}
\end{eqnarray}
with, notably, $|\dot{\Phi}_{\mathrm{c}}|\leq3A_{\mathrm{ex}}=3$ due to $|P_{\mathrm{c}}|\leq1$.

Finally, the equation of motion of $\dot{\Phi}_{\mathrm{c}}$ expressed in terms of the configuration space $(\Phi_{\mathrm{c}},\dot{\Phi}_{\mathrm{c}})$ reads
\setlength\abovedisplayskip{6pt}
\setlength\belowdisplayskip{6pt}
\begin{eqnarray}
\ddot{\Phi}_{\mathrm{c}}&=&\frac{1}{m_{\mathrm{eff},\mathbf{s}}(\dot{\Phi}_{\mathrm{c}})}\Bigg[-\alpha S(\dot{\Phi}_{\mathrm{c}})\dot{\Phi}_{\mathrm{c}}+\beta(\dot{\Phi}_{\mathrm{c}},J)\nonumber\\
&&-\frac{1}{3}\left(\frac{\partial H_{\mathrm{u},\mathbf{s}}}{\partial\Phi_{\mathrm{c}}}\right)_{P_{\mathrm{c}}=
\frac{\dot{\Phi}_{\mathrm{c}}}{3A_{\mathrm{ex}}}}\Bigg],
\label{}
\end{eqnarray}
with $S(\dot{\Phi}_{\mathrm{c}})=1-[\dot{\Phi}_{\mathrm{c}}/(3A_{\mathrm{ex}})]^{2}$ and $\beta(\dot{\Phi}_{\mathrm{c}},J)=a_{J0}S(\dot{\Phi}_{\mathrm{c}})$.

\section{\label{appc} Derivation of the Newton-like Model for Each of the three Spins}
We aim to derive an approximated and complete Newton-like description for the dynamics of each individual spin in the system. As evident from the transformation relationship, Eq. (\ref{inverssphi_to_Phi}), the angular accelerations of the three spins ($\ddot{\phi}_{1}, \ddot{\phi}_{2}, \ddot{\phi}_{3}$) are linearly related to the accelerations of the CM and RM variables ($\ddot{\Phi}_{\mathrm{c}}, \ddot{\Phi}_{-12}, \ddot{\Phi}_{-23}$) via the following matrix equation:

\setlength\abovedisplayskip{6pt}
\setlength\belowdisplayskip{6pt}
\begin{eqnarray}
\left(
     \begin{matrix}
     \ddot{\phi}_{1}\\
     \ddot{\phi}_{2}\\
     \ddot{\phi}_{3}\\
     \end{matrix}
\right)=
\left(
      \begin{matrix}
      1&\frac{2}{3}&\frac{1}{3}\\
      1&-\frac{1}{3}&\frac{1}{3}\\
      1&-\frac{1}{3}&-\frac{2}{3}\\
      \end{matrix}
\right)
\left(
      \begin{matrix}
      \ddot{\Phi}_{\mathrm{c}}\\
      \ddot{\Phi}_{-12}\\
      \ddot{\Phi}_{-23}\\
      \end{matrix}
\right).
\label{inverssphi_to_Phi_ddot}
\end{eqnarray}
Equation (\ref{inverssphi_to_Phi_ddot}) clearly shows that the Newton-like equation of motion for each spin is comprised of two fundamental components: 1. The component stemming from the TVM model for the CM variable $\Phi_{\mathrm{c}}$ (representing the collective motion); 2. The component originating from the equations for the RM variables $\Phi_{-12}$ and $\Phi_{-23}$ (representing the internal, non-collective motions). As suggested by Sec. \ref{sec:2E2}, the necessary equations for the RM variables can be obtained based on the principle that the OPA provides the effective mass of inertia for the system.

The approximated equations of motion for the RM variables' conserved dynamics can be derived in the following. Coming back to Eqs. (\ref{P_n12})-(\ref{Phi_n23}) and neglecting the second terms in Eqs. (\ref{Phi_n12}) and (\ref{Phi_n23}) that are related to $M_{x(y)}$ under $M_{x(y)}\approx0$, we have

\setlength\abovedisplayskip{6pt}
\setlength\belowdisplayskip{6pt}
\begin{eqnarray}
\dot{P}_{\mathrm{-12}}
&=&A_{\mathrm{ex}}\bigg[2\sqrt{\left(1-p_{1}^{2}\right)\left(1-p_{2}^{2}\right)}
\sin\Phi_{-12}\nonumber\\
&&+\sqrt{\left(1-p_{1}^{2}\right)\left(1-p_{3}^{2}\right)}
\sin\left(\Phi_{-12}+\Phi_{-23}\right)\nonumber\\
&&-\sqrt{\left(1-p_{2}^{2}\right)\left(1-p_{3}^{2}\right)}
\sin\Phi_{-23}\bigg],\nonumber
\end{eqnarray}

\setlength\abovedisplayskip{6pt}
\setlength\belowdisplayskip{6pt}
\begin{eqnarray}
\dot{\Phi}_{-12}\approx\left(\frac{k'}{2}\right)P_{-12},\nonumber
\end{eqnarray}

\setlength\abovedisplayskip{6pt}
\setlength\belowdisplayskip{6pt}
\begin{eqnarray}
\dot{P}_{\mathrm{-23}}
&=&A_{\mathrm{ex}}\bigg[-\sqrt{\left(1-p_{1}^{2}\right)\left(1-p_{2}^{2}\right)}
\sin\Phi_{-12}\nonumber\\
&&+2\sqrt{\left(1-p_{2}^{2}\right)\left(1-p_{3}^{2}\right)}
\sin\Phi_{\mathrm{n}-23}\nonumber\\
&&+\sqrt{\left(1-p_{1}^{2}\right)\left(1-p_{3}^{2}\right)}
\sin\left(\Phi_{-12}+\Phi_{-23}\right)\bigg],\nonumber
\end{eqnarray}
and

\setlength\abovedisplayskip{6pt}
\setlength\belowdisplayskip{6pt}
\begin{eqnarray}
\dot{\Phi}_{-23}\approx\left(\frac{k'}{2}\right)P_{-23},\nonumber
\end{eqnarray}
where, notably, $k'\approx k/2$ is estimated from the macrospin simulation, and, therefore, the effective mass is $M_{\mathrm{eff},\mathbf{s}}=2/k'\gg m_{\mathrm{eff},\mathbf{s}}$. Then, the Newton-like equations for $\Phi_{-12(23)}$ are
\setlength\abovedisplayskip{6pt}
\setlength\belowdisplayskip{6pt}
\begin{eqnarray}
\ddot{\Phi}_{-12(23)}\approx\left(\frac{1}{M_{\mathrm{eff},\mathbf{s}}}\right)
\dot{P}_{-12(23)},
\label{RM_conserved_Newton}
\end{eqnarray}
where $p_{i}$ is replaced by $P_{\mathrm{c}}$ and $P_{-12(23)}$, according to the transformation Eq. (\ref{inversep_to_P}), then substituted with $\dot{\Phi}_{\mathrm{c}}$ and $\dot{\Phi}_{-12(23)}$ using the relations $P_{\mathrm{c}}=m_{\mathrm{eff},\mathbf{s}}\dot{\Phi}_{\mathrm{c}}$ and $P_{-12(23)}=M_{\mathrm{eff},\mathbf{s}}\dot{\Phi}_{-12(23)}$, i.e.
\setlength\abovedisplayskip{6pt}
\setlength\belowdisplayskip{6pt}
\begin{eqnarray}
\left(
     \begin{matrix}
     p_{1}\\
     p_{2}\\
     p_{3}\\
     \end{matrix}
\right)=
\left(
      \begin{matrix}
      1&\frac{2}{3}&\frac{1}{3}\\
      1&-\frac{1}{3}&\frac{1}{3}\\
      1&-\frac{1}{3}&-\frac{2}{3}\\
      \end{matrix}
\right)
\left(
      \begin{matrix}
      m_{\mathrm{eff},\mathbf{s}}\dot{\Phi}_{\mathrm{c}}\\
      M_{\mathrm{eff},\mathbf{s}}\dot{\Phi}_{\mathrm{-12}}\\
      M_{\mathrm{eff},\mathbf{s}}\dot{\Phi}_{\mathrm{-23}}\\
      \end{matrix}
\right),
\label{p_to_phi}
\end{eqnarray}
and finally expressed by $\dot{\phi}_{i}$ utilizing the transformation Eq. (\ref{transphi_to_Phi}), i.e.
\setlength\abovedisplayskip{6pt}
\setlength\belowdisplayskip{6pt}
\begin{eqnarray}
\left(
     \begin{matrix}
     \dot{\Phi}_{\mathrm{c}}\\
     \dot{\Phi}_{-12}\\
     \dot{\Phi}_{-23}\\
     \end{matrix}
\right)=
\left(
      \begin{matrix}
      \frac{1}{3}&\frac{1}{3}&\frac{1}{3}\\
      1&-1&0\\
      0&1&-1\\
      \end{matrix}
\right)
\left(
      \begin{matrix}
      \dot{\phi}_{1}\\
      \dot{\phi}_{2}\\
      \dot{\phi}_{3}\\
      \end{matrix}
\right),
\label{transphi_to_Phi_dot}
\end{eqnarray}
we get the transformation between the two quantities $p_{i}$ and $\phi_{i}$ in matrix form
\setlength\abovedisplayskip{6pt}
\setlength\belowdisplayskip{6pt}
\begin{eqnarray}
\mathbf{p}=\mathbf{\Lambda}\left(m_{\mathrm{eff},\mathbf{s}}, M_{\mathrm{eff},\mathbf{s}}\right)\cdot\dot{\mathbf{\phi}},
\label{p_to_phi_matrix}
\end{eqnarray}

with the matrix of mass of inertia being
\setlength\abovedisplayskip{6pt}
\setlength\belowdisplayskip{6pt}
\begin{eqnarray}
&&\mathbf{\Lambda}\doteq\left(\frac{1}{3}\right)\times\nonumber\\
&&\left(
 \begin{matrix}
      m_{\mathrm{eff},\mathbf{s}}+2M_{\mathrm{eff},\mathbf{s}}&m_{\mathrm{eff},\mathbf{s}}-M_{\mathrm{eff},\mathbf{s}}&m_{\mathrm{eff},\mathbf{s}}-M_{\mathrm{eff},\mathbf{s}}\\
      m_{\mathrm{eff},\mathbf{s}}-M_{\mathrm{eff},\mathbf{s}}&m_{\mathrm{eff},\mathbf{s}}+2M_{\mathrm{eff},\mathbf{s}}&m_{\mathrm{eff},\mathbf{s}}-M_{\mathrm{eff},\mathbf{s}}\\
      m_{\mathrm{eff},\mathbf{s}}-M_{\mathrm{eff},\mathbf{s}}&m_{\mathrm{eff},\mathbf{s}}-M_{\mathrm{eff},\mathbf{s}}&m_{\mathrm{eff},\mathbf{s}}+2M_{\mathrm{eff},\mathbf{s}}\\
 \end{matrix}
\right)\nonumber\\
\label{matrix of mass}
\end{eqnarray}
.

Regarding the portrait of the non-conservative part for the RM variables' dynamics, an effective damping force suggested by the simplified model introduced in Sec. \ref{sec:2E2} (or see Eq. (\ref{sim_model_damp})) is here given to be $-\alpha S_{\mathrm{B}}(\dot{\Phi}_{\mathrm{-12(23)}})\dot{\Phi}_{\mathrm{-12(23)}}$. Eq. (\ref{RM_conserved_Newton}) thus becomes
\setlength\abovedisplayskip{6pt}
\setlength\belowdisplayskip{6pt}
\begin{eqnarray}
\ddot{\Phi}_{-12(23)}&\approx&\left(\frac{1}{M_{\mathrm{eff},\mathbf{s}}}\right)
\bigg[-\alpha S_{\mathrm{B}}\left(\dot{\Phi}_{-12(23)}\right)\dot{\Phi}_{-12(23)}\nonumber\\
&&+\dot{P}_{-12(23)}\left(\Phi_{-12(23)},\dot{\Phi}_{-12(23)}\right)\bigg],
\label{RM_conserved_Newton}
\end{eqnarray}
where the positive damping factor $S_{\mathrm{B}}$ can be estimated through comparing with the macrospin simulation.

\begin{figure*}
\begin{center}
\includegraphics[width=17cm]{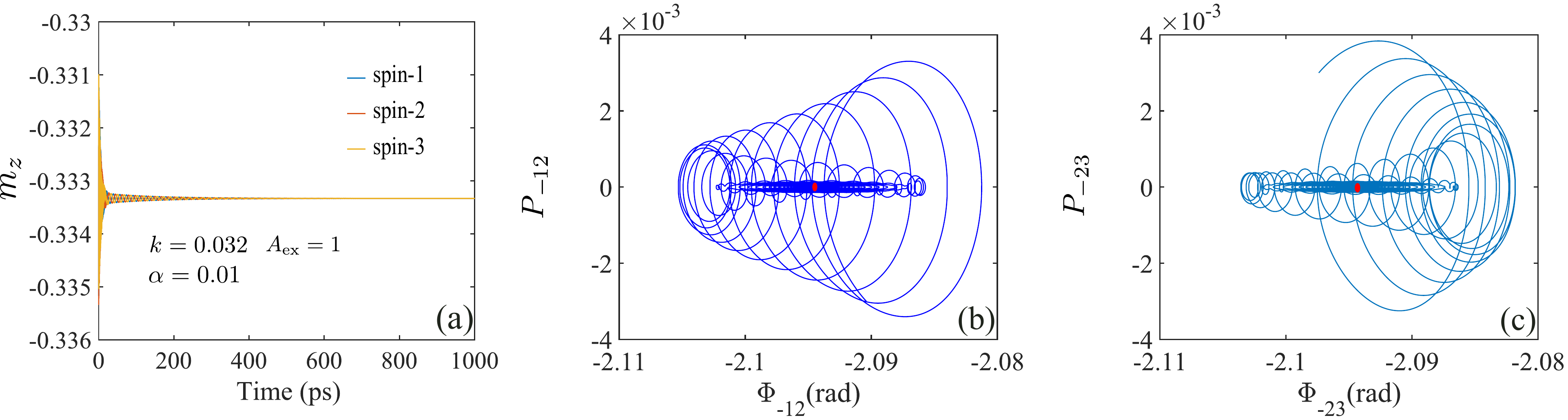}
\end{center}
\caption{(Color online) (Color online) Newton-like dynamical evolution of the relative motion (RM) coordinates toward the stable state \textbf{s}. Panels (a)-(c) present the relaxation process: (a) Time traces of $m_{z}$; (b) and (c) phase portraits of the relative coordinates $(\Phi_{-12}, P_{-12})$ and $(\Phi_{-23}, P_{-23})$, respectively. The trajectories illustrate how the internal degrees of freedom, following the lifting of the exchange-degenerate manifold by anisotropy, undergo oscillatory decay toward the equilibrium state \textbf{s}. This localized Newton-like model provides a simplified mechanical description of the RM coordinate's convergence.}
\label{Newton-like_RM}
\end{figure*}

Combined with the TVM model for the CM variable, that is, using the transformation Eq. (\ref{transphi_to_Phi_dot}), the complete Newton-like equation of motion for each spin expressed in the configuration space can thereby be obtained. By the way, we would like to point out that through the transformations Eqs. (\ref{inverssphi_to_Phi}) and (\ref{inversep_to_P}), Eq. (\ref{RM_conserved_Newton}) can also ensure the fact that the tiny perturbations of $M_{\mathrm{x(y)}}$ away from zero driven by the OPA will still be damped out into disappearance, see Fig. \ref{Newton-like_RM}, which is in good agreement with the macrospin simulated result (see Fig. \ref{Longtermdecay}(f)).

\section{\label{appd} Generalization of the TVM Model from a Single Lattice to Multiple Lattices}
\begin{figure}
\begin{center}
\includegraphics[width=7cm]{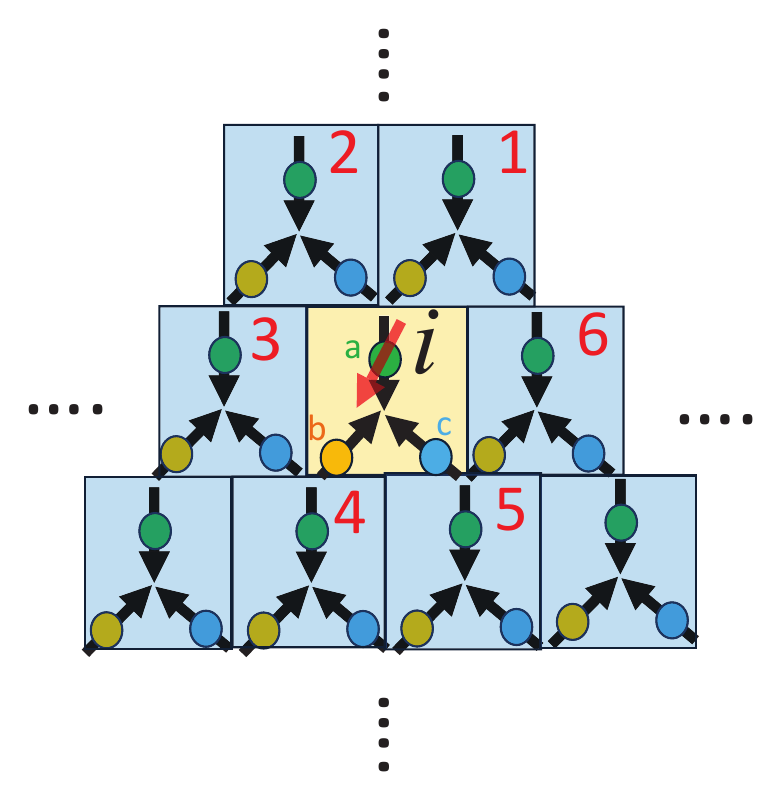}
\end{center}
\caption{(Color online) Schematic hexagonal spin structure of $\text{M}_{\mathrm{n}3}\text{Sn}$ in the (111) kagome plane. This figure schematically depicts the hexagonal spin structure of the $\text{M}_{\mathrm{n}3}\text{Sn}$ system in the (111) \textbf{kagome plane}. \textbf{I. Atomic and Sublattice Structure}. \textbf{Atoms}: Spin-carrying $\text{Mn}$ atoms are located at the corners of each hexagon, while $\text{Sn}$ atoms are at the center but are not drawn here as they carry no spin. \textbf{Lattice Definition}: \textbf{Light blue boxes} mark individual lattices, each containing the three required sublattice spins. \textbf{Sublattices}: The three sublattices are indicated by circles colored \textbf{green (a)}, \textbf{yellow (b)}, and \textbf{blue (c)}, respectively. \textbf{II. Spin Arrangement and Dynamics}. \textbf{Ground State}: Black arrows indicate a possible periodic arrangement of the spins. \textbf{Local Fluctuation}: The \textbf{transparent red arrow} denotes a tiny and local fluctuation appearing in sublattice spin a of the $i$-th lattice.}
\label{Perodic}
\end{figure}

To bridge the gap between single-lattice dynamics and atomic-scale simulations-where localized spins are governed by the LLG equation-we demonstrate the generalization of the TVM model from an isolated three-spin lattice to a system comprising multiple lattices. For simplicity, we consider a 2D large-scale system containing an extensive number of lattices, specifically the stationary hexagonal spin structure of $\mathrm{M}\mathrm{n}_{3}\mathrm{S}\mathrm{n}$ in the infinitely extended (111) kagome plane (see Fig. \ref{Perodic}) \cite{Hu2024}. This approach is equally applicable to more complex 3D systems. As shown in Fig. \ref{Perodic}, we focus on the magnetic $\mathrm{M}\mathrm{n}$ atoms, where each lattice (denoted by a square box) consists of three sub-lattices (a, b, c) forming a regular triangular arrangement.

Considering only the nearest-neighbor (NN) AFM and next-nearest-neighbor (NNN) FM exchange couplings \cite{Hu2024}, the local Hamiltonian for the sub-lattices in the $i$-th lattice (highlighted by the light yellow box in Fig. \ref{Perodic}) is given by:
\setlength\abovedisplayskip{6pt}
\setlength\belowdisplayskip{6pt}
\begin{eqnarray}
H_{i,\mathrm{a}} &=& A_{\mathrm{ex}}^{\mathrm{(A)}}(\mathbf{m}_{i\mathrm{a}} \cdot \mathbf{m}_{i\mathrm{b}} + \mathbf{m}_{i\mathrm{a}} \cdot \mathbf{m}_{i\mathrm{c}} + \mathbf{m}_{i\mathrm{a}} \cdot \mathbf{m}_{1\mathrm{b}}\nonumber\\
&&+ \mathbf{m}_{i\mathrm{a}} \cdot \mathbf{m}_{2\mathrm{c}}) + A_{\mathrm{ex}}^{\mathrm{(F)}} \sum_{j=1}^{6} \mathbf{m}_{i\mathrm{a}} \cdot \mathbf{m}_{j\mathrm{a}} + H_{\mathrm{ua}}.\nonumber\\
\label{H_periodic}
\end{eqnarray}
(and similarly for $H_{i,\mathrm{b}}$ and $H_{i,\mathrm{c}}$), where the indices 1-6 denote adjacent lattices. Under a periodic constraint where $m_{i\mathrm{a(b)(c)}} = m_{j\mathrm{a(b)(c)}}$, the total Hamiltonian for the $i$-th lattice reduces to:
\setlength\abovedisplayskip{6pt}
\setlength\belowdisplayskip{6pt}
\begin{eqnarray}
H_{i} &=& 2 A_{\mathrm{ex}}^{\mathrm{(A)}}(\mathbf{m}_{i\mathrm{a}} \cdot \mathbf{m}_{i\mathrm{b}} + \mathbf{m}_{i\mathrm{a}} \cdot \mathbf{m}_{i\mathrm{c}}+ \mathbf{m}_{i\mathrm{b}} \cdot \mathbf{m}_{i\mathrm{c}})\nonumber\\
&& + 6 A_{\mathrm{ex}}^{\mathrm{(F)}} (|\mathbf{m}_{i\mathrm{a}}|^2 + |\mathbf{m}_{i\mathrm{b}}|^2 + |\mathbf{m}_{i\mathrm{c}}|^2) + H_{\mathrm{u}i},\nonumber\\
&=& A_{\mathrm{ex}}^{\mathrm{(A)}}(|\mathbf{M}_{i}|^2 - 3) + 18 A_{\mathrm{ex}}^{\mathrm{(F)}} + H_{\mathrm{u}i},
\label{H_peri_reduce}
\end{eqnarray}
where $\mathbf{M}_i$ is the total magnetization of the $i$-th lattice.

Comparing this with Eq. (\ref{H_ex_M}), the Hamiltonian remains functionally identical, with the AFM coupling coefficient effectively doubled due to contributions from neighboring lattices. This implies that under periodic constraints, the conserved dynamic states remain RBP states. Thus, all theoretical analyses derived for the three-macrospin system remain valid for a large-scale lattice system.

A critical question arises: what happens if the periodic constraints are locally disturbed? Suppose a spin $\mathbf{m}_{i\mathrm{a}}$ deviates slightly from the periodic configuration, such that $\mathbf{m}_{i\mathrm{a}} = \mathbf{m}_{j\mathrm{a}} + \delta \mathbf{m}_{i\mathrm{a}}$ (with $|\delta \mathbf{m}_{i\mathrm{a}}| \ll 1$ and $\delta \mathbf{m}_{i\mathrm{a}} \perp \mathbf{m}_{j\mathrm{a}}$), as indicated by the red arrow in Fig. (\ref{Perodic}). Neglecting $H_{i}$ for simplicity, this fluctuation introduces an extra term $\delta H_{i} = 2 A_{\mathrm{ex}}^{\mathrm{(A)}} [\delta \mathbf{m}_{i\mathrm{a}} \cdot (\mathbf{m}_{i\mathrm{b}} + \mathbf{m}_{i\mathrm{c}})]$. This term acts as a generator that triggers an RM mode (dephasing) because the Poisson brackets $[\mathbf{m}_{i\mathrm{a}}, \delta H_{i}]$ and $[\mathbf{m}_{i\mathrm{b(c)}},\delta H_{i}]$ no longer vanish. This local fluctuation propagates through the exchange coupling, analogous to a localized wave packet described by the Schr\"{o}dinger equation.

However, under the combined effect of SOT and damping, these local excitations (RM modes) are systematically suppressed. As analyzed via the time-dependent RBP/RBUT approach in Sec. \ref{sec:2C}, the system's dissipation mechanisms act to eliminate these non-uniformities, restoring the entire system to the pure RBUT state (the state \textbf{s}). Consequently, the TVM model, originally derived for CM variables, can be robustly generalized to govern the collective motion of the entire lattice system.

To self-consistently extend the TVM framework to the micromagnetic scale—particularly in scenarios where SOT excitation may deviate from the standard 120$^{\circ}$ configuration (state $\mathbf{s}$) due to the absence of uniaxial anisotropy—we adopt a refined coarse-graining strategy. Instead of directly reducing the entire three-spin lattice into a single collective coordinate as in the preceding sections, we preserve the internal degrees of freedom by maintaining periodic boundary conditions (PBC) for each individual sub-lattice species. Specifically, the spins corresponding to the same sub-lattice index across multiple unit cells are first reduced to their respective center-of-mass (CM) coordinates. Consequently, a single micromagnetic cell is represented as a composite assembly of three distinct sub-lattice CM variables. This formulation ensures that the high-frequency relative motions between sub-lattices, inherently present at the atomic scale, are rigorously preserved within the macroscopic micromagnetic framework.

Furthermore, within a singular micromagnetic cell, the system Hamiltonian incorporates the nearest-neighbor exchange interactions between sub-lattice spins, all of which strictly adhere to the aforementioned periodic constraints. Other long-range interactions, including the Dzyaloshinskii-Moriya interaction (DMI) \cite{Yamane2019,Wu2024}, are significantly weaker than the exchange coupling between adjacent lattices; thus, they are insufficient to disrupt the periodic symmetry and effectively become diluted by the ensemble of spin lattices within the cell. Consequently, these higher-order interactions can be elegantly expressed using the collective CM coordinates of the sub-lattices. This hierarchical mapping provides the flexibility to address macroscopic phenomena, such as dynamics in poly-crystalline grains or magnetic domains. Specifically, if the macroscopic SOT excitation preserves the symmetric 120$^{\circ}$ state ($\mathbf{s}$), the framework allows for a secondary reduction, wherein the TVM model can be derived using a singular collective coordinate at an even larger scale. This multi-layered analytical approach establishes a robust foundation for modeling complex, large-scale NC-AFM devices.

\nocite{*}

\bibliography{HH_phaselocking}
\end{document}